\documentclass[a4paper,fleqn,usenatbib]{mnras}

\usepackage{newtxtext,newtxmath}

\usepackage[T1]{fontenc}
\usepackage{ae,aecompl}

\usepackage{graphicx}	
\usepackage{amsmath}	
\usepackage{pdflscape}	
\usepackage{multicol}
\usepackage{afterpage}
\usepackage{color}
\usepackage{etoolbox}
\usepackage{hyperref}
\usepackage{booktabs}
\usepackage[utf8]{inputenc}
\usepackage{ulem}

\definecolor{darkmagenta}{rgb}{0.5, 0, 0.5}
\definecolor{darkgreen}{rgb}{0, 0.6, 0.05}
\definecolor{darkred}{rgb}{0.86,0.078,0.235}

\title[Asymmetry of Lensed Arcs]{Dents in the Mirror: A Novel Probe of Dark Matter Substructure in Galaxy Clusters from the Astrometric Asymmetry of Lensed Arcs}

\author[Derek Perera et al.]{Derek Perera$^{1}$\thanks{E-mail: perer030@umn.edu}, Daniel Gilman$^{2}$, 
Liliya L. R. Williams$^{1}$, Liang Dai$^{3}$, Xiaolong Du$^{4}$, 
\newauthor{Gregor Rihtar$\check{\rm s}$i$\check{\rm c}$$^{5}$, Joaquin Becerra-Espinoza$^{6}$, Allison Keen$^{1}$}
\\
$^{1}$School of Physics and Astronomy, University of Minnesota, Minneapolis, MN, 55455, USA. \\
$^{2}$Department of Astronomy and Astrophysics, University of Chicago, Chicago, IL 60637, USA. \\
$^{3}$Department of Physics, University of California, 366 Physics North MC 7300, Berkeley, CA 94720, USA. \\
$^{4}$Department of Physics and Astronomy, University of California, Los Angeles, CA 90095, USA. \\
$^{5}$Faculty of Mathematics and Physics, Jadranska ulica 19, SI-1000 Ljubljana, Slovenia. \\
$^{6}$Department of Physics, University of California at Santa Barbara, Santa Barbara, CA 93106, USA. \\
}
\date{Accepted XXX. Received YYY; in original form ZZZ}

\pubyear{2025}

\begin{document}
\label{firstpage}
\pagerange{\pageref{firstpage}--\pageref{lastpage}}
\maketitle

\begin{abstract}
    Astrometric perturbations of lensed arcs behind galaxy clusters have been recently suggested as promising probes of small-scale ($\lesssim10^9 M_{\odot}$) dark matter substructure. Populations of cold dark matter (CDM) subhalos, predicted in hierarchical structure formation theory, can break the symmetry of arcs near the critical curve, leading to positional shifts in the observed images. We present a novel statistical method to constrain the average subhalo mass fraction ($f_{\rm sub}$) in clusters that takes advantage of this induced positional asymmetry. Focusing on CDM, we extend a recent semi-analytic model of subhalo tidal evolution to accurately simulate realistic subhalos within a cluster-scale host. We simulate the asymmetry of lensed arcs from these subhalo populations using Approximate Bayesian Computation. Using mock data, we demonstrate that our method can reliably recover the simulated $f_{\rm sub}$ to within 68\% CI in 73\% of cases, regardless of the lens model, astrometric precision, and image morphology. We show that the constraining power of our method is optimized for larger samples of well observed arcs, ideal for recent JWST observations of cluster lenses. As a preliminary test, we apply our method to the MACSJ0416 Warhol arc and AS1063 System 1. For Warhol we constrain the upper limit on $\log f_{\rm sub} < -3.48^{+1.00}_{-0.91}$, while for AS1063 System 1 we constrain $\log f_{\rm sub} = -2.44^{+0.61}_{-0.86}$ (both at 68\% CI), consistent with CDM predictions. We elaborate on our method's limitations and its future potential to place stringent constraints on dark matter properties in cluster environments.
\end{abstract}



\section{Introduction}

Current observations of the cosmic microwave background strongly support the theory that dark matter is cold and collisionless \citep{planck20}. In this standard cosmological model, cold dark matter (CDM) halos form hierarchically at all length scales of cosmological significance \citep{davis85,klypin99,moore99} and their density profiles can be universally described by the Navarro-Frenk-White (NFW) profile \citep{nfw97}. Additionally, a universal subhalo mass function (SHMF) is expected to describe CDM halos at all scales \citep{giocoli08}. This is particularly useful in studying small-scale CDM substructures within a main host halo, hereafter referred to as subhalos. CDM subhalos are predicted in the standard CDM paradigm, however this paradigm fails to explain some small-scale observations of galaxies \citep[see for review e.g.][]{delpopolo17,bullock17,sales22}. As a result, a new frontier for astrophysical probes of dark matter is the search for populations of small-scale ($\lesssim10^9 M_{\odot}$) CDM subhalos, the detection of which can help to determine the true nature of dark matter.

One such unique probe of dark matter is strong gravitational lensing, which is ideal due to its sensitivity to the gravitational potential of mass structures at all scales. Using galaxy-scale lenses, dark matter substructures have been prolifically studied, with numerous constraints having been made on dark matter properties using subhalos \citep{dalal02,vegetti14,despali17,birrer17,hsueh20,gilman20,he22,gilman24,keeley24}. For galaxy cluster-scale lenses, dark matter constraints from subhalos are much more sparse, mostly restricted to intermediate mass ($\sim 10^{11} M_{\odot})$ halos \citep{natarajan04,natarajan17} and the distribution of subhalos \citep{umetsu16}. Recently, the discovery of numerous microlensed individual stars with the Hubble Space Telescope (HST) \citep{kelly22} and James Webb Space Telescope (JWST) \citep{windhorst23,yan23,fudamoto2025} have extended the search for dark matter subhalos in galaxy clusters by providing high resolution observational data on sub-arcsecond angular scales. As a result, new methods to probe subhalos or other small-scale dark matter structures in galaxy clusters have been recently developed \citep{venumadhav17,dai18,dai20a,dai20c,williams24,broadhurst25,palencia25}. In particular, optically faint or invisible CDM subhalos may be responsible for the puzzling aspects of some high-$z$ magnified sources~\citep{diego22b,ji25,pascale25}.

Many of these new methods focus on CDM subhalo perturbations to the cluster's critical curve. Given that the critical curve indicates the region of highest magnification in the gravitational lens, the observational signatures of these CDM subhalos are primarily flux ratio anomalies within the images \citep{dai20a,ji25}. This is a similar concept as is done on galaxy scales with gravitational imaging \citep{koopmans05,vegetti09}, a method that has been successful at constraining the CDM subhalo SHMF \citep{vegetti18,ritondale19}. At cluster-scales, an underappreciated effect of these perturbations is the astrometric shifts of the lensed image positions \citep{dai18,abe23}. Any detection of these positional shifts of images would be evidence ruling out a smooth density profile on small scales, as is commonly implied from lens models \citep{limousin22}. The difficulty lies in disentangling between astrometric uncertainty from the observed image positions and a genuine perturbation. Additionally, disentangling whether a more complex lens model can account for an observed image shift, rather than a local perturbation, poses another challenge. The former can be addressed with new high resolution data from JWST, while the latter can be accounted for with a statistical or model agnostic approach.

An important and necessary prerequisite for these types of studies is the development of sophisticated models of CDM subhalos. The physical interaction of CDM subhalos within their host leads to strong tidal stripping effects \citep{hayashi03}. This tidal evolution is dependent on the host halo mass and the trajectory of the orbiting subhalo, thus making models of the process statistical in nature \citep{kravtsov04,han16}. This has motivated the creation of high resolution cosmological simulations of CDM to create precise theoretical predictions of subhalo density structure \citep{Springel2008}. These predictions allow for an ease of use of physically realistic CDM subhalo structure that can be quickly applied to a wide variety of dark matter probes \citep{taylor01,penarrubia08,benson12,du24}. 

In this paper, we present a novel method to investigate the astrometric shifts of images in cluster lenses. The fundamental goal of this work is to constrain the CDM subhalo SHMF based on the induced perturbations from populations of CDM subhalos near the lensing critical curve. To ease comparison with previous works, we focus on the mean subhalo mass fraction ($f_{\rm sub}$) as the primary parameter of interest, implicitly assuming that it is the same in both galaxy and cluster environments. This parameter can be directly calculated from the subhalo SHMF, and has been constrained in many studies using galaxy-scale lenses \citep{vegetti14,despali17,hsueh20,gilman20}, but not with cluster-scale lenses. For our method, we utilize a known result from gravitational lensing in that sources that form near a fold caustic will form symmetric image pairs across the critical curve \citep{schneider92}. Thus, astrometric perturbations from a population of CDM subhalos lying in the lens plane will manifest as an asymmetry\footnote{The induced asymmetry from CDM subhalos is akin to ``Denting the mirror'', hence the title.} in the image positions across the critical curve~\citep{dai18}. We develop a methodology to infer the hidden subhalo population using a likelihood-free inference method that uses the asymmetry of lensed image pairs as a summary statistic. After validating the modeling framework on simulated datasets, we apply it to two lensed arcs in MACSJ0416 and AS1063 to infer the projected mass fraction in dark matter subhalos. In the future, the method demonstrated in this paper will be applied to larger samples of gravitational lenses to provide stringent constraints on $f_{\rm sub}$ from cluster lenses.

This paper is organized as follows: In Section \ref{txt:theory}, we provide a review of the necessary formalism of gravitational lensing, as well as an overview of the effect of CDM subhalos on small-scale lensing. In Section \ref{txt:methodology}, we carefully describe our new method. We describe the physical models used for CDM subhalos, along with an updated semi-analytic model for the subhalo tidal evolution. We also outline the statistical methodology to constrain $f_{\rm sub}$, which is implemented within the Approximate Bayesian Computation framework. In Section \ref{txt:mockresults} we demonstrate the efficacy of our method using mock lensed arcs. In Section \ref{txt:real} we apply our method to two well observed arcs and derive the first tentative constraints on $f_{\rm sub}$ with the method. In Section \ref{txt:conclusions} we discuss our results and the prospects for future work that can be done with our method.

Throughout this work, we assume a flat $\Lambda$CDM cosmology with $\Omega_M = 0.27$, $\Omega_{\Lambda} = 0.73$, and $H_0 = 70$ km s$^{-1}$ Mpc$^{-1}$.

\section{Gravitational Lensing near Critical Curves}\label{txt:theory}

\subsection{Gravitational Lensing Formalism}\label{txt:formalism} 

Here, we briefly review the strong gravitational lensing formalism necessary for this article. We refer the reader to various review articles \citep[e.g][]{blandford86,schneider92,narayan96} for additional details.

For this paper, as is commonly done in lens modelling studies, we use the thin lens approximation, where the 3D mass distribution of the lens $\rho(\boldsymbol{\theta},z)$ is approximated as a 2D projected surface mass density lying in the lens plane at redshift $z_d$:
\begin{equation}
    \Sigma\left(\boldsymbol{\theta}\right) = \int \rho(\boldsymbol{\theta},z) dz
\end{equation}
Here, $\boldsymbol{\theta}$ is the vector position within the lens plane, and $z$ is the line of sight distance. The lensing deflection angle can then be computed by integrating all the density contributions in the lens plane:
\begin{equation}
   \boldsymbol{\alpha}\left(\boldsymbol{\theta}\right) = \frac{4 D_dD_{ds}G}{c^2D_s}\int \frac{\left(\boldsymbol{\theta}-\boldsymbol{\theta}'\right)\Sigma\left(\boldsymbol{\theta}\right)}{|\boldsymbol{\theta}-\boldsymbol{\theta}'|^{2}} d\boldsymbol{\theta}',
    \label{eq:deflection}
\end{equation}
where $D_d$, $D_{ds}$, and $D_s$ are the angular diameter distances between the observer and lens, the lens and source, and the observer and source, respectively. We note that it is often computationally simpler to work with the scalar lensing potential $\psi(\boldsymbol{\theta})$, which obeys the lensing Poisson equation:
\begin{equation}
    \nabla^2\psi(\boldsymbol{\theta}) = 2\,\frac{\Sigma(\boldsymbol{\theta})}{\Sigma_{\rm crit}}=2\kappa(\boldsymbol{\theta}),
    \label{eq:lenspoission}
\end{equation}
where $\Sigma_{\rm crit} = (c^2/4\pi\,G)(D_s/D_{ds}\,D_d)$ is the critical surface density. The gradient of $\psi(\boldsymbol{\theta})$ gives the deflection angle (equation \ref{eq:deflection}). Likewise, the dimensionless quantity $\Sigma(\boldsymbol{\theta})/\Sigma_{\rm crit}$ is known as the convergence $\kappa(\boldsymbol{\theta})$. All of this can be summarized by the lensing equation:
\begin{equation}
    \boldsymbol{\beta} = \boldsymbol{\theta} - \boldsymbol{\alpha}(\boldsymbol{\theta}),
    \label{eq:lensequation}
\end{equation}
where $\boldsymbol{\beta}$ is the source position.

The last required ingredient is the lensing Jacobian matrix:
\begin{equation}
    \mathcal{A} = \left(\delta_{ij} - \frac{\partial^2\psi\left(\boldsymbol{\theta}\right)}{\partial\theta_i\partial\theta_j}\right).
    \label{eq:jacobian}
\end{equation}
The magnification $\mu$ can be simply calculated from $\mathcal{A}$:
\begin{equation}
    \mu\left(\boldsymbol{\theta}\right) = \frac{1}{\det\left(\mathcal{A}\right)}.
    \label{eq:magn}
\end{equation}
$\mu\left(\boldsymbol{\theta}\right)$ tends to $\infty$ in the limit that $\det\left(\mathcal{A}\right)$ approaches 0. This is a consequence of the geometric optics approximation, and in reality infinite magnification is never reached. The corresponding lens plane positions $\boldsymbol{\theta}$ at this limit trace out the critical curve. Projecting these positions to the source plane using equation \ref{eq:lensequation} gives the positions of the caustic. For most gravitational lenses, the caustic (corresponding to the tangential caustic in this case) shape resembles a diamond, with vertices termed ``cusps", and smooth sides termed ``folds". For galaxy cluster lenses, the complete shape of the caustic is often more complex than this; however, on large scales the diamond shape is preserved.

\subsection{Lensing Near the Critical Curve}\label{txt:cclenstheory}

\begin{figure}
    \centering
    \includegraphics[trim={5.1cm 0.35cm 5.1cm 0.35cm},clip,width=0.49\textwidth]{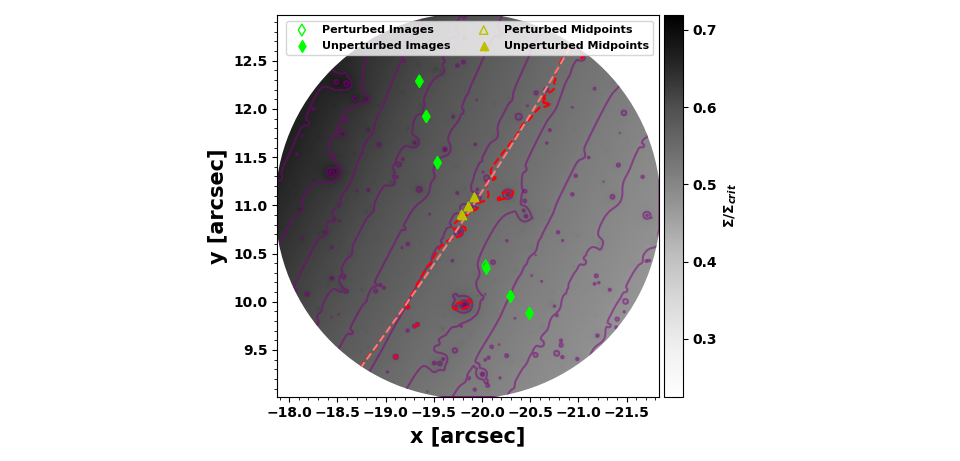}
    \includegraphics[trim={5.1cm 0.35cm 5.1cm 0.35cm},clip,width=0.49\textwidth]{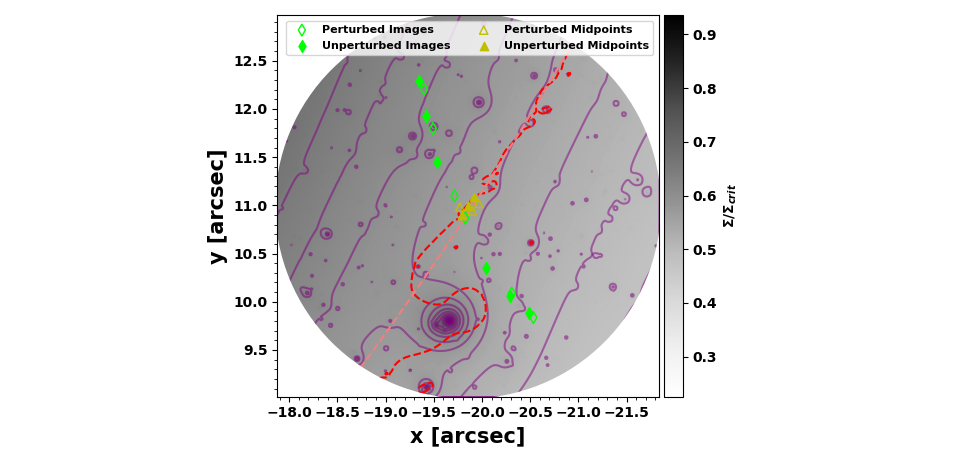}
\caption{Example realizations of dark matter subhalo populations sampled with the same $f_{\rm sub}$. This illustrates the scatter of the asymmetry from realizations sampled with the same $f_{\rm sub}$. The circular window is the 2" aperture that the SHMF is sampled within surrounding the perpendicular arc in Figure \ref{fig:perpendiculararc}. The initial unperturbed image positions and midpoints are shown as closed green diamonds and closed yellow triangles respectively. The perturbed image positions and midpoints from the dark matter subhalo population are shown as open green diamonds and open yellow triangles respectively. The unperturbed and perturbed critical curve are shown as dashed light red and dashed bright red lines. Concentrations of matter represent evolved dark matter subhalos. Purple contours trace the density profile, with subhalos easily visible. {\it Top:} The presented realization is an example arc with low asymmetry, as can be seen by the low displacement of the perturbed midpoints. {\it Bottom:} The presented realization is an example arc with high asymmetry, as can be seen by the large displacement of the perturbed midpoints. Both realizations are made with $\log \Sigma_{\rm sub} = -1.5$ and $\log f_{\rm bound} = -1.0$, thus having $\log f_{\rm sub} = -2.6$ (see Section \ref{txt:methodology}). The measured asymmetry metric (see Section \ref{txt:asymmetry}) is $\xi = -2.43$ and $\xi = -0.17$ for the top and bottom panels, respectively.}  
\label{fig:asymrealizations}
\end{figure}

In this section, we derive a property of lensed images near a cluster critical curve that we will use to probe dark matter substructure. In particular, we show that the midpoints of image pairs will collect along a straight line, unless there are perturbations to the lens model on angular scales smaller than the image separation. The top and bottom panels of Figure \ref{fig:asymrealizations} illustrate this effect for two regions of a strong lensing cluster with dark subhalos injected near the critical curve. In the presence of small-scale perturbations by cluster subhalos, the midpoints (yellow triangles) of lensed images (green diamonds) deviate from a straight line. 


Cluster lens modeling analyses often model extended lensed arcs using identified counterimaged ``knots'' within the arc as constraints. Modeling the relative positions of lensed knots improves the cluster lens model by constraining the deflection field on angular scales comparable to the image separation \citep[e.g.][]{bergamini23,perera24b}. These knots likely correspond to bright sub-galactic structures within the source galaxy \citep[e.g. HII regions in the case of MACSJ1149][]{hwilliams24}. Due to these knots forming on sub-arcsecond scales, they are treated as point images with lower positional uncertainty than a typical cluster scale image. For this paper, we represent lensed arcs as a collection of lensed point image knots across a critical curve.

Under this setup, the addition of dark matter substructure in the lens plane near the critical curve will perturb both the cluster-scale critical curve and the point image knots. The strength of this perturbation is dependent on a multitude of factors, such as the amount of substructure, the density profile of each subhalo, and the relative positions of subhalos. As a result, the degree of perturbation to the image positions of lensed knots is a highly stochastic variable. As an example, the deviation from a straight line among lensed images is more pronounced for the bottom panel than the top panel in Figure \ref{fig:asymrealizations}, despite both examples having the same projected mass in subhalos (i.e. same $f_{\rm sub}$) though different realizations. 

The prediction that image pairs collect along a straight line is a well known result in gravitational lensing near a fold caustic \citep{blandford86,schneider92}. We derive this property below. We begin with the lensing Jacobian as defined in equation \ref{eq:jacobian}. It is more useful to write the Jacobian in terms of the convergence and shear tensor components ($\gamma_1$ and $\gamma_2$):
\begin{equation}
    \mathcal{A}\left(\boldsymbol{\theta}\right) = \left( \begin{array}{cc} 1 - \kappa(\boldsymbol{\theta}) - \gamma_1(\boldsymbol{\theta}) & -\gamma_2(\boldsymbol{\theta}) \\
 -\gamma_2(\boldsymbol{\theta}) & 1 - \kappa(\boldsymbol{\theta}) + \gamma_1(\boldsymbol{\theta}) \end{array} \right)
\end{equation}
Where the eigenvalues are $1-\kappa \mp \gamma$, with $\gamma = \sqrt{\gamma_1^2 + \gamma_2^2}$. Recall that from equation \ref{eq:magn}, the lens plane positions $\boldsymbol{\theta}$ where $\det\mathcal{A} = 0$ trace out the critical curve. Therefore, it is possible to rewrite $\mathcal{A}$ in a coordinate system where the origin is located somewhere on the critical curve, and the axes correspond to the principal directions of $\mathcal{A}$. For clarity, let's define the two principal directions: the tangential direction being tangent to the critical curve at any position, and the critical direction being perpendicular to the critical curve at any position.

Since we are in this principal coordinate system, $\mathcal{A}$ is diagonal, so $\gamma_2 = 0$. Likewise, the first eigenvalue of $\mathcal{A}$ disappears such that $1-\kappa-\gamma_1 = 0$. Importantly, this is the eigenvalue corresponding to the critical direction.  This expression can be simplified if we consider values very close to the critical curve (i.e. close to the origin of the principal coordinates). To do so, let's do a Taylor expansion such that:
\begin{equation}
    \kappa(\boldsymbol{\theta}) = \kappa_0 + \boldsymbol{\theta} \cdot \nabla \kappa(\boldsymbol{\theta}=0)
\end{equation}
and
\begin{equation}
      \gamma_1(\boldsymbol{\theta}) = \gamma_{1,0} + \boldsymbol{\theta} \cdot \nabla \gamma_1(\boldsymbol{\theta}=0)  
\end{equation}
where $\kappa_0$ and $\gamma_{1,0}$ are $\kappa$ and $\gamma_1$ evaluated at the origin, respectively. Under this expansion, we can show that the eigenvalues of $\mathcal{A}$ become:
\begin{equation}
    1 - \kappa - \gamma_1 = -\boldsymbol{\theta} \cdot \left(\nabla \kappa(\boldsymbol{\theta}=0) + \nabla \gamma_1(\boldsymbol{\theta}=0)\right)
\end{equation}
and
\begin{equation}
    1 - \kappa + \gamma_1 = 2\left(1-\kappa_0\right) - \boldsymbol{\theta} \cdot \left(\nabla \kappa(\boldsymbol{\theta}=0) - \nabla \gamma_1(\boldsymbol{\theta}=0)\right)
\end{equation}

Since $\mathcal{A}$ is diagonal, $\det\mathcal{A}$ is the product of these two expressions, which to first order can be written as:
\begin{equation}
    \det\mathcal{A} = 2\left(1-\kappa_0\right)\left[\boldsymbol{\theta} \cdot \left(-\nabla \kappa(\boldsymbol{\theta}=0) - \nabla \gamma_1(\boldsymbol{\theta}=0)\right)\right] + \mathcal{O}(2)
\end{equation}
Therefore, when close to the critical curve, the critical curve positions must satisfy $\boldsymbol{\theta} \cdot \left(-\nabla \kappa(\boldsymbol{\theta}=0) - \nabla \gamma_1(\boldsymbol{\theta}=0)\right) = 0$. Noticing that the quantity $-\nabla \kappa(\boldsymbol{\theta}=0) - \nabla \gamma_1(\boldsymbol{\theta}=0)$ is simply the gradient of the critical direction eigenvalue, this shows that locally the critical curve can be approximated as a straight line. Importantly, this is an approximation only in the immediate vicinity of the critical curve. In Appendix \ref{txt:lineartest} we empirically test the limits of this approximation, on scales relevant for this paper.

With the critical curve linearity established, we now show from the fold catastrophe that sources forming near the caustic fold produce images symmetric across the critical curve. From equation \ref{eq:lensequation}, we know that the source position $\boldsymbol{\beta}$ depends on the lens plane position. Let's consider a source that forms images very close to the critical curve, such that the image position can be approximated as a small displacement $\delta\boldsymbol{\theta}$ from the critical curve position $\boldsymbol{\theta_c}$. Once again, this applies only in the limit where images are very close to the critical curve, so we invoke a Taylor expansion for the source position:
\begin{equation}
    \boldsymbol{\beta_i}\left(\boldsymbol{\theta_c} +\delta\boldsymbol{\theta}\right) = \boldsymbol{\beta_i}(\boldsymbol{\theta_c}) + \sum_j \frac{\partial\boldsymbol{\beta_i}}{\partial\boldsymbol{\theta_j}}\bigg\rvert_{\boldsymbol{\theta_c}}\delta\boldsymbol{\theta_j} + \frac{1}{2}\sum_{j,k}\frac{\partial^2\boldsymbol{\beta_i}}{\partial\boldsymbol{\theta}_j\partial\boldsymbol{\theta}_k}\bigg\rvert_{\boldsymbol{\theta_c}}\delta\boldsymbol{\theta_j}\delta\boldsymbol{\theta_k}\label{eq:order2expansion}
\end{equation}
Usefully, $\mathcal{A}(\boldsymbol{\theta}) =\partial\boldsymbol{\beta}/\partial\boldsymbol{\theta}$. Since we are close to the critical curve, we can use the same approximations as before, namely that the origin is located at the critical curve position ($\boldsymbol{\theta_c}=0$). We define the components of the displacement $\delta\boldsymbol{\theta} = (\delta\theta_1,\delta\theta_2)$ and source position $\boldsymbol{\beta} = (\beta_1,\beta_2)$. The subscripts 1 and 2 indicate the tangential and critical directions in the principal coordinate system\footnote{These subscripts are not the same as those of the shear tensor components. $\gamma_{1,0}$ in this coordinate system does not change to $\gamma_{2,0}$ when calculating $\beta_2$. Instead, the eigenvalue is what changes in this coordinate frame, where $1-\kappa-\gamma_{1,0} = 0$. }. The leading order terms can be written as:
\begin{equation}
    \beta_1 = (1 - \kappa_0 + \gamma_{1,0})\delta\theta_1
    \label{eq:beta1}
\end{equation}
and
\begin{equation}
    \beta_2 = \frac{1}{2}\frac{\partial^2\beta_2}{\partial\theta_2^2}\bigg\rvert_{0}\delta\theta_2^2
    \label{eq:beta2}
\end{equation}
This is the general definition of the fold catastrophe. Thus, the two image positions relative to the critical curve position at $\boldsymbol{\theta_c} = 0$ can be easily solved for as:
\begin{equation}
    \delta\theta_1 = \frac{\beta_1}{1-\kappa_0+\gamma_{1,0}}
\end{equation}
and 
\begin{equation}
    \delta\theta_2 = \pm\sqrt{2\beta_2\left(\frac{\partial^2\beta_2}{\partial\theta_2^2}\bigg\rvert_0 \right)^{-1}}\label{eq:midpointposition}
\end{equation}
Since $\delta\theta_2$ is the critical direction (perpendicular to the critical curve), this shows that the formed images near the caustic fold are in fact symmetric on either side. As a consequence, the midpoints of these images will form at the critical curve. It should be noted that including higher order terms in equation \ref{eq:order2expansion} will cause the image midpoints to shift slightly off the critical curve. We demonstrate that this additional shift is negligible in most cases in Appendix \ref{txt:higherorder}.

The focus of this paper is to statistically model the perturbations of lensed arcs away from this symmetry caused by dark matter subhalos to constrain properties of dark matter. In the following section, we describe the full statistical inference framework used in this paper, along with a summary statistic that we use to quantify the degree of asymmetry along a curve of image midpoints.

\section{Methodology}\label{txt:methodology}

In this Section, we describe our analysis method to simulate asymmetric lensed arcs. The goal of this work is to constrain the subhalo mass fraction ($f_{\rm sub}$) from the degree of observed asymmetry in the midpoints of image pairs of lensed arcs. To do so, we develop a statistical analysis based on Approximate Bayesian Computation (ABC). The procedure can be broken down into four steps: (1) Simulate lensed arcs near the caustic fold from a background smooth cluster-scale lens profile (Section \ref{txt:clusterscale}), (2) Inject dark matter subhalos around the lensed images (Section \ref{txt:subbhalopop}), (3) Define a summary statistic to quantify the asymmetry of lensed arcs (Section \ref{txt:asymmetry}), and (4) Use Approximate Bayesian Computing (ABC) to evaluate the likelihood function and infer $f_{\rm sub}$ for a given lensed arc (Section \ref{txt:ABC}). For this work, we focus only on CDM subhalo populations in the lens plane, and do not include subhalos along the line of sight.

\subsection{Simulating Galaxy Cluster Lensed Arcs}\label{txt:clusterscale}

\subsubsection{Cluster Lens Profile}

\begin{figure}
    \centering
    \includegraphics[trim={5.1cm 0.35cm 5.1cm 0.35cm},clip,width=0.49\textwidth]{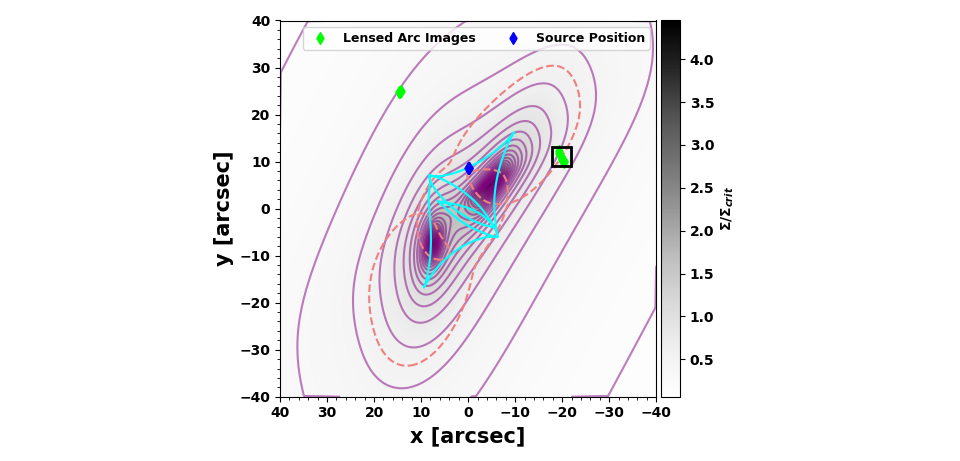}
\caption{The cluster lens surface mass density distribution used as the large scale macrolens for this work. The lens is made of 3 NSIEs whose properties are listed in Table \ref{tab:macrolens}. The morphology is designed to imitate a merging cluster. The lens is placed at a redshift $z_d = 0.25$. 3 sources (blue diamonds) are placed near to the caustic (cyan lines) folds at a redshift $z_s = 1$, forming 3 images (green diamonds) per source. 2 images per source form very close to the critical curve (light red dashed lines). The pairs of images that form near the critical curve simulate knots belonging to the same source galaxy, which imitates the common description of a lensed arc in lens modelling. The black box denotes the window in which we simulate subhalo populations. }  
\label{fig:mainlens}
\end{figure}

To demonstrate the methodology presented in this work, we create a catalog of simulated cluster lens systems. These systems are comprised of a macro cluster lens profile, plus subhalos. We model the macro cluster-scale lens profile utilizing 3 main halos at a lens redshift $z_d = 0.25$. Two of the halos are separated by 0.5" (1.97 kpc) such that their mass profile is effectively combined. This is done to mimic a slightly non-elliptical profile. The third profile is located a distance of 17.5" (68.8 kpc) from the other combined halos. Each halo is modelled as a Non-Singular Isothermal Ellipsoid (NSIE), which has an analytical lens potential \citep{hinshaw87,halkola06}:
\begin{equation}
    \psi\left(\boldsymbol{\theta}\right) = \frac{4\pi D_{ds}\sigma_v^2}{D_s c^2}\sqrt{R_c^2 + q\theta_1^2 + \frac{\theta_2^2}{q}}
\end{equation}
where $\sigma_v$ is the velocity dispersion, $R_c$ is the angular core radius\footnote{The physical core radius for the NSIE is equal to $D_d R_c$.}, $q$ is the axis ratio, and the vector position $\boldsymbol{\theta}$ has components $(\theta_1,\theta_2)$. The total mass of the cluster-scale lens profile is $1.3 \times 10^{14} M_{\odot}$. Large clusters such as this are the most efficient lenses akin to the Frontier Fields \citep{lotz17}, motivating our lens profile as reasonable and realistic. Properties of each NSIE halo are presented in Table \ref{tab:macrolens}. Figure \ref{fig:mainlens} shows the surface mass density of the cluster lens. From here, we refer to this profile as the fiducial macrolens.

\begin{table}
    \caption{Macrolens parameters for the 3 main cluster-scale NSIEs that make up our fiducial model. The density distribution is presented in Figure \ref{fig:mainlens}.}
    \centering
    \begin{tabular}{cccccc}
    \hline
        Profile & Position [x",y"] & $\sigma_v$ [km s$^{-1}$]  & $R_c$ [kpc] & ($q$,PA) \\
    \hline
        NSIE 1 & (-5,5) & 700 & 5.0 & (0.30,$30^{\circ}$) \\
        NSIE 2 & (-4.5,5) & 500 & 2.5 & (0.30,$50^{\circ}$)\\
        NSIE 3 &  (7.5,-7.5) & 650 & 3.3 & (0.40,$10^{\circ}$) \\
    \hline
    \end{tabular}
    \label{tab:macrolens}
\end{table}

\subsubsection{Lensed Arcs Near Caustic Folds}\label{txt:lensedarc}

\begin{figure}
    \centering
    \includegraphics[trim={5.1cm 0.35cm 5.1cm 0.35cm},clip,width=0.49\textwidth]{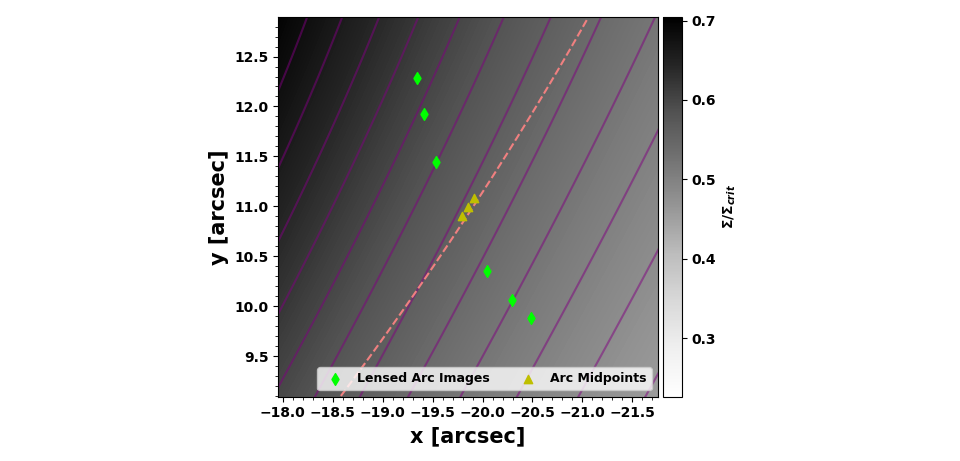}
    \includegraphics[trim={5.1cm 0.35cm 5.1cm 0.35cm},clip,width=0.49\textwidth]{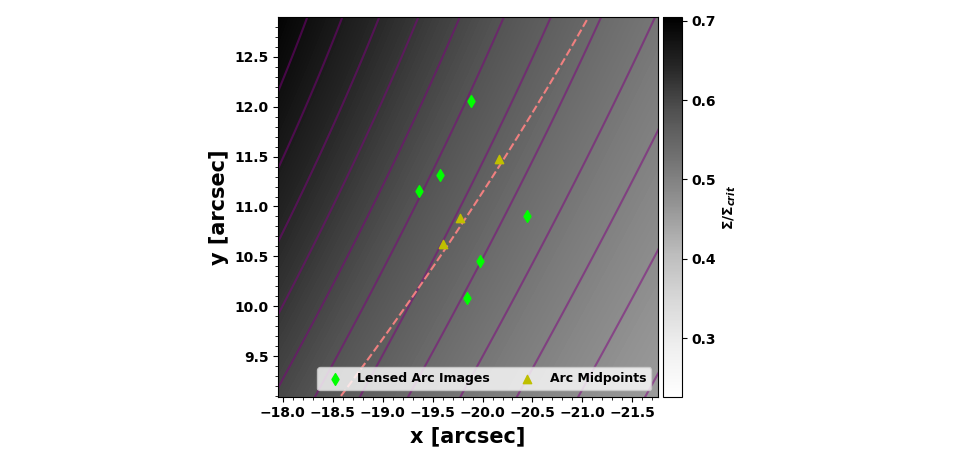}
\caption{ View of the Arc region highlighted by the black box in Figure \ref{fig:mainlens}. The 3 knots (green diamonds) are highlighted here, representing the knots in the same source galaxy that has been lensed across the critical curve (light red dashed line). As expected in lensing theory, the midpoints of each image pair (yellow triangles) form along the critical curve. Since the source is located on a caustic fold, the critical curve can be approximated as a straight line, and thus the midpoints form in a straight line. The window shown here is the region that we simulate subhalo populations. The top panel shows a perpendicular arc while the bottom panel shows a parallel arc. Since subhalo populations will affect these arcs differently, we treat them as independent cases.}  
\label{fig:perpendiculararc}
\end{figure}

In Section \ref{txt:cclenstheory} we described the general properties of lensed images near the critical curve. We also showed that sources that form near the caustic folds of cluster-scale gravitational lenses will appear as fully symmetric lensed arcs across a linear critical curve for a smooth lens profile \citep{blandford86,schneider92,dai18}. Here, we focus specifically on the corresponding approximations that we adopt.

The addition of dark matter substructures in the region within $\sim1-2$" of the critical curve will displace the smooth critical curve and can produce asymmetries in the arcs. It should be noted that the subhalo mass fraction is constrained to be on the order of $\sim1\%$ \citep{gilman19,dai20a, ji25}, but these constraints come mainly from galaxy-scale lens systems. The addition of dark matter subhalos will not cause lensing perturbations on scales beyond the $\sim1-2$" window. Therefore, we only simulate subhalos within the immediate vicinity of lensed arcs.

According to gravitational lensing theory, the midpoints of images that form near a smooth critical curve lie along the critical curve \citep{blandford86,venumadhav17,dai18} as we showed in Section \ref{txt:cclenstheory}. This symmetric setup can be broken with the inclusion of dark matter subhalos near the critical curve, as shown in Figure \ref{fig:asymrealizations}. Recall that images in these cases where the source is very near to the caustic manifest as extended arcs and are instead represented as lensed point image knots. Because of this, the image midpoints are easily calculated as the midpoints of the counterimaged knots. We note that this scenario where the midpoints lie along the critical curve is restricted to sources that form near to the caustic, and is not generally true for all source positions.

When considering angular scales significantly smaller than the radius of curvature of the macrolens near the caustic fold, the forward lens-projected (i.e. mapped from the source to lens plane) critical curve can be approximated as a straight line, and we expect unperturbed image midpoints to collect along this straight line. In real examples of arcs, this appears to be relatively common \citep[e.g.][]{kaurov19,chen19,diego23c}. In fact, this is an intuitively expected setup for lensed arcs, given that in clusters the caustic fold spans a significantly larger region than cusps. It should also be noted that this assumption regarding the image midpoints is commonly used as a proxy for the true location of the smooth critical curve \citep{kelly22,broadhurst25}, emphasizing the utility of this approximation.

It is important to note that there can be a degeneracy when evaluating lensed arcs along the critical curve. Deviations from a smooth fold can occur due to the presence of galaxy scale perturbers, such as cluster member galaxies \citep[e.g.][]{dai20a}. In this case, the midpoints will still lie along the critical curve, but the critical curve will no longer be a straight line. Likewise, large subhalos ($\gtrsim10^{10} M_{\odot}$) may also contribute, albeit rarely, to the variance in the critical curve linearity. Furthermore, sources forming slightly away from the caustic fold will produce images across an intrinsically curved smooth critical curve, thus not allowing the straight line critical curve approximation. As such, there exists a degeneracy between deviations from the symmetry of arcs stemming from either dark matter subhalos or lens model complexities. For this work, we restrict the analysis to lensed arcs that are unaffected by cluster members or large subhalos and are subject to only the smooth macrolens near a caustic fold. It turns out that this need not be a strict requirement, but we caution anyway that our results should only be applied to lensed arcs that are consistent with this description. We examine in greater detail the limits of this assumption in Appendix \ref{txt:lineartest}. In future works, these contributions away from linearity should be included in the statistical analysis.

Since the straight line critical curve assumption depends on the size of the lensed arc, we consider two simulated arcs in our analysis. We place 3 point sources (representing the knots) close to the caustic fold, forming 3 pairs of counterimages within 2" from the smooth, straight critical curve. The orientations of the arcs is either "perpendicular" or "parallel". Perpendicular arcs are oriented with the lensed images at increasing distances from the smooth critical curve, whose midpoints will span a relatively short distance along the critical curve. Parallel arcs are oriented with the lensed images at roughly the same distance from the smooth critical curve, whose midpoints will span a large distance along the critical curve. The setups are shown in Figure \ref{fig:perpendiculararc}. The effect of dark matter subhalo perturbations will be different for perpendicular and parallel arcs: perpendicular arcs are generally more susceptible to higher asymmetries. For our simulated arcs, the perpendicular and parallel cases span 0.23" and 1.02" along the critical curve, respectively.

The perpendicular and parallel arcs depicted in Figure \ref{fig:perpendiculararc} have a source redshift $z_s = 1.0$. These arcs, together with the lens redshift $z_d = 0.25$ and fiducial macrolens, form our fiducial model which we use to demonstrate the proof of concept of astrometric asymmetries resulting from dark matter subhalo populations. Our fiducial model represents the simplest possible version of a lensed arc (i.e. 3 observed image knots). We note that in practice it is necessary to roughly approximate the morphology of the observed arc of interest in order to derive realistic constraints on the underlying subhalo population. We elaborate more on this in Section \ref{txt:sampleresults}.

\subsection{Generating Populations of Dark Matter Subhalos}\label{txt:subbhalopop}

\subsubsection{Subhalo Mass Function}\label{txt:SHMF}

\begin{figure}
    \centering
    \includegraphics[trim={0cm 0.15cm 0cm 0.35cm},clip,width=0.49\textwidth]{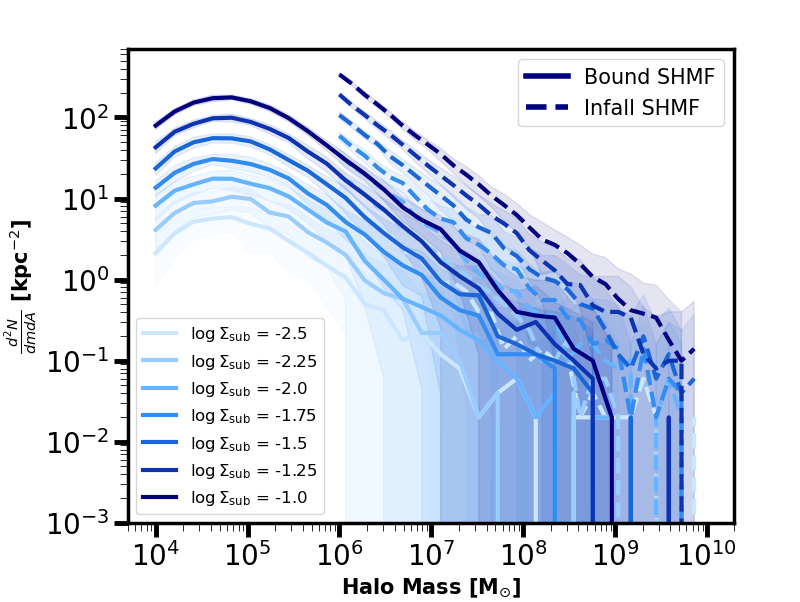}
\caption{The infall (dashed) and bound (solid) subhalo mass functions that we sample from at each realization for our simulation. For illustrative purposes, darker shades of blue correspond to increasing $\Sigma_{\rm sub}$. The infall SHMF is restricted to subhalos in the mass range $6 < \log (m/M_{\odot}) < 10$. The bound SHMF is calculated from the tidal stripping model presented in \protect\cite{du25}. Shaded regions indicate the $1\sigma$ scatter in sampling the SHMF. }  
\label{fig:SHMF}
\end{figure}

We use the open-souce software {\tt{pyHalo}}\footnote{https://github.com/dangilman/pyHalo} \citep{gilman20} to populate the region near lensed images with cluster subhalos distributed with a constant surface mass density near a critical curve. We draw cluster subhalo masses from the following SHMF:
\begin{equation}
    \frac{d^2N}{dm dA} = \frac{\Sigma_{\rm sub}}{m_0} \left(\frac{m}{m_0}\right)^{-\alpha}
\end{equation}
where $\Sigma_{\rm sub}$ is the SHMF normalization (in units of kpc$^{-2}$ throughout unless otherwise stated), $\alpha=1.9$ \citep{Springel2008,giocoli08}, and the pivot mass $m_0 = 10^8 M_{\odot}$. The SHMF represents the infall SHMF (thus $m$ is equal to the infall subhalo mass), which is the mass function of accreted halos prior to tidal evolution. The infall SHMF is predicted by CDM to be a universal property for subhalos \citep{giocoli08}, whereas the evolved SHMF (or equivalently the bound SHMF) depends on the cluster host properties and subhalo infall trajectories \citep{han16}. Figure \ref{fig:SHMF} shows the SHMF for various $\Sigma_{\rm sub}$. We only sample subhalos between the infall mass range $6 < \log (m/M_{\odot}) < 10$, consistent with previous works at cluster scales \citep{dai20a,williams24}. Subhalos above this range are rare, and likely would host a visible galaxy that we would identify as a cluster member. Below this range, halos produce negligible contributions to the deflection field \citep{dai18}. We examine how changes in this mass range would affect our results in Appendix \ref{txt:individualsubhalos}. Given the low redshifts of the clusters we consider, most perturbers are expected to be subhalos, and we leave the inclusion of line-of-sight halos to future work. 

After infall, subhalos lose mass to tidal stripping, and retain a fraction $f_{\rm{bound}}$ of their infall mass, with a final mass $m_{\rm{bound}} = f_{\rm{bound}} m$. In principle, a subhalo population will have a distribution of $f_{\rm{bound}}$, depending on individual subhalo orbits, concentrations, and infall times \citep{han16,du25}. Predicting the distribution of $f_{\rm{bound}}$ for a given cluster requires developing a tidal evolution model for cluster subhalos to account for these factors, which could follow the approach used for galaxy-scale lens systems \citep{du25}. 

Lacking a detailed tidal evolution model for cluster subhalos, we define $\bar{f}_{\rm bound}$ as the mean bound mass fraction for cluster subhalos. In our inference procedure, we treat the mean bound mass fraction $\bar{f}_{\rm bound}$ as a free parameter, alongside the normalization of the infall mass function $\Sigma_{\rm{sub}}$. We set the mean bound mass fraction with a log-uniform prior $\log\left(\bar{f}_{\rm bound}\right) \sim \mathcal{U}\left(-1.25,-0.75\right)$, comparable to the distribution of bound masses for a group-scale lens \citep{du25}. For each realization, the subhalos will have a normally distributed $f_{\rm bound}$, with mean $\bar{f}_{\rm bound}$ and standard deviation of each log-distribution to be 0.5. This corresponds to a mean bound mass fraction range between $\sim  5-20\%$. For $\Sigma_{\rm sub}$, we set a wide log-uniform prior $\log\left(\Sigma_{\rm sub}\right) \sim \mathcal{U}\left(-3.5,-0.1\right)$.

\subsubsection{Subhalo Density Profiles}


\begin{figure}
    \centering
    \includegraphics[trim={0cm 0cm 0cm 0cm},clip,width=0.49\textwidth]{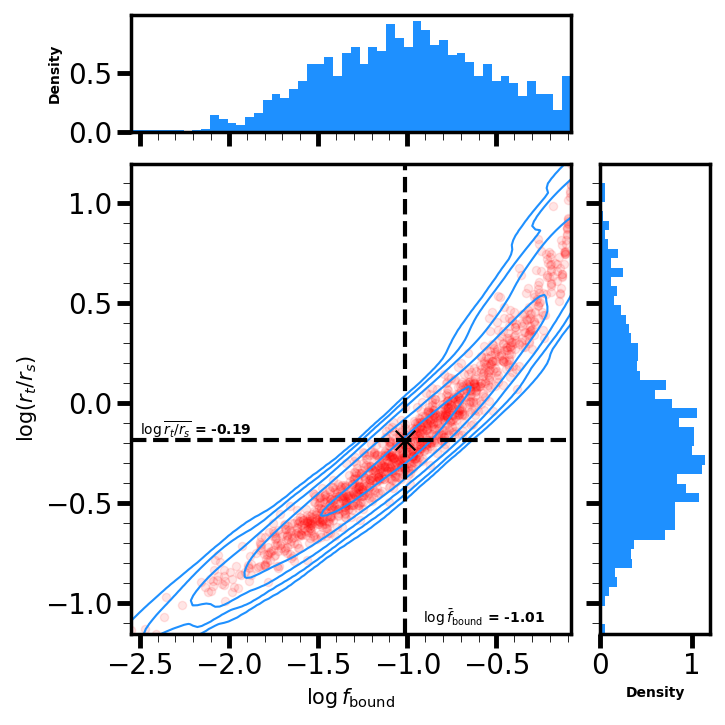}
    \includegraphics[trim={0cm 0cm 0cm 0cm},clip,width=0.49\textwidth]{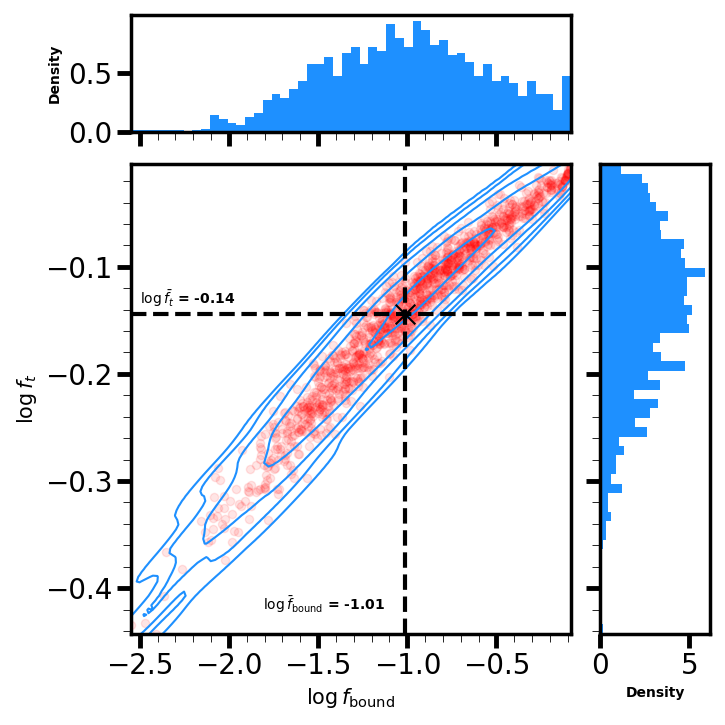}
\caption{ Summary plots of the tidal stripping process and how it manifests in the subhalo density profiles. For this example, the subhalo population is simulated to have a Gaussian bound mass fraction distribution: $\log f_{\rm bound} \sim \mathcal{N}\left(-1.0,0.5\right)$. We note that this mean bound mass fraction is the center of the log-uniform prior we set for $\bar{f}_{\rm bound}$. From this distribution, the tidal evolution tracks from \protect\cite{du25} are used to calculate the each subhalo's corresponding truncation radius $r_t$, and normalization $f_t$, with some scatter. These tidal tracks can be visualized as a tight relation between $f_{\rm bound}$ and $r_t$ ({\it Top}) and  $f_{\rm bound}$ and $f_t$ ({\it Bottom}). In both panels, blue contours are logarithmically spaced 2D density distributions for the parameters, red dots are an example subhalo population (for illustrative purposes $\log \Sigma_{\rm sub} = -1.0$ is shown, although the tidal tracks are the same for any value of $\Sigma_{\rm sub}$), and dashed lines are the means of the parameter distributions (top and right histograms).}  
\label{fig:subhalotidaltrack}
\end{figure}

Once the bound mass is known, and normally distributed according to the defined $f_{\rm bound}$ distribution described in Section \ref{txt:SHMF}, the tidal tracks \citep{errani21,du24} are used to calculate each subhalo's density profile. We simulate each subhalo as a tidally stripped NFW. For CDM, we model this as a truncated NFW:
\begin{equation}
    \rho_{CDM}(r) = \frac{\rho_s}{\frac{r}{r_s}\left(1 + \frac{r}{r_s}\right)^2}\frac{f_t}{1 + \left(\frac{r}{r_t}\right)^2}
\end{equation}
where $\rho_s$ is the scale density, $r_s$ is the scale radius, and $f_t$ and $r_t$ are truncation parameters determined by the bound mass fraction $f_{\rm bound}$ after tidal stripping \citep{du24}.

The effect of the tidal stripping process manifests in the density profile as an effective rescaling and truncation. The tidal tracks that govern this can be found in \cite{du24}, where our parametrization of the subhalo density profile matches their Nuker model with $\alpha=1$, $\beta=3$, $\gamma=1$, and $\delta=2$ (see equation 8 of \cite{du24}). As in \cite{du25}, we do not employ the exact relations between $f_{\rm bound}$, $f_t$, and $r_t$, due to redshift dependency of the halo mass. Instead, following \cite{du25}, we first convert $f_{\rm bound}$ to $f_{\rm bound,mx} \equiv m_{\rm bound}/m_{\rm mx,0}$, and then adopt the relation between $f_{\rm bound,mx}$, $f_t$ and $r_t$. This relation has been shown to be more universal and is independent of the virial mass definition across different redshifts. Here, $m_{\rm mx,0}$ is the mass enclosed within $r_{\rm max,0}$ at the time of infall, where $r_{\rm max,0}$ is the radius at which the circular velocity $v_c$ reaches its maximum value:
\begin{equation}
    m_{\rm mx,0} = 5.88 \rho_s  r_s^3.
\end{equation}
In this procedure, $r_s$ and $\rho_s$ are calculated with respect to $M_{200}$ at the critical density of the Universe at the infall redshift. We derive the distribution of infall redshifts for a $10^{14} M_{\odot}$ host halo using {\tt galacticus} \citep{benson12}. The universal tidal track to compute the corresponding $m_{\rm bound}$ is obtained from the mass enclosed within the radius where the maximum $v_c$ is reached. This allows us to directly calculate the specified $f_{\rm bound}$ distribution from a universal tidal track at any infall redshift. The resulting relations between $f_{\rm bound}$, $f_t$, and $r_t$ that we use in this work are shown in Figure \ref{fig:subhalotidaltrack}. We note that the distribution of 3D radii of subhalos near the critical curve affects the distribution of the bound mass fraction for a given subhalo population. This projection effect is accounted for in multiple ways. We account for it in our modeling framework by setting the mean bound mass fraction as a free parameter which we marginalize over. It is also accounted for when calculating the infall redshift distribution.

\subsection{Quantifying the Asymmetry of Lensed Arcs}\label{txt:asymmetry}

As established in Section \ref{txt:lensedarc}, we are restricting this analysis to arcs that contain image midpoints that form along a straight critical curve. We quantify the asymmetry of a lensed arc based on the deviation of the image midpoints from linearity using the Pearson correlation coefficient $\rho_{\rm mid}$. $|\rho_{\rm mid}| = 1$ indicates perfect linearity, while $|\rho_{\rm mid}| = 0$ indicates no linearity. We define the asymmetry metric to be
\begin{equation}
    \xi = \log\left(1-|\rho_{\rm mid}|\right).
    \label{eq:asymmetry}
\end{equation}
The domain of this metric extends from $-\infty$ (perfect symmetry) to 0 (unambiguous asymmetry). In reality, $\xi$ never reaches $-\infty$, and the initial asymmetry is some finite and low value dependent on how far along the critical curve the arc spans, with $\xi$ increasing for longer arcs. As an order of magnitude, $\xi \lesssim -3$ for a smooth critical curve. We discuss this further in Appendix \ref{txt:lineartest}. The line of midpoints shown in Figure \ref{fig:asymrealizations} depicts $\xi = -2.43$ and $\xi = -0.17$ for the top and bottom panels, respectively.

It is important to note that the use of $\xi$ as our asymmetry metric is a measure of the degree of asymmetry in an arc, rather than the specific image perturbations that produce the asymmetry. This is important because it allows $\xi$ to be independent of the specific lens model. The only assumption to use $\xi$ as a metric is that the critical curve is locally a straight line. 

The primary consideration for this asymmetry metric is astrometric uncertainty on the image positions, which presents a degeneracy for asymmetric arcs. In general, it is difficult to determine whether the measured asymmetry of an arc is caused by perturbations from a dark matter subhalo population or just astrometric uncertainty in the identification of knot centroids. To account for this in our simulations, we include astrometric uncertainty $\delta_{xy}$ in our forward modelling, which we discuss in greater detail in Section \ref{txt:ABC}.

We also note that our definition of $\xi$ is not the only way to quantify asymmetry. One example alternative that we do not explore would be to fit a line to the image midpoints, and calculate the $\chi^2$ of the fit, using $\chi^2$ as the asymmetry metric. As we mentioned previously, since the level of asymmetry is on subarcsecond scales, the logarithmic nature of $\xi$ allows us to better examine the structure of induced asymmetry from a dark matter subhalo population. We emphasize that the use of $\xi$ in this work is a choice, and that other metrics do exist and can be used in the future.

\subsection{Bayesian Inference of the Cluster Mass Fraction}\label{txt:ABC}

The goal of this work is to constrain the mean subhalo mass fraction, which we define as the ratio of the expected average bound mass in subhalos from the SHMF to the macrolens mass within the simulation aperture ($M_{\rm ap}$):
\begin{equation}
    f_{\rm sub} = \frac{1}{M_{\rm ap}}\int dA_{\rm ap}\int^{M_{\rm high}}_{M_{\rm low}} \frac{\Sigma_{\rm sub}}{m_0}\left(\frac{m}{m_0}\right)^{-\alpha}m dm \times \bar{f}_{\rm bound}
\end{equation}
where $A_{\rm ap}$ is the projected area of the lens plane that we are simulating dark matter subhalos (i.e. the area of the simulation aperture). $M_{\rm low}$ and $M_{\rm high}$ are the low and high subhalo mass limits that we sample the SHMF between, respectively. As mentioned in Section \ref{txt:SHMF}, these are set to be $\log (M_{\rm low}/M_{\odot}) = 6$ and $\log (M_{\rm high}/M_{\odot}) = 10$. It is worth noting that $f_{\rm sub}$ is calculated directly by integrating each sampled SHMF. We fix $\alpha=1.9$ for this analysis, as it is a robust prediction from N-body simulations in CDM \citep{giocoli08,Springel2008}. $f_{\rm sub}$ can be analytically written as:
\begin{equation}
    f_{\rm sub} = \frac{10\pi R_{\rm ap}^2 m_0^{0.9}}{M_{\rm ap}}\left(M_{\rm high}^{0.1} - M_{\rm low}^{0.1}\right) \Sigma_{\rm sub} \bar{f}_{\rm bound}\label{eq:fsub}
\end{equation}
where we define the simulation aperture to be circular with radius $R_{\rm ap}$.

Our approach will be to infer the joint posterior distribution of $\Sigma_{\rm{sub}}$ and $\bar{f}_{\rm{bound}}$, and then use Equation \ref{eq:fsub} to translate this to an inference of $f_{\rm{sub}}$. To infer $f_{\rm{sub}}$, we will first compute a posterior distribution $P\left(\boldsymbol{q}|\boldsymbol{D}\right)\propto P\left(\boldsymbol{q}\right)\mathcal{L}\left(\boldsymbol{D}|\boldsymbol{q}\right)$, where $P\left(\boldsymbol{q}\right)$ is the prior, $\boldsymbol{q}=\left(\Sigma_{\rm{sub}},f_{\rm{bound}}\right)$ specifies the subhalo mass function, and 
$\mathcal{L}\left(\boldsymbol{D}|\boldsymbol{q}\right)$ is the likelihood function. The likelihood function is:
\begin{equation}
    \mathcal{L}\left(\boldsymbol{D}|\boldsymbol{q}\right) = \int P\left(\boldsymbol{D}|\boldsymbol{m},\boldsymbol{M}\right) P\left(\boldsymbol{m},\boldsymbol{M}|\boldsymbol{q}\right) d\boldsymbol{M}d\boldsymbol{m} \label{eq:likelihood} 
\end{equation}
where $\boldsymbol{m}$ are the realizations of dark matter subhalos, $\boldsymbol{M}$ are the macrolens parameters, and $\boldsymbol{D}$ are the observed image positions that make up the lensed arc.

Evaluating equation \ref{eq:likelihood} is challenging due to the high dimensionality of $\boldsymbol{m}$ and $\boldsymbol{M}$, and the need for many realizations of $\boldsymbol{m}$ to effectively sample the parameter space. Moreover, we are exceedingly unlikely to match the exact image positions for a random draw of $\boldsymbol{m}$ and $\boldsymbol{M}$. However, as discussed in Sections \ref{txt:cclenstheory} and \ref{txt:asymmetry}, we do not necessarily need to reproduce the exact cluster lens model that matches the image positions to constrain small-scale structure. Instead, we can use the deviations from a straight line of lensed image midpoints, as quantified by $\rho_{\rm{mid}}$, to measure the amount of small-scale structure in the lens model.

We have identified an informative summary statistic, $\xi$, defined in Equation \ref{eq:asymmetry}, in the context of a Bayesian inference problem with an intractable likelihood function. This motivates our use of an Approximate Bayesian Computing (ABC) algorithm to approximate Equation \ref{eq:likelihood}. In our ABC framework, we generate a set of model-predicted image positions ${\boldsymbol{D^{\prime}}}\left(\boldsymbol{m}, \boldsymbol{M}\right)$, from which we compute $S^{\prime} \equiv \xi\left(\boldsymbol{D^{\prime}}\right)$ with Equation \ref{eq:asymmetry}. The model-predicted image positions are computed by solving the lens equation (equation \ref{eq:lensequation}) forward for a given realization's deflection field. Similarly, we compute a summary statistic from the observed data $S\equiv \xi\left(\boldsymbol{D}\right)$. We then define a metric distance $\rho\left(S,S^{\prime}\right)=|S-S^{\prime}|$, and an importance weight $w\left(\rho\right)$. The ABC algorithm approximates Equation \ref{eq:likelihood} as
\begin{equation}
\mathcal{L}\left(\boldsymbol{D} |\boldsymbol{q}\right) \approx \int w\left(\rho\right) P\left(\boldsymbol{m}|\boldsymbol{q}\right) d \boldsymbol{M} d \boldsymbol{m} \label{eq:abclike}
\end{equation}
It should be understood that $\rho$ depends on the observations $\boldsymbol{D}$, and model parameters $\boldsymbol{m}$ and $\boldsymbol{M}$ through the model-predicted datasets $\boldsymbol{D^{\prime}}$. We implement a rejection sampling ABC algorithm, with $w=1$ when $\rho_S < \epsilon$, and zero otherwise, with $\epsilon$ being a tolerance threshold. The ABC rejection algorithm rejects model proposals unless they are ``close'' to the observed data in the summary statistic. 

Even in this framework, however, the use of ABC is still intractable. This is primarily due to the modeling of the macrolens parameters $\boldsymbol{M}$. In principle, at each sampling iteration a different set of $\boldsymbol{M}$ would be generated from a lens model fit, as has been done in similar analyses at galaxy scales \citep{gilman20}. On cluster scales, this is computationally intractable, as lens models for clusters often take multiple days to run. We therefore proceed with the approximation of fixing the background macrolens during the ABC process. This turns out to a be a reasonable approximation for this problem since the variations in the macrolens will not affect the asymmetry of an arc, as quantified by $\xi$.

The idea here is to reduce the dimensionality of the problem and to ensure that we are adequately capturing the relevant information from each realization, which in our case is the asymmetry (equation \ref{eq:asymmetry}). Thus, we replace the formal likelihood function with an approximation based on $w\left(\rho\right)$, which depends on the model parameters $\boldsymbol{m}$ and $\boldsymbol{M}$ through the model-predicted datasets $\boldsymbol{D^{\prime}}$. In our analysis procedure, we set $\epsilon$ implicitly by generating a large number $N$ of model-predicted datasets $\boldsymbol{D^{\prime}}$, and accepting the top 100 samples. We verify that our ABC analysis yields converged posterior distributions in Appendix \ref{txt:convergencetest}. Our ABC analysis proceeds as follows (in order):
\begin{itemize}
    \item First, we set the priors on  $\Sigma_{\rm sub}$ and $\bar{f}_{\rm bound}$, which make up $p(\boldsymbol{q})$. We sample the SHMF normalization from a log-uniform prior with $\log\left(\Sigma_{\rm sub}\right) \sim \mathcal{U}\left(-3.5,-0.1\right)$. As mentioned in Section \ref{txt:SHMF}, subhalos are sampled within infall mass range $6 < \log_{10} (m/M_{\odot}) < 10$.  Likewise, as discussed in Section \ref{txt:SHMF}, we set the mean bound mass fraction with a log-uniform prior $\log\left(\bar{f}_{\rm bound}\right) \sim \mathcal{U}\left(-1.25,-0.75\right)$.
    \item A subhalo population realization $\boldsymbol{m}$ is generated according to the sampled $\Sigma_{\rm sub}$ and tidal evolution model described in Section \ref{txt:subbhalopop}. Subhalos are modelled only within a $R_{\rm ap} = 2$" circular aperture centered on the arc region. For the fiducial model, this is depicted in Figure \ref{fig:perpendiculararc}. This aperture sets the value of $M_{\rm ap}$, allowing us to calculate $f_{\rm sub}$ for the realization.
    \item The perturbed image positions are calculated for this realization by solving the lens equation (equation \ref{eq:lensequation}) forward. At this step, we also add astrometric uncertainties $\delta_{xy}$ to each of the model-predicted image positions. We model $\delta_{xy}$ as a Gaussian and representative of the expected positional uncertainty in the image positions. As a test of our method, we evaluate this procedure for $\delta_{xy}$ of 0.01", 0.02", and 0.03", as we discuss more in Section \ref{txt:paramspace}. In practice, one should use the estimated $\delta_{xy}$ from the measurement of the image positions. Following this, the asymmetry metric $\xi$ is evaluated. Examples of this are shown in Figure \ref{fig:asymrealizations}.
    \item The preceding steps are repeated for $N$ realizations, taken to be 20000 in this work. We obtain an approximation of the posterior distribution by accepting the 100 samples corresponding to the lowest distance metrics $\rho$ (a tolerance threshold of 0.5\%). We emphasize that this statistical framework accounts for the relative likelihood that $\xi$ is dominated by deflection from a single massive subhalo by averaging over many realizations. We explicitly examine this effect in Appendix \ref{txt:individualsubhalos}.
\end{itemize}

Once the posteriors for $\Sigma_{\rm sub}$ and $\bar{f}_{\rm bound}$ are calculated, they can be easily recast for a posterior on $f_{\rm sub}$ using equation \ref{eq:fsub}. A necessary subtlety to consider is the fact that log-uniform sampling of $\Sigma_{\rm sub}$ and $\bar{f}_{\rm bound}$ does not yield a uniform distribution of $f_{\rm{sub}}$. Instead, uniform sampling of the $\Sigma_{\rm{sub}}$/$\bar{f}_{\rm{bound}}$ prior gives an effective prior on $\bar{f}_{\rm{sub}}$. This effective prior is shown in Figure \ref{fig:mockindividual}, and is mostly uniform throughout the $f_{\rm sub}$ parameter space, except at the edges. The inferred $f_{\rm{sub}}$ from the ABC-derived posterior distribution corresponds to an effective posterior distribution, $\tilde{p}\left(f_{\rm sub}|\boldsymbol{D}\right)$. To calculate $p\left(f_{\rm sub}|\boldsymbol{D}\right)$, we divide $\tilde{p}\left(f_{\rm sub}|\boldsymbol{D}\right)$ by the effective prior. This operation ensures that the posterior distribution $p\left(f_{\rm{sub}} | \boldsymbol{D}\right)$ will be uniform (or unconstrained) with uninformative or no data. In Section \ref{txt:mockresults}, we use simulated datasets to determine how the constraining power of the method is affected by the choice of $\delta_{xy}$, and the type of arc (perpendicular or parallel as defined in Section \ref{txt:lensedarc}).  

\subsection{Joint Constraints from Multiple Galaxy Clusters}
\label{txt:combinedpdf}

The inference procedure described in the previous section computes the likelihood function for the projected mass in cluster subhalos for a single collection of image knots in a lensed arc. In this work, and moving forward, will apply this methodology to multiple cluster lens systems to obtain more precise inferences of subhalo abundance. To do so, we must generalize the definitions of subhalo abundance and tidal evolution to account for variations in the cluster virial mass, and the radius where we make the measurement. 

We can use the self-similarity of halo substructure in CDM to generalize our model to a cluster of any virial mass and density profile. Following \citet{han16}, we can write
\begin{eqnarray}
\nonumber \Sigma_{\rm{sub,pop}}=\Sigma_{\rm{sub}}\times\tilde{\kappa}_{\rm{host}}\left(M_{\rm{host}},R/R_s\right) \\ \bar{f}_{\rm{bound,pop}} = \bar{f}_{\rm{bound}}\times\left(R/R_s\right)^{\gamma}
\end{eqnarray}
where $R$ is the projected distance from the cluster mass center where we observed lensed images, $R_s$ is the scale radius of the host, and $\gamma\sim1$ is a parameter that encodes the amount of tidal stripping experienced by subhalos with different projected distances to the cluster center. The function $\tilde{\kappa}_{\rm{host}}\left(M_{\rm{host}},R/R_s\right)$ varies in proportion to the projected mass density of the host halo, and captures the radial and cluster-mass dependence of the projected surface mass density of subhalos. The term $\left(R/R_s\right)^{\gamma}$ encodes the radial dependence of the mean bound mass fraction. Note that tidal stripping of CDM subhalos appears approximately independent of the subhalo infall mass \citep[e.g.][]{du25}.

For an inference that combines observations from several galaxy clusters, the terms $\tilde{\kappa}_{\rm{host}}\left(M_{\rm{host}},R/R_s\right)$ and $\left(R/R_s\right)^{\gamma}$ must be evaluated for each individual system. The parameters $\Sigma_{\rm{sub,pop}}$ and $\bar{f}_{\rm{bound,pop}}$ then become the hierarchical parameters inferred from the data. In this approach, we calculate the likelihood $\mathcal{L}\left(\boldsymbol{D_{i}}|\Sigma_{\rm{sub,pop}},\bar{f}_{\rm{bound,pop}}\right)$ following the methodology outlined in the previous section for each dataset $\boldsymbol{D_i}$ by sampling $\Sigma_{\rm{sub}}$ and $\bar{f}_{\rm{bound}}$ and scaling by $\tilde{\kappa}\left(M_{\rm{host}},R/R_s\right)$ and $\left(R/R_s\right)^{\gamma}$, respectively. The posterior distribution for the collection of $N$ cluster arcs $\boldsymbol{D}$ is then 
\begin{eqnarray}
\nonumber P\left(\Sigma_{\rm{sub,pop}},\bar{f}_{\rm{bound,pop}}|\boldsymbol{D}\right) \propto P\left(\Sigma_{\rm{sub,pop}},\bar{f}_{\rm{bound,pop}}\right) \\ 
\times\prod_{i=1}^{N}\mathcal{L}\left(\boldsymbol{D_{i}}|\Sigma_{\rm{sub,pop}},\bar{f}_{\rm{bound,pop}}\right)\label{eq:sigmasubpop}.
\end{eqnarray}

This approach lays the groundwork for future inferences of the cluster subhalo mass function from multiple systems with variations in $M_{\rm{host}}$ and $R_s$ among the sample. In this first analysis, we will apply the methodology to two cluster arcs, AS1063, and MACSJ0416, and derive an inference of $f_{\rm{sub}}$. If the total mass in dark matter $M_{\rm{ap}}$ (see Equation \ref{eq:fsub}) varies in proportion to $\tilde{\kappa}\left(M_{\rm{host}},R/R_s\right)$, and assuming dark matter dominates the surface mass density near the cluster critical curve, the subhalo mass fraction between the different arcs will differ only by the factor $\left(R/R_s\right)^{\gamma}$. When combining inferences of $f_{\rm{sub}}$, we assume that we are measuring the mass fraction at approximately the same $R/R_s$, such that we can directly multiply $f_{\rm{sub}}$ likelihoods obtained from each arc. This approximation is sufficient to first order, as estimates from CLASH measure $M_{\rm 200}$ and $R_{\rm 200}$ for both clusters to be $\sim 10^{15} M_{\odot}$ and $\sim 2$ Mpc, respectively \citep{umetsu14}. Future analyses will account for the scaling with cluster mass and radius described by Equation \ref{eq:sigmasubpop}. 

\section{Demonstration of the Forward Model using Mock Arcs}\label{txt:mockresults}

\begin{figure}
    \centering
    \includegraphics[trim={0cm 0.35cm 0cm 0.35cm},clip,width=0.49\textwidth]{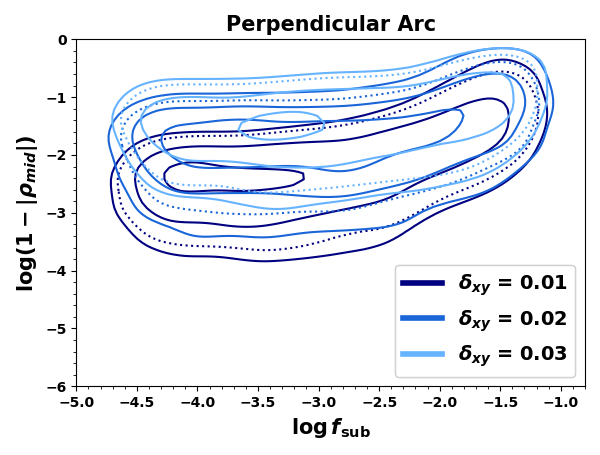}
    \includegraphics[trim={0cm 0.35cm 0cm 0.35cm},clip,width=0.49\textwidth]{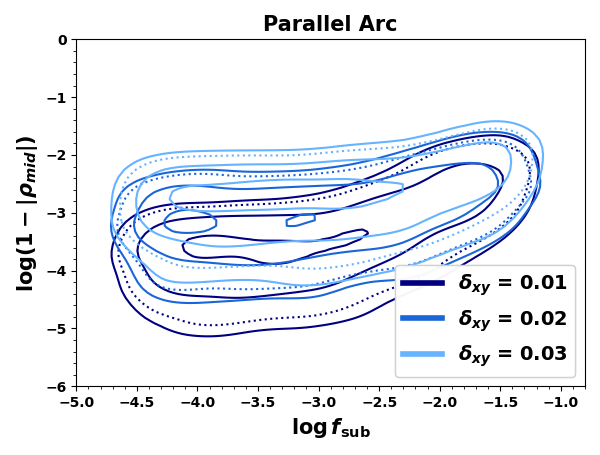}
\caption{2D density distributions of the parameter space of our simulations. The contours trace out the probability of a given asymmetry $\xi = \log\left(1-|\rho_{\rm mid}|\right)$ being observed for a simulated mass fraction $f_{\rm sub}$. The distributions are shown for the fiducial perpendicular ({\it Top}) and parallel ({\it Bottom}) arcs. Darker blue contours indicate lower astrometric uncertainty $\delta_{xy}$ (in arcsec). Solid contours are spaced by 25\% confidence intervals. Dotted contours indicate the 68\% confidence interval. In general, the greater $\delta_{xy}$, the larger the asymmetry can be. The comparison of the two plots shows that parallel arcs generally exhibit less asymmetry than perpendicular arcs. }  
\label{fig:parameterspace}
\end{figure}

\begin{table}
    \caption{Macrolens parameters for the 3 main cluster-scale NSIEs that make up Mock (No Gal.). This macrolens represents the true macrolens distribution for the mock arcs, allowing the fiducial model macrolens parameters (Table \ref{tab:macrolens}) to effectively represent the result from a lens model.}
    \centering
    \begin{tabular}{cccccc}
    \hline
        Profile & Position [x",y"] & $\sigma_v$ [km s$^{-1}$]  & $R_c$ [kpc] & ($q$,PA) \\
    \hline
        NSIE 1 & (-4.7,5.1) & 734 & 4.5 & (0.34,$31^{\circ}$) \\
        NSIE 2 & (-4.8,4.9) & 487 & 2.4 & (0.27,$48^{\circ}$)\\
        NSIE 3 &  (7.7,-7.4) & 650 & 3.8 & (0.41,$10^{\circ}$) \\
    \hline
    \end{tabular}
    \label{tab:mocknogal}
\end{table}

\begin{figure*} 
\begin{multicols}{2}
    \includegraphics[trim={0cm 0.35cm 0cm 0.0cm},clip,width=\linewidth]{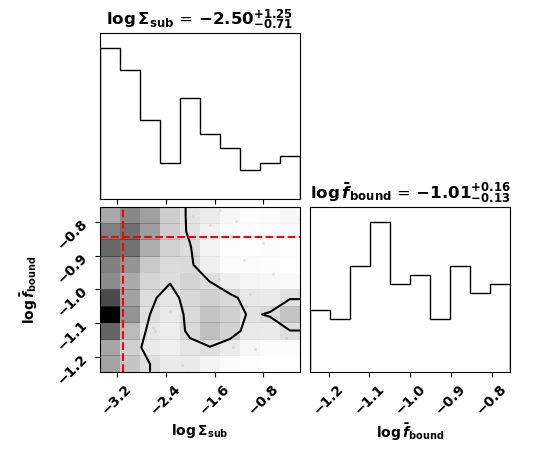}\par 
    \includegraphics[trim={0cm 0.35cm 0cm 0.0cm},clip,width=\linewidth]{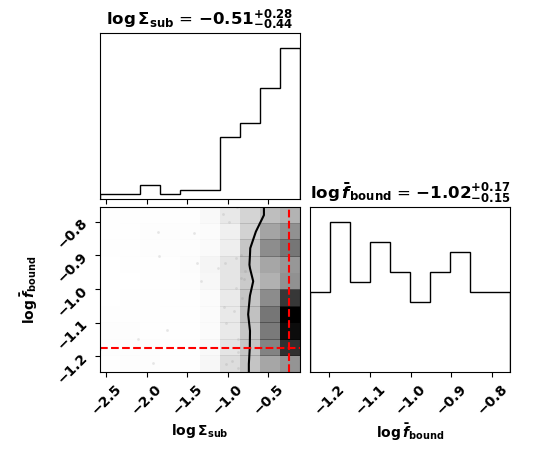}\par   
    \end{multicols}
\begin{multicols}{2}
    \includegraphics[trim={0cm 0.35cm 0cm 0.35cm},clip,width=\linewidth]{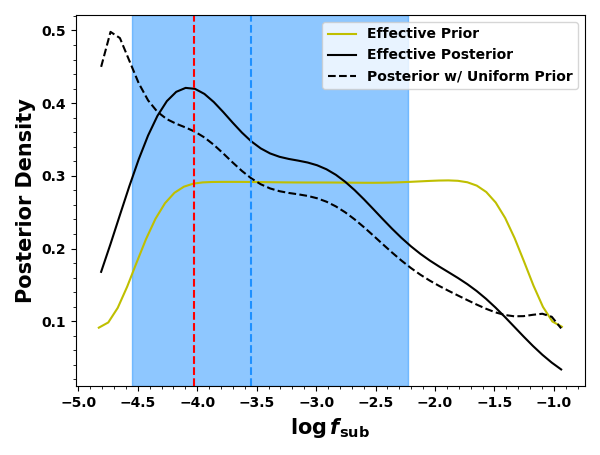}\par
    \includegraphics[trim={0cm 0.35cm 0cm 0.35cm},clip,width=\linewidth]{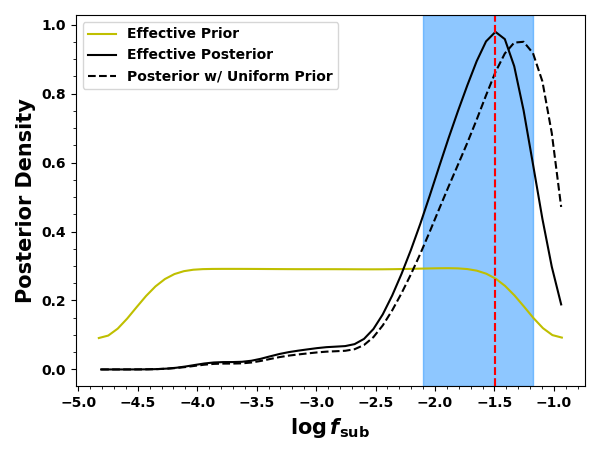}\par
\end{multicols}
\caption{Example posterior constraints for two perpendicular mock arcs, assuming $\delta_{xy} = 0.01$". The left column shows the results for a mock arc exhibiting low asymmetry ($\xi = -2.39$), while the right column shows the results for a mock arc exhibiting high asymmetry ($\xi = -0.17$). The top row shows corner plots for the joint constraint on $\Sigma_{\rm sub}$ and $\bar{f}_{\rm bound}$. The solid contour indicates the 68\% confidence interval, and the dashed red lines indicate the true $\Sigma_{\rm sub}$ and $\bar{f}_{\rm bound}$. The bottom row shows the resulting posterior on $f_{\rm sub}$, calculated with equation \ref{eq:fsub}. The yellow line is the effective prior on $f_{\rm sub}$ resulting from the uniform sampling of $\Sigma_{\rm sub}$ and $f_{\rm bound}$. Under this prior, the effective posterior on $f_{\rm sub}$ ($\tilde{P}(f_{\rm sub}|\boldsymbol{D})$) is the solid black line. We divide $\tilde{P}(f_{\rm sub}|\boldsymbol{D})$ by the effective prior to obtain the posterior $P(f_{\rm sub}|\boldsymbol{D})$ assuming a log-uniform prior on $f_{\rm sub}$ (dashed black line). The vertical dashed blue and red lines indicate the posterior median and true $f_{\rm sub}$, respectively. We note that in the bottom right panel these two lines overlap. The shaded blue region indicates the 68\% confidence interval.}
\label{fig:mockindividual}
\end{figure*}

To demonstrate the ABC method described in Section \ref{txt:ABC}, we test how well we can recover the simulated $f_{\rm sub}$ from mock arcs. Throughout this section, we use the term ``fiducial'' to describe the arcs and macrolens that we perform the ABC method on, with the fiducial macrolens shown in Figure \ref{fig:mainlens} and fiducial arcs (perpendicular and parallel) shown in Figure \ref{fig:perpendiculararc}. For the remainder of this paper, constraints from posteriors are taken to be the posterior medians with 68\% confidence intervals, unless otherwise stated.

\subsection{Disentangling Astrometric Uncertainties from Astrometric Perturbations}\label{txt:paramspace}

We begin by first examining the parameter space explored by our method. As mentioned in Section \ref{txt:ABC}, we utilize two free parameters in our analysis, $\Sigma_{\rm sub}$ and $\bar{f}_{\rm bound}$, both sampled with log-uniform priors: $\log\left(\Sigma_{\rm sub}\right) \sim \mathcal{U}\left(-3.5,-0.1\right)$ and $\log\left(\bar{f}_{\rm bound}\right) \sim \mathcal{U}\left(-1.25,-0.75\right)$. After computing the model-predicted image positions, we then inject some $\delta_{xy}$ to the model-predicted image positions and calculate $\xi$ from each realization. For the demonstrations we show throughout this section, we do this for three scenarios of $\delta_{xy}$: 0.01", 0.02", and 0.03". $\delta_{xy} = 0.03$" represents the JWST precision, as is typically assumed on lensed images. As mentioned in Section \ref{txt:cclenstheory}, knots in lensed arcs are typically treated as point images, and thus can have greater precision. We test this using 0.02" and 0.01" astrometric uncertainties. In practice, these uncertainties are only possible for lensed knots that span less than a few pixels. We note that an astrometric uncertainty of 0.01" is not an unreasonable assumption for a lensed image knot, as it is effectively claiming confident positional identification of an image to pixel scale.

We illustrate the parameter space of our simulations in Figure \ref{fig:parameterspace}. Three primary trends are important to appreciate in our simulations. First is that parallel arcs on average exhibit less asymmetry than perpendicular arcs. Second is that as $\delta_{xy}$ increases, the average asymmetry increases. This is unsurprising, but importantly highlights that the constraining power of our method decreases with increasing astrometric uncertainty. Third is that for $\xi \lesssim -3$ the contours become increasingly horizontal, implying a reduction in the constraining power of the method on $f_{\rm sub}$. This limit is consistent with the approximate criterion for a smooth critical curve as established in Section \ref{txt:asymmetry}. Lastly, we note that for the distributions in Figure \ref{fig:parameterspace}, horizontal cuts correspond to the posterior for $f_{\rm sub}$. This figure visualizes the relationship between $f_{\rm sub}$ and $\xi$, which will be useful in interpreting our results in later sections. 

\subsection{Calculating Posteriors from Individual Mock Arcs}\label{txt:indivresults}

Before analyzing the efficacy of our method on a large sample of mock arcs, we first present the results from individual mock arcs. The results presented here are designed to imitate the practical usage of our method for any individual arc.

First, we generate a mock macrolens profile based on a random perturbation of the fiducial model shown in Figure \ref{fig:mainlens}. This mock macrolens profile is represented by the parameters shown in Table \ref{tab:mocknogal}. This allows us to consider our fiducial macrolens as an approximation for the mock macrolens, which mimics the reality of lens modelling (i.e. our fiducial model is the macromodel of the true mass distribution generated from a lens model). We call this mock macrolens ``Mock (No Gal.)'' or MNG for short. Like our fiducial model, MNG does not contain any cluster member galaxies.

As an example demonstration of the method, we consider both a low and high asymmetry perpendicular arc with $\xi = -2.39$ and $\xi = -0.17$, respectively. Both arcs are generated from MNG with $\delta_{xy} = 0.01$". The low and high asymmetry mock arcs are generated with a true $\log f_{\rm sub}$ of -4.02 and -1.49, respectively. The goal of our method is to recover these true $f_{\rm sub}$ with the posterior.

We present the posteriors from our method for these arcs in Figure \ref{fig:mockindividual}. Figure \ref{fig:mockindividual} shows all three distributions discussed in Section \ref{txt:ABC}: the effective prior, $\tilde{P}\left(f_{\rm sub}|\boldsymbol{D}\right)$, and $P\left(f_{\rm sub}|\boldsymbol{D}\right)$. For the remainder of this paper, posteriors mentioned assume a uniform prior on $f_{\rm sub}$, and hence correctly represent $P\left(f_{\rm sub}|\boldsymbol{D}\right)$.

For both the low and high asymmetry mock arc, our method calculates the posterior to accurately constrain the true $f_{\rm sub}$ to within 68\% confidence interval. For the low asymmetry arc, our method constrains $\log f_{\rm sub} = -3.54^{+1.31}_{-1.00}$ at an upper limit. The predictive power is increased for the high asymmetry arc, where we constrain $f_{\rm sub} = -1.49^{+0.31}_{-0.61}$. The corner plots shown in Figure \ref{fig:mockindividual}, show the joint posteriors for $\Sigma_{\rm sub}$ and $\bar{f}_{\rm bound}$. As can be seen, $\Sigma_{\rm sub}$ and $\bar{f}_{\rm bound}$ are fairly unconstrained by the method, with some improvement for high asymmetry arcs. This motivates the continued use of $f_{\rm sub}$ as the best constrained parameter.

The demonstration presented here is for just two example individual arcs. To show that our method is accurate and consistent, we repeat this exercise for a large sample of mock arcs with different $\delta_{xy}$, which we discuss in the next section.

\subsection{Evaluating the Accuracy of our Method with a Sample of Mock Arcs}\label{txt:sampleresults}

\begin{figure}
    \centering
    \includegraphics[trim={0cm 0.35cm 0cm 0.35cm},clip,width=0.49\textwidth]{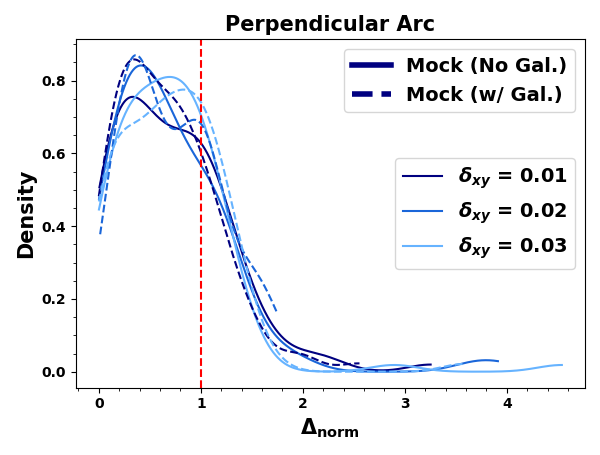}
    \includegraphics[trim={0cm 0.35cm 0cm 0.35cm},clip,width=0.49\textwidth]{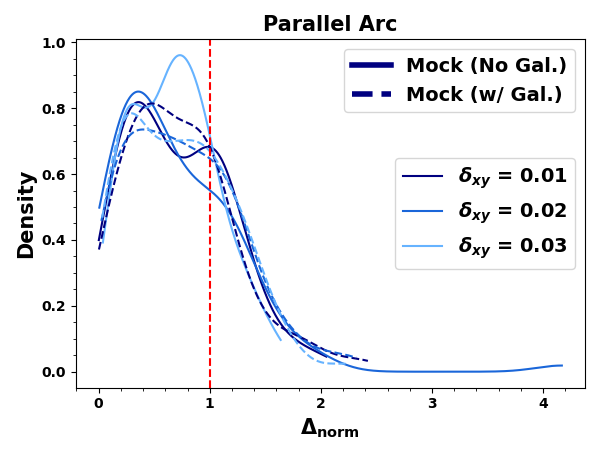}
\caption{ PDF distributions of the effective z-score $\Delta_{\rm norm}$ for perpendicular ({\it Top}) and parallel ({\it Bottom}) arcs. Solid and dashed lines indicate the distributions for 100 simulated arcs based on the ``Mock (No Gal.)'' and ``Mock (w/ Gal.)'' macrolenses, respectively. Darker shades of blue indicate lower astrometric uncertainty ($\delta_{xy}$ given in arcsec). The dashed red line indicates $\Delta_{\rm norm} = 1$.}  
\label{fig:mockdemonstration}
\end{figure}

The accuracy and reliability of our method can be best shown when repeating the test in Section \ref{txt:indivresults} for a large sample of mock arcs. To do this, we generalize the methodology even further.

First, we include a second mock macrolens, to further demonstrate that our method is impartial to the macrolens. This second mock macrolens is the same as the fiducial macrolens, but including a nearby cluster member. This mock macrolens is shown in Figure \ref{fig:arcwgal}. We call this mock macrolens ``Mock (w/ Gal.)'', or MWG for short. Thus, we use two mock macrolenses in this analysis, MNG and MWG.

For both mock macrolenses, simulated lensed arcs (both perpendicular and parallel) are generated for different combinations of $\Sigma_{\rm sub}$ and $\bar{f}_{\rm bound}$, from which we calculate $f_{\rm sub}$. The simulated perpendicular and parallel arcs roughly span the same length along the critical curve as the fiducial arc that we perform the ABC method on. In this way, our fiducial lensed arcs are approximating each mock arc to within $\sim$0.1" in arc span. The importance of the arcs roughly spanning the same angular scales is exemplified in Figure \ref{fig:parameterspace}, where the sampled parameter space is shown to be dependent on how long the arc spans the critical curve.  We generate samples of mock arcs for for three scenarios of $\delta_{xy}$: 0.01", 0.02", and 0.03", as discussed previously. 

In summary, we generate 12 samples of mock arcs for each combination of 1) Mock macrolens (MNG or MWG), 2) arc type (perpendicular or parallel), and 3) $\delta_{xy}$ (0.01", 0.02", or 0.03"). Each sample contains 100 mock arcs. For each sample, we seek to quantify how many of the mock arcs are accurately constrained by our method.

To quantify how well our ABC method recovers the true mock $f_{\rm sub}$, we define an effective z-score:
\begin{equation}\label{eq:effzscore}
    \Delta_{\rm norm} = \left\{ \begin{array}{lrc} \frac{|x_{\rm meas} - x_{\rm mock}|}{x_{+68} - x_{\rm meas}} & \mbox{for} & x_{\rm meas} - x_{\rm mock} < 0 \\
    \frac{|x_{\rm meas} - x_{\rm mock}|}{x_{\rm meas} - x_{-68}} & \mbox{for} & x_{\rm meas} - x_{\rm mock} > 0
\end{array}\right.
\end{equation}
where $x$ is the parameter of interest, in this case $f_{\rm sub}$. $x_{\rm meas}$ is the measured parameter from the posterior, which we define to be the posterior median. $x_{\rm mock}$ is the true known mock parameter. $x_{+68}$ and $x_{-68}$ are the $\pm68\%$ confidence intervals of the posterior. Defining $\Delta_{\rm norm}$ in this way accounts for the fact that the posterior may not be Gaussian, rendering a typical z-score ineffective. We define the threshold of recovery of the true $f_{\rm sub}$ to be when $\Delta_{\rm norm} < 1$, as this indicates that the ABC method constrained $f_{\rm sub}$ to within the 68\% confidence interval.

We show the distributions of $\Delta_{\rm norm}$ for perpendicular and parallel mock arcs in Figure \ref{fig:mockdemonstration}. In both cases, the ABC method remains consistent in its effectiveness regardless of $\delta_{xy}$ and background macrolens. On average, the ABC method recovers the true $f_{\rm sub}$ to within 68\% confidence 73\% and 72\% of the time for perpendicular and parallel arcs, respectively. These results also illustrate that the true macrolens distribution in the vicinity of the arc need not be reconstructed to complete accuracy, since there is no trend in the results with the mock macrolens. This emphasizes that reconstructing the morphology of the arc is more important than the macrolens that produces it. This has a useful consequence for lens modelling. The RMS of the reconstructed images in real lens models are often significantly larger than the observed $\delta_{xy}$. However, this is not a concern for our method, as long as the morphology of the arc is preserved in the model. From these results, we conclude that our ABC method is able to reconstruct the true $f_{\rm sub}$ in the majority of cases, with minimal effects from the accuracy of the macrolens or $\delta_{xy}$. This allows us to apply this method to real arcs, and have $\sim73\%$ confidence that we can recover $f_{\rm sub}$ to within the 68\% confidence interval. 

\subsection{Forecasting Constraints from Mock Observations}\label{txt:sampleobs}

\begin{figure}
    \centering
    \includegraphics[trim={0cm 0.35cm 0cm 0.35cm},clip,width=0.49\textwidth]{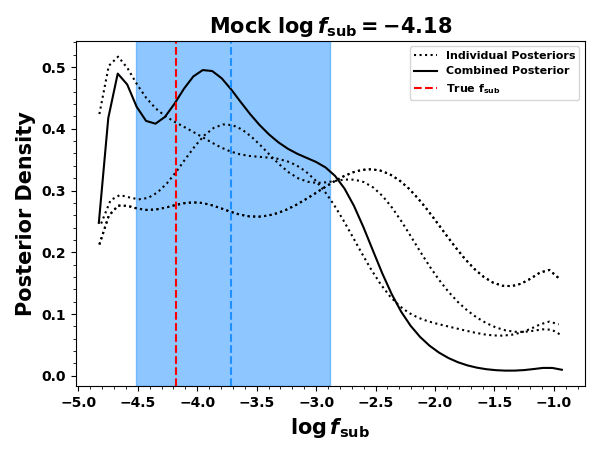}
    \includegraphics[trim={0cm 0.35cm 0cm 0.35cm},clip,width=0.49\textwidth]{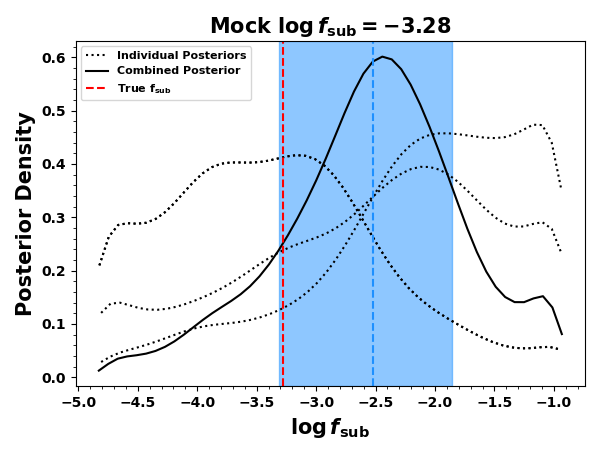}
    \includegraphics[trim={0cm 0.35cm 0cm 0.35cm},clip,width=0.49\textwidth]{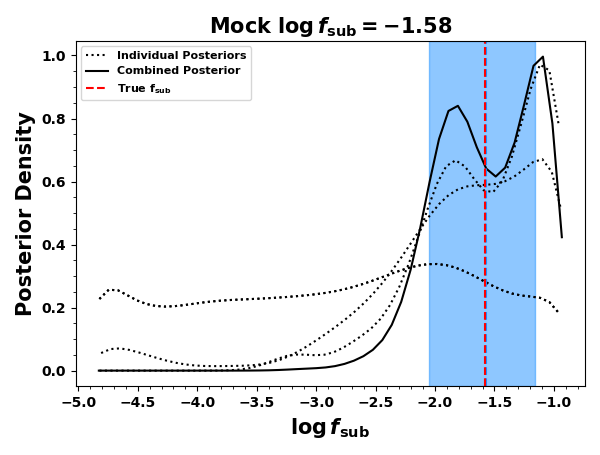}
\caption{ Combined global posteriors for three mock arcs generated with $\log f_{\rm sub} = -4.18$ ({\it Top}), $\log f_{\rm sub} = -3.28$ ({\it Middle}), $\log f_{\rm sub} = -1.58$ ({\it Bottom}). Dotted black lines indicate individual posteriors for each mock arc, while the solid black lines indicate the combined posterior. Dashed vertical red and blue lines indicate the true $f_{\rm sub}$ and combined posterior median, respectively. The blue shaded region is the combined 68\% confidence interval.  }  
\label{fig:sampleobs}
\end{figure}

With the method established as being effective, we now focus on forecasting what a full constraint on $f_{\rm sub}$ would look like using sample mock observations. Since $f_{\rm sub}$ is directly calculated from the SHMF, we can multiply inferences together to get a combined global constraint. In practice, combining the likelihoods from our method from multiple lensed arcs will yield a tighter and more accurate constraint than with individual arcs. Multiplying likelihoods together implies that the physical conditions between clusters are the same. We justify this assumption as sufficient since the clusters have approximately the same mass and the arcs appear at approximately the same radius from the center, so we are looking at the same radial distribution of subhalos (and hence the same projected mass density). 

For this test, we generate three mock arcs with the same true $f_{\rm sub}$, all with the same $\Sigma_{\rm sub}$ and $\bar{f}_{\rm bound}$. We only consider three mock arcs here to simulate the simplest scenario for a forecasted combined posterior. We note that including more arcs will result in even tighter constraints. The mock arcs are randomly generated as a combination of macrolens (MNG or MWG), arc type (perpendicular or parallel) and $\delta_{xy}$ (0.01", 0.02" or 0.03"). This allows broad representation for the diversity of lensed arc types and observational resolution. For each of the three mock arcs, we calculate individually the posterior on $f_{\rm sub}$ with our method. We then multiply the likelihoods together to get the global combined posterior. We are able to conclude that our forecasts have good constraining power if the global posterior is able to accurately recover the true $f_{\rm sub}$ to within the 68\% confidence interval and exhibit a smaller 68\% confidence interval.

We conduct this test for three different versions of true $f_{\rm sub}$ ($\log f_{\rm sub} = $ -4.18, -3.28, and -1.58), spanning the rough range of our effective prior. The results are shown in Figure \ref{fig:sampleobs}. In all three cases, the true $f_{\rm sub}$ is recovered to within the 68\% confidence interval. Furthermore, the global posterior overcomes the main limitations of constraining $f_{\rm sub}$ with individual arcs, namely broad posteriors with little constraining power and the failed constraints with $\Delta_{\rm norm} > 1$,  which as we showed in Section \ref{txt:sampleresults}, constitute 26\% of the mock sample. For the sample observations with true $\log f_{\rm sub} = $ -4.18, -3.28, and -1.58, our method constrains with the global posterior $\log f_{\rm sub}$ to be $-3.71^{+0.83}_{-0.80}$, $-2.52^{+0.67}_{-0.79}$, and $-1.57^{+0.41}_{-0.48}$, respectively. Lastly, the 68\% confidence interval range, which we define as $x_{+68}-x_{-68}$, is reduced (from the mean of the sample of individuals to the global posterior) from 2.1 to 1.6, 1.9 to 1.5, and 1.6 to 0.9, for $\log f_{\rm sub} = $ -4.18, -3.28, and -1.58, respectively. Since the 68\% confidence interval range is smaller for the global constraint and the global posterior accurately recovers the true $f_{\rm sub}$, we conclude that our method forecasts accurate constraints with a good constraining power.

With the efficacy of our method sufficiently demonstrated with mock arcs, we now focus in the next section on applying it to a sample of real lensed arcs.

\section{Constraining the Subhalo Mass Fraction with Real Arcs}\label{txt:real}

In this Section, we apply the statistical method described above to two well observed arcs. We consider both a perpendicular and parallel arc to evaluate the types of constraints that can be made on $f_{\rm sub}$ with both. It should be noted that many other candidate arcs are suitable to be used with our method; however, we focus on just two for this work to simply demonstrate our method on real arcs. We will extend our analysis to a larger sample of lensed arcs in a future work, while also incorporating line-of-sight halos and variations in the radial distribution of cluster subhalos in different systems. Additionally, for this exercise we only consider a single lens macromodel for each arc, and leave a more thorough analysis of the macrolens uncertainty to future works. However, as we described in Section \ref{txt:sampleresults}, the morphology of the arc is more important for our method than the actual predicted macrolens density distribution because suitable macrolens models are all smooth on the small angular scales relevant for the image knots. In Section \ref{txt:data} we introduce the two arcs and the datasets that we utilize. We then discuss the observed asymmetry in both arcs and how to interpret them in Section \ref{txt:upperlimits}. Lastly, we discuss our resulting constraints on $f_{\rm sub}$ and necessary considerations and cautions in their interpretation in Section \ref{txt:obsresults}.

\subsection{Data}\label{txt:data}

\begin{figure*} 
\begin{multicols}{2}
    \includegraphics[trim={5.9cm 0.25cm 5.9cm 0.25cm},clip,width=\linewidth]{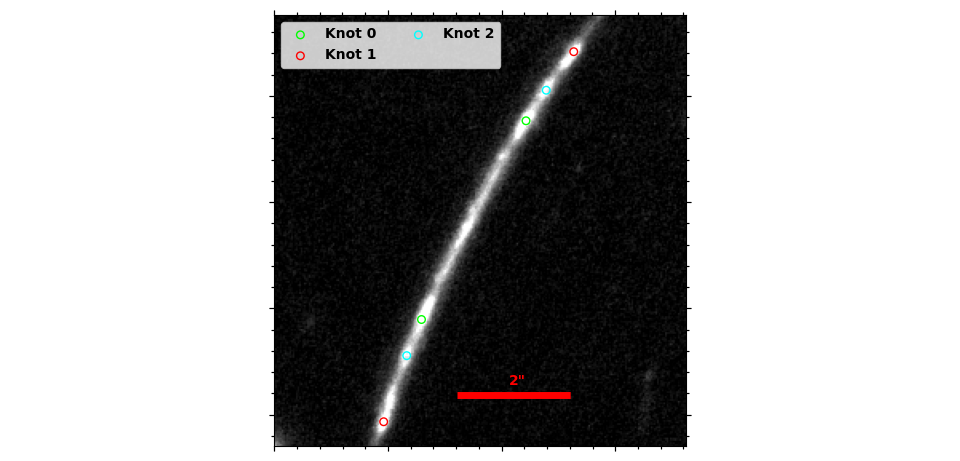}\par 
    \includegraphics[trim={5.9cm 0.25cm 5.9cm 0.25cm},clip,width=\linewidth]{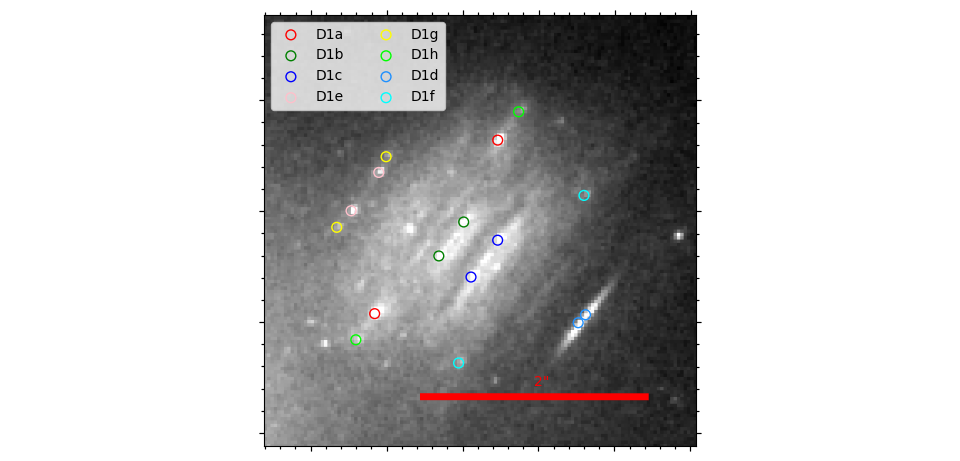}\par 
    \end{multicols}
\begin{multicols}{2}
    \includegraphics[trim={5.2cm 0.45cm 5.7cm 0.45cm},clip,width=\linewidth]{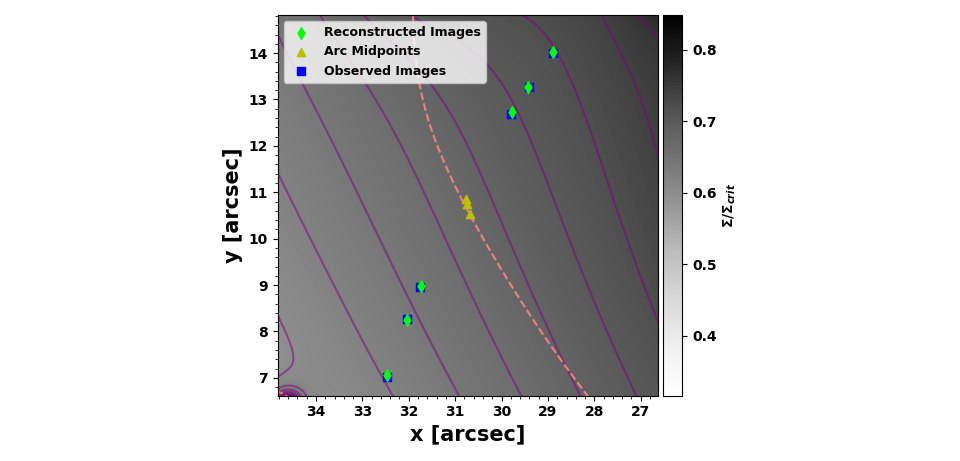}\par    
    \includegraphics[trim={5.2cm 0.45cm 5.7cm 0.45cm},clip,width=\linewidth]{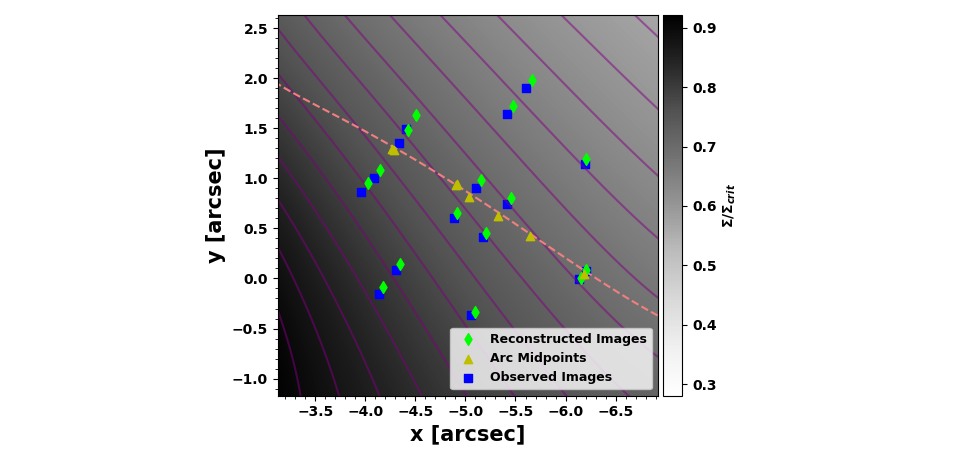}\par    
\end{multicols}
\caption{The two arcs we consider in this work: AS1063 System 1 ({\it Left Column}) and the MACSJ0416 Warhol Arc ({\it Right Column}). The top row shows images for both arcs, with an HST F606W image for AS1063 System 1 \protect\citep{lotz17} and a JWST F090W image \protect\citep{windhorst23} for the Warhol arc. In each case, North is up and East is left. Open circles indicate the observed image positions, with knots sharing color representing counterimaged pairs. The bottom row shows the modeling window used for our analysis for each arc. The window depicted is the same for each arc as their image in the top row. For AS1063 System 1 and the Warhol Arc, we use the lens models from \protect\cite{bergamini19} and \protect\cite{rihtarsic25} as the macrolens density, respectively. The blue squares show the observed image positions, which correspond to the open circles in the image in the top row. Green diamonds and yellow triangles indicate the reconstructed images from the lens model and predicted midpoints, respectively. The light red dashed line is the macrolens critical curve. Purple contours show the logarithmically spaced contours of the macrolens density profile. The $x$ and $y$ positions are presented in arcseconds with respect to RA,Dec = (342.1832, -44.5309) and (64.0382, -24.0675) for AS1063 System 1 and the Warhol Arc, respectively.}
\label{fig:observedarcs}
\end{figure*}

\begin{table}
    \caption{Observed image positions of the 3 counterimaged knots in AS1063 System 1. Image IDs are given as X.Y, where X is the counterimage knot pair, and Y is the negative (Y=0) or positive (Y=1) parity image in the pair.}
    \centering
    \begin{tabular}{ccc}
    \hline
        Image ID & RA [deg] & Dec [deg] \\
    \hline
        0.0 & 342.1948208 & -44.5273528 \\
        0.1 & 342.1955875 & -44.5283917 \\
        1.0 & 342.1944708 & -44.5269917 \\
        1.1 & 342.1958650 & -44.5289261 \\
        2.0 & 342.1946725 & -44.5271931 \\
        2.1 & 342.1956958 & -44.5285806 \\
    \hline
    \end{tabular}
    \label{tab:AS1063}
\end{table}

\begin{table}
    \caption{Observed image positions of the 8 counterimaged knots in Warhol Arc. Image IDs are given as D1X.Y, where X is the counterimage knot pair, and Y is the negative (Y=2) or positive (Y=1) parity image in the pair. IDs with an asterisk are newly identified images.}
    \centering
    \begin{tabular}{ccc}
    \hline
        Image ID & RA [deg] & Dec [deg]  \\
    \hline
        D1a.1 & 64.0365705 & -24.0670440 \\
        D1a.2 & 64.0369079 & -24.0674788 \\
        D1b.1 & 64.0366637 & -24.0672491 \\
        D1b.2 & 64.0367319 & -24.0673342 \\
        D1c.1 & 64.0365705 & -24.0672947 \\
        D1c.2 & 64.0366436 & -24.0673870 \\
        D1e.1 & 64.0368968 & -24.0671251 \\
        D1e.2 & 64.0369727 & -24.0672211 \\
        D1g.1* & 64.0368767 & -24.0670857 \\
        D1g.2* & 64.0370120 & -24.0672625 \\
        D1h.1* & 64.0365131 & -24.0669730 \\
        D1h.2* & 64.0369595 & -24.0675443 \\
        D1d.1 & 64.0363295 & -24.0674814 \\
        D1d.2 & 64.0363495 & -24.0675014 \\
        D1f.1 & 64.0363338 & -24.0671824 \\
        D1f.2 & 64.0366776 & -24.0676023 \\
    \hline
    \end{tabular}
    \label{tab:warhol}
\end{table}

We consider two arcs in this work: Abell S1063 (hereafter AS1063) System 1 and the MACS J0416.1-2403 (hereafter MACSJ0416) Warhol Arc. AS1063 and MACSJ0416 have lens redshifts $z_d$ of 0.348 \citep{guzzo09} and 0.396 \citep{postman12}, respectively. The spectroscopically confirmed source redshifts $z_s$ for AS1063 System 1 and the Warhol arc are 1.229 \citep{balestra13} and 0.9397 \citep{caminha17}, respectively. Both clusters are well studied and were extensively observed with the Hubble Frontier Fields (HFF) program \citep{lotz17}. Figure \ref{fig:observedarcs} shows an HST image of AS1063 System 1 and a JWST image of the MACSJ0416 Warhol Arc.

The primary aspects of the observed data that we require for our analysis are a lens model and the image positions of the knots in the arc. For the former, we choose a parametric lens model from the literature. In the latter, we estimate the image positions based on the brightest pixel in the knot.

For AS1063, there have been numerous lens models published in the last decade \citep{diego16b,bergamini19,limousin22}. For this work, we adopt the model from \cite{bergamini19} (hereafter B19) to act as the macrolens for this analysis. The B19 lens model utilizes the parametric lens inversion algorithm {\tt LensTool} \citep{kneib96,jullo07}, which models the density with smooth elliptical potentials for cluster-scale halos and cluster member galaxies, making it suitable for this analysis. B19 achieves a lens plane RMS of 0.55" with HFF data. 

AS1063 System 1 is a large perpendicular arc that spans $\sim0.3$" along the critical curve, and $\sim7$" perpendicularly from its critical curve. There are three bright counterimaged knots that we use to measure the observed asymmetry. In previous lensing analyses, however, only a single pair of counterimages has been used to constrain the lens model. Therefore, we measure the image positions of the knots ourselves by simply taking them to be the brightest pixel position of each knot's flux centroid. We note that this is not the most rigorous procedure to measure the image positions; however, it serves as a reasonable initial approximation for this analysis. We also emphasize that this is roughly consistent with the typical image identification procedure that occurs in lens modelling studies, where new image candidates are identified by color and morphology, then confirmed by their consistency in being reproduced by existing lens models \citep{lotz17}. Therefore, we implement this simple procedure as a sufficient approximation, and reserve more sophisticated follow-up image identification procedures for lensed arcs to future work. With the observed image positions (presented in Figure \ref{fig:observedarcs} and Table \ref{tab:AS1063}), we calculate the source positions by backprojecting the observed image positions with the B19 lens model and taking the mean of the two counterimages to be the model source position. This is then forward projected with the lens model to calculate the reconstructed image position. The reconstructed image positions are the images that we then apply our analysis on (i.e. to generate many realizations of $\hat{\xi}$ to sample the posterior with Equation \ref{eq:abclike}).

For MACSJ0416, the procedure is largely the same as for AS1063. MACSJ0416 has a variety of recent lens models, some of which have made use of recent JWST imaging allowing for $>400$ lensed image constraints \citep[e.g.][]{bergamini23,cha23,diego23a,perera24b,rihtarsic25,limousin25}. We adopt the model from \cite{rihtarsic25} (hereafter R25) for the macrolens for this arc. R25 makes use of JWST imaging from the CAnadian NIRISS Unbiased Cluster Survey (CANUCS) along with pre-JWST multiple image catalogues \citep{richard21,bergamini23} and also utilizes {\tt LensTool} for a lens plane RMS of 0.52". 

The Warhol Arc \citep{kaurov19,chen19} in MACSJ0416 is a large parallel arc that extends $\sim2.3$" along the critical curve. It has been the subject of considerable interest in the search for lensed transients, and thus has had numerous follow-up observations conducted on it \citep{kelly22,yan23}. Recently, \cite{palencia25} examined the effect of compact dark matter objects on the spatial distribution of transient events, establishing the Warhol Arc as a strong candidate to use to study the nature of dark matter. R25 identifies six lensed knots that they use as constraints in their model. For our analysis, we include an additional two lensed knots, which are clearly visible in recent JWST imaging (see Figure 2 of \cite{yan23}). This gives us a total of eight imaged knots to use for our analysis. Like with AS1063 System 1, we measure each image's position ourselves by taking them as the brightest pixel position. The same procedure is then repeated to calculate the R25 reconstructed image positions for the eight knots. The image positions for the Warhol Arc are shown in Figure \ref{fig:observedarcs} and Table \ref{tab:warhol}.

\subsection{Measuring Asymmetry in Observed Arcs}\label{txt:upperlimits}

With the observed image positions, we measure the asymmetry metric for these arcs to be $\xi = -1.05$ and $\xi = -2.48$ for  AS1063 System 1 and the Warhol arc, respectively. As mentioned in Section \ref{txt:asymmetry}, it is necessary to interpret these observed $\xi$ in the context of the astrometric uncertainty. Since we are only simply measuring the image positions by adopting the brightest pixel positions (using HST F606W for AS1063 System 1 and JWST F090W and the Warhol arc), the astrometric uncertainty is likely to be greater than pixel-level precision. However, even with a rigorous procedure to identify the images, observational challenges remain. For example, astrophysical transient events or microlensing may lead to false identifications of the image positions. For these reasons, we make the following assumptions prior to conducting our statistical analysis.

First, we assume that in both arcs, the brightest pixel of each knot corresponds to its image position. Thus, we implicitly are ignoring transient and microlensing effects, and are assuming a very simple surface brightness for the source. This is likely a fair assumption for AS1063 System 1; however, for the Warhol arc this is a simplification. The Warhol arc has known lensed transients \citep{yan23}, some of which are likely to be contributing to the flux of the identified knots. Thus, the treatment of the transients in Warhol is ignored for now, and left for a future analysis of this arc.

Second, and following the discussion in Section \ref{txt:paramspace}, we assume $\delta_{xy} = 0.01$" for both arcs. Given that we are assuming the brightest pixel is the image position, this assumption ascribes a high level of confidence to the image identifications. We choose to do this in order to forecast the types of constraints that can be made using these two arcs. We emphasize that future studies for these arcs should utilize a more rigorous image identification procedure than the simplified one we use here, and that that procedure should inform $\delta_{xy}$. 

Our last assumption is that the macromodel critical curve for both arcs is sufficiently straight according to the rough criterion established in Section \ref{txt:asymmetry}. We assume that we can use the reconstructed images to established the linearity of the model critical curve, using only the smooth macrolens model for the cluster. As such, we measure $\xi$ for the reconstructed images, where we find $\xi$ of -3.14 and -2.65 for AS1063 System 1 and the Warhol arc, respectively. In Section \ref{txt:asymmetry} we established an order of magnitude threshold of $\xi \lesssim -3$ for a straight critical curve. This condition is met for AS1063 System 1, but not for the Warhol arc. This is likely due to the fact that the Warhol arc is a very long parallel arc, spanning >2" along the critical curve. As a result, it is likely no longer consistent with the necessary approximation for a straight critical curve as detailed in Section \ref{txt:cclenstheory}. To investigate whether curvature along the tangential direction of the critical curve would bias our inferences of small-scale structure, we show in Appendix \ref{txt:lineartest} that bias from non-straight critical curves occurs for cases with far greater $\xi$ than considered here. 


To summarize the discussion in this Section, we are assuming that our simplified image identification procedure provides $\delta_{xy} = 0.01$". Additionally, we assume that the B19 and R25 model critical curves are sufficiently straight for both arcs, which they appear to be based on the test we do in Appendix \ref{txt:lineartest}. We acknowledge that these assumptions are strong and encourage future studies with our method to employ more rigorous procedures for image identification. Since this work is the first to attempt this statistical analysis on real arcs, we caution against interpreting our results as stringent constraints on $f_{\rm sub}$ and instead suggest viewing them as a preliminary demonstration of our method in action as applied to real data. In future work, as we will discuss further in Section \ref{txt:conclusions}, we plan to employ more sophisticated methodologies and apply the method to a larger sample of arcs to place a stronger constraint on $f_{\rm sub}$. With this established, we present our results in the following section.

\subsection{Results}\label{txt:obsresults}

\begin{figure}
    \centering
    \includegraphics[trim={0cm 0.35cm 0cm 0.35cm},clip,width=0.49\textwidth]{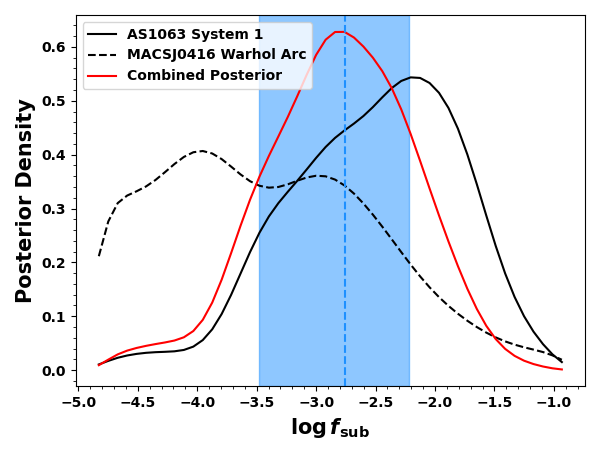}
\caption{Posterior constraints on $f_{\rm sub}$ derived with our statistical method from AS1063 System 1 (solid black line) and the Warhol arc (dashed black line). The constraints from the Warhol arc should be treated as an upper limit. The solid red line indicates the combined posterior from the two arcs assuming the asymmetry detection in Warhol is genuine. The vertical dashed blue line indicates the combined posterior median, while the shaded blue region indicates the combined posterior's 68\% confidence interval. If we combine the two arcs, we constrain $\log f_{\rm sub} = -2.76^{+0.55}_{-0.72}$. We emphasize that these results are tentative, especially for AS1063 System 1, and elaborate on the limitations further in Section \ref{txt:obsresults}.}  
\label{fig:fullconstraints}
\end{figure}

Following the statistical method outlined in Section \ref{txt:methodology}, we derive constraints on $f_{\rm sub}$ using AS1063 System 1 and the Warhol arc. With the conditions presented in Section \ref{txt:upperlimits}, we present the posteriors for the two arcs in Figure \ref{fig:fullconstraints}.

Beginning with AS1063 System 1, we constrain at 68\% CI $\log f_{\rm sub} = -2.44^{+0.61}_{-0.86}$. This is the first constraint on $f_{\rm sub}$ using our presented methodology. That being said, it is important to scrutinize the limitations of this constraint. The primary limitation concerns the underlying assumption that asymmetry is solely caused by astrometric perturbations from dark matter subhalos. It is important to realize that AS1063 System 1 is a very large perpendicular arc extending $\sim$7" perpendicularly from the critical curve, which is roughly twice the length of the fiducial perpendicular arc that we consider earlier in the paper. Therefore, it is less clear whether subhalos will be the dominant perturbing effect for the arc. Due to its size, it is possible that macrolens scale perturbations may contribute more to the arc's asymmetry. To test this further, future studies should make use of multiple macrolens density models, rather than the single model (B19) that we use here. Furthermore, the large extent of this arc may no longer be suitable for the approximations of the image midpoints lying exactly on the critical curve (equation \ref{eq:midpointposition}). We examine this possibility in greater detail in Appendix \ref{txt:higherorder}. Nonetheless, our constraint is consistent with that found in previous studies \citep{vegetti14,despali17,hsueh20}.

For the Warhol arc, we infer at 68\% CI the upper limit of $\log f_{\rm sub}$ to be $-3.48^{+1.00}_{-0.91}$. As we saw with mock arcs, the relatively low asymmetry of this arc cause $f_{\rm sub}$ to be significantly less constrained. 

Lastly, we implicitly assume that $f_{\rm sub}$ in both clusters results from the same underlying physics. As discussed in Section \ref{txt:combinedpdf}, we assume that both clusters have the same radial distribution of subhalos, have approximately the same mass, and that we are probing the cluster profile the same value of $R/R_s$. These assumptions are sufficient to first order based on CLASH measurements \citep{umetsu14}. This allows us to combine the two posteriors to achieve a joint constraint on $f_{\rm sub}$, much like in Section \ref{txt:sampleobs}. We show this combined posterior in Figure \ref{fig:fullconstraints}. We find that $\log f_{\rm sub} = -2.76^{+0.55}_{-0.72}$. Future studies with our method will deliver more precise inferences from a greater number of arcs.

\section{Conclusions}\label{txt:conclusions}

We present a new method to constrain the dark matter subhalo mass fraction within galaxy clusters. Our method uses an Approximate Bayesian Computation framework that simulates the degree of astrometric asymmetry induced by populations of dark matter subhalos near the critical curve of a cluster lens. The performance of our method is then extensively tested using mock lensed arcs generated from different macrolens profiles, arc morphologies, and astrometric uncertainties. After validating the modeling assumptions,  we apply the method to two well studied arcs to illustrate its efficacy on constraining $f_{\rm sub}$ with real data. Our main results are as follows:

\begin{itemize}
    \item Using a sample of 100 mock lensed arcs, we find that our method can recover the true $f_{\rm sub}$ to within the 68\% confidence interval $\sim73$\% of the time. This success rate is consistent for different macrolens profiles, arc morphologies, and astrometric uncertainties, illustrating its reliability for many types of lensed arcs in different observing scenarios. The constraining power of our method is stronger for mock observations with lower astrometric uncertainty, thus motivating future studies of cluster lenses to improve image identification techniques.
    \item With mock observations of samples of 3 lensed arcs, the combined posterior from the sample is able to recover the true $f_{\rm sub}$ to within the 68\% confidence interval consistently for different degrees of astrometric asymmetry. In each case, the constraining power is increased with multiple observations, demonstrating the reliability and accuracy of our method for future applications to larger samples of lensed arcs.
    \item We apply our method to two real arcs. We make use of recent parametric lens models from {\tt LensTool} for this exercise, and remeasure the arc image positions. We note that for this exercise, we make several simplifications and thus view our results as demonstrations of our method as applied to real data, rather than strict constraints on $f_{\rm sub}$. It is necessary to conduct more careful follow-up analyses for these two systems in the future. Even though our results are preliminary, it is still interesting to compare them to other results in the literature. For AS1063 System 1, we constrain $\log f_{\rm sub} = -2.44^{+0.61}_{-0.86}$ at 68\% CI, which is consistent with previous constraints \citep{vegetti14,despali17,hsueh20}. For the Warhol arc, we constrain the upper limit of $\log f_{\rm sub}$ to be $-3.48^{+1.00}_{-0.91}$ at 68\% CI. We find the combined constraint to be $\log f_{\rm sub} = -2.76^{+0.55}_{-0.72}$ at 68\%. We expect that future studies from a larger sample of lensed arcs will yield reliable, tight constraints on $f_{\rm sub}$.
\end{itemize}

There are many different directions of future work that we plan to pursue using this method. The easiest is to apply our method to a larger sample of lensed arcs, with the goal of obtaining a tight constraint on $f_{\rm sub}$. This work outlines this process using two well observed and modeled cluster lenses that allow for a quick application of our method. With recent high quality observations of cluster lenses from JWST, some cluster lenses have only very recently been updated in their lens models (e.g. the lenses observed in the Strong LensIng
and Cluster Evolution (SLICE) program \citep{cerny25}). Some candidate lensed arcs that appear to be usable with our method are the Quyllur arc in El Gordo \citep{diego23c}, Abell 68 System 1 \citep{cerny25}, Abell 2744 Systems 65 and 77 \citep{furtak23}, AS1063 System 2 \citep{diego16b}, SMACS J0723.3–7327 Systems 5 and 7 \citep{mahler23b}, and the $z\sim10$ Cosmic Gems arc \citep{bradley25}. This is not an exhaustive list, and many others will be discovered with future observations allowing for higher precision identifications of lensed knots in arcs.

One of the major simplifications in our application of our method to AS1063 System 1 and the Warhol arc was the usage of the brightest pixel positions of knots in arcs as the image positions. For this work, this is consistent with what is done in lens modelling studies, and in those cases it is sufficient since lens model reconstructions typically have worse precision than the resolution of HST and JWST. As we emphasized, this is an initial approximation that we make in order to illustrate the method on real data. In reality, more sophisticated image identification procedures are required. Ideally, image positions should be confirmed from more exact techniques, such as spectroscopy. Furthermore, some knots may not be compact enough such that the brightest pixel position is an accurate description of the image position. To account for this, the astrometric uncertainty of image positions should be determined based on the flux centroid of each knot. These suggested improvements should be considered in future studies. 

Furthermore, we only consider a single lens macromodel for our analysis of AS1063 System 1 and the Warhol arc. A necessary improvement in future works will be to repeat our method for an ensemble of lens models. Combining the posteriors from the ensemble will also yield tighter constraints on $f_{\rm sub}$ for each arc. This is especially suitable for the Warhol arc, which has been the subject of recent interest in lens modelling with a variety of methods, including parametric \citep{rihtarsic25,limousin25}, free-form \citep{cha23,perera24b}, and hybrid \citep{diego23a,cha26}. Additionally, it was recently shown that lens models of MACSJ0416 continue to exhibit wide variability in mass reconstruction, despite the large increase in image constraints \citep{perera25}. To address this, it is necessary to marginalize our method over the variety of lens models available for the Warhol arc in order to eliminate bias from individual macrolens model assumptions. As we showed in this work, our method is resistant to macrolens profile, although somewhat different inferences will be made depending on the lens model's reconstruction of the arc. For all these reasons, we recommend future studies to reexamine the Warhol arc and apply its myriad lens models with our method.

An immediate extension of our analysis in the context of CDM is the inclusion of line-of-sight halos. Another direction of future work is to apply our method to different alternative models of dark matter. In this work, we only consider standard cold dark matter subhalos. The only changes in our method when using a different dark matter model would be the subhalo density profile (which would need to be re-calibrated to the respective dark matter model properties), and the subhalo mass function. In the warm dark matter paradigm, the primary changes would be the inclusion of free-streaming effects in the calculation of the tidal evolution, and the addition of a power-law term in the SHMF that includes the half-mode mass \citep{lovell20}. Applying our method for cluster lenses to constrain the half-mode mass of warm dark matter would compliment previous galaxy-scale lens constraints \citep{gilman20}. In the self-interacting dark matter (SIDM) framework, stringent upper limits on self-interaction cross section from cluster-scale lenses have been placed \citep{jauzac16,andrade22}. SIDM halos are more cored than standard CDM halos \citep{nadler23} and have been modeled analytically \citep{hou25} or as cored NFWs \citep{gilman21,gilman23}. The SHMF can be re-parameterized based on the fraction of subhalos undergoing core-collapse as a function of mass \citep{gilman23}. Like with warm dark matter, using our method to constrain the SIDM interaction cross section would compliment recent galaxy-scale constraints \citep[e.g. ][]{kong24,tajalli25}. For wave dark matter, constraints on the axion mass from cluster lenses are relatively sparse, although there have been some recent proposed methods \citep{broadhurst25}. Wave dark matter is distinct from warm dark matter and SIDM in that it can be modeled a Gaussian random field whose fluctuations are related to the axion mass \citep{schive14}. Reworking our method to constrain the wave dark matter axion mass would be one of the first constraints at cluster-scales.

The method demonstrated in this paper has a multitude of applications and improvements that will be expanded on in future work. For now, this paper demonstrates the accuracy and reliability of the method on mock data. We have also performed first use case on two arcs to derive a constraint on $f_{\rm sub}$.


\section*{Acknowledgements}
The authors would like to thank Claudia Scarlata, John H. Miller Jr., Sung Kei Li, Patrick Kelly, Ashish Kumar Meena, and Jose Diego for useful discussions and suggestions regarding this work. The authors also thank Thomas Collett for inspiring the title of this work. DP acknowledges the Minnesota Supercomputing Institute for providing the computational resources necessary to perform this work. The collaborative effort for this work was established at the 2025 University of Hong Kong workshop, ``Dark Matter Under the Gravitational Lens''. L.D. acknowledges research grant support from the Alfred P. Sloan Foundation (Award Number FG-2021-16495), and is partially supported by the Office of Science, Office of High Energy Physics of the U.S. Department of Energy (Award Number DE-SC-0025293). D.G. acknowledges support from the Brinson Foundation through a Brinson Prize Fellowship grant. G.R. acknowledges support from the ERC Grant FIRSTLIGHT, the Slovenian national research agency ARIS through grants N1-0238 and P1-0188 and the European Space Agency through Prodex Experiment Arrangement No. 4000146646. 


\section*{Data Availability}
Data generated from this article will be shared upon reasonable request to the corresponding author.

\bibliographystyle{mnras}
\bibliography{references}

@ARTICLE{Springel2008,
       author = {{Springel}, V. and {Wang}, J. and {Vogelsberger}, M. and {Ludlow}, A. and {Jenkins}, A. and {Helmi}, A. and {Navarro}, J.~F. and {Frenk}, C.~S. and {White}, S.~D.~M.},
        title = "{The Aquarius Project: the subhaloes of galactic haloes}",
      journal = {\mnras},
         year = 2008,
        month = dec,
       volume = {391},
       number = {4},
        pages = {1685-1711},
          doi = {10.1111/j.1365-2966.2008.14066.x},
archivePrefix = {arXiv},
       eprint = {0809.0898},
 primaryClass = {astro-ph},
       adsurl = {https://ui.adsabs.harvard.edu/abs/2008MNRAS.391.1685S}
}

@ARTICLE{natarajan17,
       author = {{Natarajan}, Priyamvada and {Chadayammuri}, Urmila and {Jauzac}, Mathilde and {Richard}, Johan and {Kneib}, Jean-Paul and {Ebeling}, Harald and {Jiang}, Fangzhou and {van den Bosch}, Frank and {Limousin}, Marceau and {Jullo}, Eric and {Atek}, Hakim and {Pillepich}, Annalisa and {Popa}, Cristina and {Marinacci}, Federico and {Hernquist}, Lars and {Meneghetti}, Massimo and {Vogelsberger}, Mark},
        title = "{Mapping substructure in the HST Frontier Fields cluster lenses and in cosmological simulations}",
      journal = {\mnras},
         year = 2017,
        month = jun,
       volume = {468},
       number = {2},
        pages = {1962-1980},
          doi = {10.1093/mnras/stw3385},
archivePrefix = {arXiv},
       eprint = {1702.04348},
 primaryClass = {astro-ph.GA},
       adsurl = {https://ui.adsabs.harvard.edu/abs/2017MNRAS.468.1962N}
}

@ARTICLE{nfw97,
       author = {{Navarro}, Julio F. and {Frenk}, Carlos S. and {White}, Simon D.~M.},
        title = "{A Universal Density Profile from Hierarchical Clustering}",
      journal = {\apj},
         year = 1997,
        month = dec,
       volume = {490},
       number = {2},
        pages = {493-508},
          doi = {10.1086/304888},
archivePrefix = {arXiv},
       eprint = {astro-ph/9611107},
 primaryClass = {astro-ph},
       adsurl = {https://ui.adsabs.harvard.edu/abs/1997ApJ...490..493N}
}

@ARTICLE{broadhurst25,
       author = {{Broadhurst}, Tom and {Li}, Sung Kei and {Alfred}, Amruth and {Diego}, Jose M. and {Morilla}, Paloma and {Kelly}, Patrick L. and {Sun}, Fengwu and {Oguri}, Masamune and {Williams}, Hayley and {Windhorst}, Rogier and {Zitrin}, Adi and {Abe}, Katsuya T. and {Chen}, Wenlei and {Dai}, Liang and {Fudamoto}, Yoshinobu and {Kawai}, Hiroki and {Lim}, Jeremy and {Liu}, Tao and {Meena}, Ashish K. and {Palencia}, Jose M. and {Smoot}, George F. and {Williams}, Liliya L.~R.},
        title = "{Dark Matter Distinguished by Skewed Microlensing in the ``Dragon Arc''}",
      journal = {\apjl},
         year = 2025,
        month = jan,
       volume = {978},
       number = {1},
          eid = {L5},
        pages = {L5},
          doi = {10.3847/2041-8213/ad9aa8},
archivePrefix = {arXiv},
       eprint = {2405.19422},
 primaryClass = {astro-ph.CO},
       adsurl = {https://ui.adsabs.harvard.edu/abs/2025ApJ...978L...5B}
}

@ARTICLE{andrade22,
       author = {{Andrade}, Kevin E. and {Fuson}, Jackson and {Gad-Nasr}, Sophia and {Kong}, Demao and {Minor}, Quinn and {Roberts}, M. Grant and {Kaplinghat}, Manoj},
        title = "{A stringent upper limit on dark matter self-interaction cross-section from cluster strong lensing}",
      journal = {\mnras},
         year = 2022,
        month = feb,
       volume = {510},
       number = {1},
        pages = {54-81},
          doi = {10.1093/mnras/stab3241},
archivePrefix = {arXiv},
       eprint = {2012.06611},
 primaryClass = {astro-ph.CO},
       adsurl = {https://ui.adsabs.harvard.edu/abs/2022MNRAS.510...54A}
}

@ARTICLE{schive14,
       author = {{Schive}, Hsi-Yu and {Liao}, Ming-Hsuan and {Woo}, Tak-Pong and {Wong}, Shing-Kwong and {Chiueh}, Tzihong and {Broadhurst}, Tom and {Hwang}, W. -Y. Pauchy},
        title = "{Understanding the Core-Halo Relation of Quantum Wave Dark Matter from 3D Simulations}",
      journal = {\prl},
         year = 2014,
        month = dec,
       volume = {113},
       number = {26},
          eid = {261302},
        pages = {261302},
          doi = {10.1103/PhysRevLett.113.261302},
archivePrefix = {arXiv},
       eprint = {1407.7762},
 primaryClass = {astro-ph.GA},
       adsurl = {https://ui.adsabs.harvard.edu/abs/2014PhRvL.113z1302S}
}

@ARTICLE{bergamini23,
       author = {{Bergamini}, P. and {Grillo}, C. and {Rosati}, P. and {Vanzella}, E. and {Me{\v{s}}tri{\'c}}, U. and {Mercurio}, A. and {Acebron}, A. and {Caminha}, G.~B. and {Granata}, G. and {Meneghetti}, M. and {Angora}, G. and {Nonino}, M.},
        title = "{A state-of-the-art strong-lensing model of MACS J0416.1{\ensuremath{-}}2403 with the largest sample of spectroscopic multiple images}",
      journal = {\aap},
         year = 2023,
        month = jun,
       volume = {674},
          eid = {A79},
        pages = {A79},
          doi = {10.1051/0004-6361/202244834},
archivePrefix = {arXiv},
       eprint = {2208.14020},
 primaryClass = {astro-ph.CO},
       adsurl = {https://ui.adsabs.harvard.edu/abs/2023A&A...674A..79B}
}

@ARTICLE{kelly22,
       author = {{Kelly}, Patrick L. and {Chen}, Wenlei and {Alfred}, Amruth and {Broadhurst}, Thomas J. and {Diego}, Jose M. and {Emami}, Najmeh and {Filippenko}, Alexei V. and {Keen}, Allison and {Kei Li}, Sung and {Lim}, Jeremy and {Meena}, Ashish K. and {Oguri}, Masamune and {Scarlata}, Claudia and {Treu}, Tommaso and {Williams}, Hayley and {Williams}, Liliya L.~R. and {Zhou}, Rui and {Zitrin}, Adi and {Foley}, Ryan J. and {Jha}, Saurabh W. and {Kaiser}, Nick and {Mehta}, Vihang and {Rieck}, Steven and {Salo}, Laura and {Smith}, Nathan and {Weisz}, Daniel R.},
        title = "{Flashlights: More than A Dozen High-Significance Microlensing Events of Extremely Magnified Stars in Galaxies at Redshifts z=0.7-1.5}",
      journal = {arXiv e-prints},
         year = 2022,
        month = nov,
          eid = {arXiv:2211.02670},
        pages = {arXiv:2211.02670},
          doi = {10.48550/arXiv.2211.02670},
archivePrefix = {arXiv},
       eprint = {2211.02670},
 primaryClass = {astro-ph.CO},
       adsurl = {https://ui.adsabs.harvard.edu/abs/2022arXiv221102670K}
}

@ARTICLE{yan23,
       author = {{Yan}, Haojing and {Ma}, Zhiyuan and {Sun}, Bangzheng and {Wang}, Lifan and {Kelly}, Patrick and {Diego}, Jos{\'e} M. and {Cohen}, Seth H. and {Windhorst}, Rogier A. and {Jansen}, Rolf A. and {Grogin}, Norman A. and {Beacom}, John F. and {Conselice}, Christopher J. and {Driver}, Simon P. and {Frye}, Brenda and {Coe}, Dan and {Marshall}, Madeline A. and {Koekemoer}, Anton and {Willmer}, Christopher N.~A. and {Robotham}, Aaron and {D'Silva}, Jordan C.~J. and {Summers}, Jake and {Nonino}, Mario and {Pirzkal}, Nor and {Ryan}, Russell E. and {Ortiz}, Rafael and {Tompkins}, Scott and {Bhatawdekar}, Rachana A. and {Cheng}, Cheng and {Zitrin}, Adi and {Willner}, S.~P.},
        title = "{JWST's PEARLS: Transients in the MACS J0416.1-2403 Field}",
      journal = {\apjs},
         year = 2023,
        month = dec,
       volume = {269},
       number = {2},
          eid = {43},
        pages = {43},
          doi = {10.3847/1538-4365/ad0298},
archivePrefix = {arXiv},
       eprint = {2307.07579},
 primaryClass = {astro-ph.GA},
       adsurl = {https://ui.adsabs.harvard.edu/abs/2023ApJS..269...43Y}
}

@ARTICLE{diego23a,
       author = {{Diego}, Jose M. and {Adams}, Nathan J. and {Willner}, Steven P. and {Harvey}, Tom and {Broadhurst}, Tom and {Cohen}, Seth H. and {Jansen}, Rolf A. and {Summers}, Jake and {Windhorst}, Rogier A. and {D'Silva}, Jordan C.~J. and {Koekemoer}, Anton M. and {Coe}, Dan and {Conselice}, Christopher J. and {Driver}, Simon P. and {Frye}, Brenda and {Grogin}, Norman A. and {Marshall}, Madeline A. and {Nonino}, Mario and {Ortiz}, Rafael and {Pirzkal}, Nor and {Robotham}, Aaron and {Ryan}, Russell E. and {Willmer}, Christopher N.~A. and {Yan}, Haojing and {Sun}, Fengwu and {Hainline}, Kevin and {Berkheimer}, Jessica and {Polletta}, Maria del Carmen and {Zitrin}, Adi},
        title = "{JWST's PEARLS: 119 multiply imaged galaxies behind MACS0416, lensing properties of caustic crossing galaxies, and the relation between halo mass and number of globular clusters at z = 0.4}",
      journal = {\aap},
         year = 2024,
        month = oct,
       volume = {690},
          eid = {A114},
        pages = {A114},
          doi = {10.1051/0004-6361/202349119},
archivePrefix = {arXiv},
       eprint = {2312.11603},
 primaryClass = {astro-ph.CO},
       adsurl = {https://ui.adsabs.harvard.edu/abs/2024A&A...690A.114D}
}

@ARTICLE{cha23,
       author = {{Cha}, Sangjun and {Jee}, M. James},
        title = "{Model-independent Mass Reconstruction of the Hubble Frontier Field Clusters with MARS Based on Self-consistent Strong-lensing Data}",
      journal = {\apj},
         year = 2023,
        month = jul,
       volume = {951},
       number = {2},
          eid = {140},
        pages = {140},
          doi = {10.3847/1538-4357/acd111},
archivePrefix = {arXiv},
       eprint = {2301.08765},
 primaryClass = {astro-ph.GA},
       adsurl = {https://ui.adsabs.harvard.edu/abs/2023ApJ...951..140C}
}

@ARTICLE{bergamini19,
       author = {{Bergamini}, P. and {Rosati}, P. and {Mercurio}, A. and {Grillo}, C. and {Caminha}, G.~B. and {Meneghetti}, M. and {Agnello}, A. and {Biviano}, A. and {Calura}, F. and {Giocoli}, C. and {Lombardi}, M. and {Rodighiero}, G. and {Vanzella}, E.},
        title = "{Enhanced cluster lensing models with measured galaxy kinematics}",
      journal = {\aap},
         year = 2019,
        month = nov,
       volume = {631},
          eid = {A130},
        pages = {A130},
          doi = {10.1051/0004-6361/201935974},
archivePrefix = {arXiv},
       eprint = {1905.13236},
 primaryClass = {astro-ph.GA},
       adsurl = {https://ui.adsabs.harvard.edu/abs/2019A&A...631A.130B}
}

@ARTICLE{limousin22,
       author = {{Limousin}, Marceau and {Beauchesne}, Benjamin and {Jullo}, Eric},
        title = "{Dark matter in galaxy clusters: Parametric strong-lensing approach}",
      journal = {\aap},
         year = 2022,
        month = aug,
       volume = {664},
          eid = {A90},
        pages = {A90},
          doi = {10.1051/0004-6361/202243278},
archivePrefix = {arXiv},
       eprint = {2202.02992},
 primaryClass = {astro-ph.CO},
       adsurl = {https://ui.adsabs.harvard.edu/abs/2022A&A...664A..90L}
}

@ARTICLE{richard21,
       author = {{Richard}, Johan and {Claeyssens}, Ad{\'e}la{\"\i}de and {Lagattuta}, David and {Guaita}, Lucia and {Bauer}, Franz Erik and {Pello}, Roser and {Carton}, David and {Bacon}, Roland and {Soucail}, Genevi{\`e}ve and {Lyon}, Gonzalo Prieto and {Kneib}, Jean-Paul and {Mahler}, Guillaume and {Cl{\'e}ment}, Benjamin and {Mercier}, Wilfried and {Variu}, Andrei and {Tamone}, Am{\'e}lie and {Ebeling}, Harald and {Schmidt}, Kasper B. and {Nanayakkara}, Themiya and {Maseda}, Michael and {Weilbacher}, Peter M. and {Bouch{\'e}}, Nicolas and {Bouwens}, Rychard J. and {Wisotzki}, Lutz and {de la Vieuville}, Geoffroy and {Martinez}, Johany and {Patr{\'\i}cio}, Vera},
        title = "{An atlas of MUSE observations towards twelve massive lensing clusters}",
      journal = {\aap},
         year = 2021,
        month = feb,
       volume = {646},
          eid = {A83},
        pages = {A83},
          doi = {10.1051/0004-6361/202039462},
archivePrefix = {arXiv},
       eprint = {2009.09784},
 primaryClass = {astro-ph.GA},
       adsurl = {https://ui.adsabs.harvard.edu/abs/2021A&A...646A..83R}
}

@ARTICLE{kaurov19,
       author = {{Kaurov}, Alexander A. and {Dai}, Liang and {Venumadhav}, Tejaswi and {Miralda-Escud{\'e}}, Jordi and {Frye}, Brenda},
        title = "{Highly Magnified Stars in Lensing Clusters: New Evidence in a Galaxy Lensed by MACS J0416.1-2403}",
      journal = {\apj},
         year = 2019,
        month = jul,
       volume = {880},
       number = {1},
          eid = {58},
        pages = {58},
          doi = {10.3847/1538-4357/ab2888},
archivePrefix = {arXiv},
       eprint = {1902.10090},
 primaryClass = {astro-ph.GA},
       adsurl = {https://ui.adsabs.harvard.edu/abs/2019ApJ...880...58K}
}

@ARTICLE{chen19,
       author = {{Chen}, Wenlei and {Kelly}, Patrick L. and {Diego}, Jose M. and {Oguri}, Masamune and {Williams}, Liliya L.~R. and {Zitrin}, Adi and {Treu}, Tommaso L. and {Smith}, Nathan and {Broadhurst}, Thomas J. and {Kaiser}, Nick and {Foley}, Ryan J. and {Filippenko}, Alexei V. and {Salo}, Laura and {Hjorth}, Jens and {Selsing}, Jonatan},
        title = "{Searching for Highly Magnified Stars at Cosmological Distances: Discovery of a Redshift 0.94 Blue Supergiant in Archival Images of the Galaxy Cluster MACS J0416.1-2403}",
      journal = {\apj},
         year = 2019,
        month = aug,
       volume = {881},
       number = {1},
          eid = {8},
        pages = {8},
          doi = {10.3847/1538-4357/ab297d},
archivePrefix = {arXiv},
       eprint = {1902.05510},
 primaryClass = {astro-ph.GA},
       adsurl = {https://ui.adsabs.harvard.edu/abs/2019ApJ...881....8C}
}

@ARTICLE{caminha17,
       author = {{Caminha}, G.~B. and {Grillo}, C. and {Rosati}, P. and {Balestra}, I. and {Mercurio}, A. and {Vanzella}, E. and {Biviano}, A. and {Caputi}, K.~I. and {Delgado-Correal}, C. and {Karman}, W. and {Lombardi}, M. and {Meneghetti}, M. and {Sartoris}, B. and {Tozzi}, P.},
        title = "{A refined mass distribution of the cluster MACS J0416.1-2403 from a new large set of spectroscopic multiply lensed sources}",
      journal = {\aap},
         year = 2017,
        month = apr,
       volume = {600},
          eid = {A90},
        pages = {A90},
          doi = {10.1051/0004-6361/201629297},
archivePrefix = {arXiv},
       eprint = {1607.03462},
 primaryClass = {astro-ph.GA},
       adsurl = {https://ui.adsabs.harvard.edu/abs/2017A&A...600A..90C}
}

@ARTICLE{williams24,
       author = {{Williams}, Liliya L.~R. and {Kelly}, Patrick L. and {Treu}, Tommaso and {Amruth}, Alfred and {Diego}, Jose M. and {Li}, Sung Kei and {Meena}, Ashish K. and {Zitrin}, Adi and {Broadhurst}, Thomas J. and {Filippenko}, Alexei V.},
        title = "{Flashlights: Properties of Highly Magnified Images Near Cluster Critical Curves in the Presence of Dark Matter Subhalos}",
      journal = {\apj},
         year = 2024,
        month = feb,
       volume = {961},
       number = {2},
          eid = {200},
        pages = {200},
          doi = {10.3847/1538-4357/ad1660},
archivePrefix = {arXiv},
       eprint = {2304.06064},
 primaryClass = {astro-ph.CO},
       adsurl = {https://ui.adsabs.harvard.edu/abs/2024ApJ...961..200W}
}

@ARTICLE{venumadhav17,
       author = {{Venumadhav}, Tejaswi and {Dai}, Liang and {Miralda-Escud{\'e}}, Jordi},
        title = "{Microlensing of Extremely Magnified Stars near Caustics of Galaxy Clusters}",
      journal = {\apj},
         year = 2017,
        month = nov,
       volume = {850},
       number = {1},
          eid = {49},
        pages = {49},
          doi = {10.3847/1538-4357/aa9575},
archivePrefix = {arXiv},
       eprint = {1707.00003},
 primaryClass = {astro-ph.CO},
       adsurl = {https://ui.adsabs.harvard.edu/abs/2017ApJ...850...49V}
}

@ARTICLE{dai18,
       author = {{Dai}, Liang and {Venumadhav}, Tejaswi and {Kaurov}, Alexander A. and {Miralda-Escud}, Jordi},
        title = "{Probing Dark Matter Subhalos in Galaxy Clusters Using Highly Magnified Stars}",
      journal = {\apj},
         year = 2018,
        month = nov,
       volume = {867},
       number = {1},
          eid = {24},
        pages = {24},
          doi = {10.3847/1538-4357/aae478},
archivePrefix = {arXiv},
       eprint = {1804.03149},
 primaryClass = {astro-ph.CO},
       adsurl = {https://ui.adsabs.harvard.edu/abs/2018ApJ...867...24D}
}

@ARTICLE{dai20a,
       author = {{Dai}, Liang and {Kaurov}, Alexander A. and {Sharon}, Keren and {Florian}, Michael and {Miralda-Escud{\'e}}, Jordi and {Venumadhav}, Tejaswi and {Frye}, Brenda and {Rigby}, Jane R. and {Bayliss}, Matthew},
        title = "{Asymmetric surface brightness structure of caustic crossing arc in SDSS J1226+2152: a case for dark matter substructure}",
      journal = {\mnras},
         year = 2020,
        month = jan,
       volume = {495},
       number = {3},
        pages = {3192-3208},
          doi = {10.1093/mnras/staa1355},
archivePrefix = {arXiv},
       eprint = {2001.00261},
 primaryClass = {astro-ph.GA},
       adsurl = {https://ui.adsabs.harvard.edu/abs/2020MNRAS.495.3192D}
}

@ARTICLE{ji25,
       author = {{Ji}, Lingyuan and {Dai}, Liang},
        title = "{Effects of Subhalos on Interpreting Highly Magnified Sources Near Lensing Caustics}",
      journal = {\apj},
         year = 2025,
        month = feb,
       volume = {980},
       number = {2},
          eid = {190},
        pages = {190},
          doi = {10.3847/1538-4357/ada76a},
archivePrefix = {arXiv},
       eprint = {2407.09594},
 primaryClass = {astro-ph.GA},
       adsurl = {https://ui.adsabs.harvard.edu/abs/2025ApJ...980..190J}
}

@ARTICLE{abe23,
       author = {{Abe}, Katsuya T. and {Kawai}, Hiroki and {Oguri}, Masamune},
        title = "{Analytic approach to astrometric perturbations of critical curves by substructures}",
      journal = {\prd},
         year = 2024,
        month = apr,
       volume = {109},
       number = {8},
          eid = {083517},
        pages = {083517},
          doi = {10.1103/PhysRevD.109.083517},
archivePrefix = {arXiv},
       eprint = {2311.18211},
 primaryClass = {astro-ph.CO},
       adsurl = {https://ui.adsabs.harvard.edu/abs/2024PhRvD.109h3517A}
}

@ARTICLE{dai20c,
       author = {{Dai}, Liang and {Miralda-Escud{\'e}}, Jordi},
        title = "{Gravitational Lensing Signatures of Axion Dark Matter Minihalos in Highly Magnified Stars}",
      journal = {\aj},
         year = 2020,
        month = feb,
       volume = {159},
       number = {2},
          eid = {49},
        pages = {49},
          doi = {10.3847/1538-3881/ab5e83},
archivePrefix = {arXiv},
       eprint = {1908.01773},
 primaryClass = {astro-ph.CO},
       adsurl = {https://ui.adsabs.harvard.edu/abs/2020AJ....159...49D}
}

@ARTICLE{diego22b,
       author = {{Diego}, J.~M. and {Pascale}, M. and {Kavanagh}, B.~J. and {Kelly}, P. and {Dai}, L. and {Frye}, B. and {Broadhurst}, T.},
        title = "{Godzilla, a monster lurks in the Sunburst galaxy}",
      journal = {\aap},
         year = 2022,
        month = sep,
       volume = {665},
          eid = {A134},
        pages = {A134},
          doi = {10.1051/0004-6361/202243605},
archivePrefix = {arXiv},
       eprint = {2203.08158},
 primaryClass = {astro-ph.GA},
       adsurl = {https://ui.adsabs.harvard.edu/abs/2022A&A...665A.134D}
}

@ARTICLE{kong24,
       author = {{Kong}, Demao and {Yang}, Daneng and {Yu}, Hai-Bo},
        title = "{Cold Dark Matter and Self-interacting Dark Matter Interpretations of the Strong Gravitational Lensing Object JWST-ER1}",
      journal = {\apjl},
         year = 2024,
        month = apr,
       volume = {965},
       number = {2},
          eid = {L19},
        pages = {L19},
          doi = {10.3847/2041-8213/ad394b},
archivePrefix = {arXiv},
       eprint = {2402.15840},
 primaryClass = {astro-ph.GA},
       adsurl = {https://ui.adsabs.harvard.edu/abs/2024ApJ...965L..19K}
}

@ARTICLE{lotz17,
       author = {{Lotz}, J.~M. and {Koekemoer}, A. and {Coe}, D. and {Grogin}, N. and {Capak}, P. and {Mack}, J. and {Anderson}, J. and {Avila}, R. and {Barker}, E.~A. and {Borncamp}, D. and {Brammer}, G. and {Durbin}, M. and {Gunning}, H. and {Hilbert}, B. and {Jenkner}, H. and {Khandrika}, H. and {Levay}, Z. and {Lucas}, R.~A. and {MacKenty}, J. and {Ogaz}, S. and {Porterfield}, B. and {Reid}, N. and {Robberto}, M. and {Royle}, P. and {Smith}, L.~J. and {Storrie-Lombardi}, L.~J. and {Sunnquist}, B. and {Surace}, J. and {Taylor}, D.~C. and {Williams}, R. and {Bullock}, J. and {Dickinson}, M. and {Finkelstein}, S. and {Natarajan}, P. and {Richard}, J. and {Robertson}, B. and {Tumlinson}, J. and {Zitrin}, A. and {Flanagan}, K. and {Sembach}, K. and {Soifer}, B.~T. and {Mountain}, M.},
        title = "{The Frontier Fields: Survey Design and Initial Results}",
      journal = {\apj},
         year = 2017,
        month = mar,
       volume = {837},
       number = {1},
          eid = {97},
        pages = {97},
          doi = {10.3847/1538-4357/837/1/97},
archivePrefix = {arXiv},
       eprint = {1605.06567},
 primaryClass = {astro-ph.GA},
       adsurl = {https://ui.adsabs.harvard.edu/abs/2017ApJ...837...97L}
}

@ARTICLE{postman12,
       author = {{Postman}, Marc and {Coe}, Dan and {Ben{\'\i}tez}, Narciso and {Bradley}, Larry and {Broadhurst}, Tom and {Donahue}, Megan and {Ford}, Holland and {Graur}, Or and {Graves}, Genevieve and {Jouvel}, Stephanie and {Koekemoer}, Anton and {Lemze}, Doron and {Medezinski}, Elinor and {Molino}, Alberto and {Moustakas}, Leonidas and {Ogaz}, Sara and {Riess}, Adam and {Rodney}, Steve and {Rosati}, Piero and {Umetsu}, Keiichi and {Zheng}, Wei and {Zitrin}, Adi and {Bartelmann}, Matthias and {Bouwens}, Rychard and {Czakon}, Nicole and {Golwala}, Sunil and {Host}, Ole and {Infante}, Leopoldo and {Jha}, Saurabh and {Jimenez-Teja}, Yolanda and {Kelson}, Daniel and {Lahav}, Ofer and {Lazkoz}, Ruth and {Maoz}, Dani and {McCully}, Curtis and {Melchior}, Peter and {Meneghetti}, Massimo and {Merten}, Julian and {Moustakas}, John and {Nonino}, Mario and {Patel}, Brandon and {Reg{\"o}s}, Enik{\"o} and {Sayers}, Jack and {Seitz}, Stella and {Van der Wel}, Arjen},
        title = "{The Cluster Lensing and Supernova Survey with Hubble: An Overview}",
      journal = {\apjs},
         year = 2012,
        month = apr,
       volume = {199},
       number = {2},
          eid = {25},
        pages = {25},
          doi = {10.1088/0067-0049/199/2/25},
archivePrefix = {arXiv},
       eprint = {1106.3328},
 primaryClass = {astro-ph.CO},
       adsurl = {https://ui.adsabs.harvard.edu/abs/2012ApJS..199...25P}
}

@ARTICLE{windhorst23,
       author = {{Windhorst}, Rogier A. and {Cohen}, Seth H. and {Jansen}, Rolf A. and {Summers}, Jake and {Tompkins}, Scott and {Conselice}, Christopher J. and {Driver}, Simon P. and {Yan}, Haojing and {Coe}, Dan and {Frye}, Brenda and {Grogin}, Norman and {Koekemoer}, Anton and {Marshall}, Madeline A. and {O'Brien}, Rosalia and {Pirzkal}, Nor and {Robotham}, Aaron and {Ryan}, Russell E. and {Willmer}, Christopher N.~A. and {Carleton}, Timothy and {Diego}, Jose M. and {Keel}, William C. and {Porto}, Paolo and {Redshaw}, Caleb and {Scheller}, Sydney and {Wilkins}, Stephen M. and {Willner}, S.~P. and {Zitrin}, Adi and {Adams}, Nathan J. and {Austin}, Duncan and {Arendt}, Richard G. and {Beacom}, John F. and {Bhatawdekar}, Rachana A. and {Bradley}, Larry D. and {Broadhurst}, Tom and {Cheng}, Cheng and {Civano}, Francesca and {Dai}, Liang and {Dole}, Herv{\'e} and {D'Silva}, Jordan C.~J. and {Duncan}, Kenneth J. and {Fazio}, Giovanni G. and {Ferrami}, Giovanni and {Ferreira}, Leonardo and {Finkelstein}, Steven L. and {Furtak}, Lukas J. and {Gim}, Hansung B. and {Griffiths}, Alex and {Hammel}, Heidi B. and {Harrington}, Kevin C. and {Hathi}, Nimish P. and {Holwerda}, Benne W. and {Honor}, Rachel and {Huang}, Jia-Sheng and {Hyun}, Minhee and {Im}, Myungshin and {Joshi}, Bhavin A. and {Kamieneski}, Patrick S. and {Kelly}, Patrick and {Larson}, Rebecca L. and {Li}, Juno and {Lim}, Jeremy and {Ma}, Zhiyuan and {Maksym}, Peter and {Manzoni}, Giorgio and {Meena}, Ashish Kumar and {Milam}, Stefanie N. and {Nonino}, Mario and {Pascale}, Massimo and {Petric}, Andreea and {Pierel}, Justin D.~R. and {Polletta}, Maria del Carmen and {R{\"o}ttgering}, Huub J.~A. and {Rutkowski}, Michael J. and {Smail}, Ian and {Straughn}, Amber N. and {Strolger}, Louis-Gregory and {Swirbul}, Andi and {Trussler}, James A.~A. and {Wang}, Lifan and {Welch}, Brian and {B. Wyithe}, J. Stuart and {Yun}, Min and {Zackrisson}, Erik and {Zhang}, Jiashuo and {Zhao}, Xiurui},
        title = "{JWST PEARLS. Prime Extragalactic Areas for Reionization and Lensing Science: Project Overview and First Results}",
      journal = {\aj},
         year = 2023,
        month = jan,
       volume = {165},
       number = {1},
          eid = {13},
        pages = {13},
          doi = {10.3847/1538-3881/aca163},
archivePrefix = {arXiv},
       eprint = {2209.04119},
 primaryClass = {astro-ph.CO},
       adsurl = {https://ui.adsabs.harvard.edu/abs/2023AJ....165...13W}
}

@ARTICLE{jauzac16,
       author = {{Jauzac}, M. and {Eckert}, D. and {Schwinn}, J. and {Harvey}, D. and {Baugh}, C.~M. and {Robertson}, A. and {Bose}, S. and {Massey}, R. and {Owers}, M. and {Ebeling}, H. and {Shan}, H.~Y. and {Jullo}, E. and {Kneib}, J. -P. and {Richard}, J. and {Atek}, H. and {Cl{\'e}ment}, B. and {Egami}, E. and {Israel}, H. and {Knowles}, K. and {Limousin}, M. and {Natarajan}, P. and {Rexroth}, M. and {Taylor}, P. and {Tchernin}, C.},
        title = "{The extraordinary amount of substructure in the Hubble Frontier Fields cluster Abell 2744}",
      journal = {\mnras},
         year = 2016,
        month = dec,
       volume = {463},
       number = {4},
        pages = {3876-3893},
          doi = {10.1093/mnras/stw2251},
archivePrefix = {arXiv},
       eprint = {1606.04527},
 primaryClass = {astro-ph.CO},
       adsurl = {https://ui.adsabs.harvard.edu/abs/2016MNRAS.463.3876J}
}

@ARTICLE{rihtarsic25,
       author = {{Rihtar{\v{s}}i{\v{c}}}, G. and {Brada{\v{c}}}, M. and {Desprez}, G. and {Harshan}, A. and {Noirot}, G. and {Estrada-Carpenter}, V. and {Martis}, N.~S. and {Abraham}, R.~G. and {Asada}, Y. and {Brammer}, G. and {Iyer}, K.~G. and {Matharu}, J. and {Mowla}, L. and {Muzzin}, A. and {Sarrouh}, G.~T.~E. and {Sawicki}, M. and {Strait}, V. and {Willott}, C.~J. and {Gledhill}, R. and {Markov}, V. and {Tripodi}, R.},
        title = "{CANUCS: Constraining the MACS J0416.1-2403 strong lensing model with JWST NIRISS, NIRSpec, and NIRCam}",
      journal = {\aap},
         year = 2025,
        month = apr,
       volume = {696},
          eid = {A15},
        pages = {A15},
          doi = {10.1051/0004-6361/202451117},
archivePrefix = {arXiv},
       eprint = {2406.10332},
 primaryClass = {astro-ph.CO},
       adsurl = {https://ui.adsabs.harvard.edu/abs/2025A&A...696A..15R}
}

@ARTICLE{guzzo09,
       author = {{Guzzo}, L. and {Schuecker}, P. and {B{\"o}hringer}, H. and {Collins}, C.~A. and {Ortiz-Gil}, A. and {de Grandi}, S. and {Edge}, A.~C. and {Neumann}, D.~M. and {Schindler}, S. and {Altucci}, C. and {Shaver}, P.~A.},
        title = "{The REFLEX galaxy cluster survey. VIII. Spectroscopic observations and optical atlas,}",
      journal = {\aap},
         year = 2009,
        month = may,
       volume = {499},
       number = {2},
        pages = {357-369},
          doi = {10.1051/0004-6361/200810838},
archivePrefix = {arXiv},
       eprint = {0907.5457},
 primaryClass = {astro-ph.CO},
       adsurl = {https://ui.adsabs.harvard.edu/abs/2009A&A...499..357G}
}

@ARTICLE{balestra13,
       author = {{Balestra}, I. and {Vanzella}, E. and {Rosati}, P. and {Monna}, A. and {Grillo}, C. and {Nonino}, M. and {Mercurio}, A. and {Biviano}, A. and {Bradley}, L. and {Coe}, D. and {Fritz}, A. and {Postman}, M. and {Seitz}, S. and {Scodeggio}, M. and {Tozzi}, P. and {Zheng}, W. and {Ziegler}, B. and {Zitrin}, A. and {Annunziatella}, M. and {Bartelmann}, M. and {Benitez}, N. and {Broadhurst}, T. and {Bouwens}, R. and {Czoske}, O. and {Donahue}, M. and {Ford}, H. and {Girardi}, M. and {Infante}, L. and {Jouvel}, S. and {Kelson}, D. and {Koekemoer}, A. and {Kuchner}, U. and {Lemze}, D. and {Lombardi}, M. and {Maier}, C. and {Medezinski}, E. and {Melchior}, P. and {Meneghetti}, M. and {Merten}, J. and {Molino}, A. and {Moustakas}, L. and {Presotto}, V. and {Smit}, R. and {Umetsu}, K.},
        title = "{CLASH-VLT: spectroscopic confirmation of a z = 6.11 quintuply lensed galaxy in the Frontier Fields cluster RXC J2248.7-4431}",
      journal = {\aap},
         year = 2013,
        month = nov,
       volume = {559},
          eid = {L9},
        pages = {L9},
          doi = {10.1051/0004-6361/201322620},
archivePrefix = {arXiv},
       eprint = {1309.1593},
 primaryClass = {astro-ph.CO},
       adsurl = {https://ui.adsabs.harvard.edu/abs/2013A&A...559L...9B}
}

@ARTICLE{perera24b,
       author = {{Perera}, Derek and {Williams}, Liliya L.~R. and {Liesenborgs}, Jori and {Kelly}, Patrick L. and {Taft}, Sarah H. and {Li}, Sung Kei and {Jauzac}, Mathilde and {Diego}, Jose M. and {Natarajan}, Priyamvada and {Steinhardt}, Charles L. and {Faisst}, Andreas L. and {Rich}, R. Michael and {Limousin}, Marceau},
        title = "{BUFFALO wild wings: a high-precision free-form lens model of MACSJ0416 with constraints on dark matter from substructure and highly magnified arcs}",
      journal = {\mnras},
         year = 2025,
        month = jan,
       volume = {536},
       number = {3},
        pages = {2690-2713},
          doi = {10.1093/mnras/stae2753},
archivePrefix = {arXiv},
       eprint = {2407.15978},
 primaryClass = {astro-ph.GA},
       adsurl = {https://ui.adsabs.harvard.edu/abs/2025MNRAS.536.2690P}
}

@ARTICLE{hwilliams24,
       author = {{Williams}, Hayley and {Kelly}, Patrick and {Chen}, Wenlei and {Diego}, Jose M. and {Oguri}, Masamune and {Filippenko}, Alexei V.},
        title = "{Sp1149. II. Spectroscopy of H II Regions near the Critical Curve of MACS J1149 and Cluster Lens Models}",
      journal = {\apj},
         year = 2024,
        month = jun,
       volume = {967},
       number = {2},
          eid = {92},
        pages = {92},
          doi = {10.3847/1538-4357/ad4354},
archivePrefix = {arXiv},
       eprint = {2309.16769},
 primaryClass = {astro-ph.GA},
       adsurl = {https://ui.adsabs.harvard.edu/abs/2024ApJ...967...92W}
}

@ARTICLE{davis85,
       author = {{Davis}, M. and {Efstathiou}, G. and {Frenk}, C.~S. and {White}, S.~D.~M.},
        title = "{The evolution of large-scale structure in a universe dominated by cold dark matter}",
      journal = {\apj},
         year = 1985,
        month = may,
       volume = {292},
        pages = {371-394},
          doi = {10.1086/163168},
       adsurl = {https://ui.adsabs.harvard.edu/abs/1985ApJ...292..371D}
}

@ARTICLE{klypin99,
       author = {{Klypin}, Anatoly and {Kravtsov}, Andrey V. and {Valenzuela}, Octavio and {Prada}, Francisco},
        title = "{Where Are the Missing Galactic Satellites?}",
      journal = {\apj},
         year = 1999,
        month = sep,
       volume = {522},
       number = {1},
        pages = {82-92},
          doi = {10.1086/307643},
archivePrefix = {arXiv},
       eprint = {astro-ph/9901240},
 primaryClass = {astro-ph},
       adsurl = {https://ui.adsabs.harvard.edu/abs/1999ApJ...522...82K}
}

@ARTICLE{moore99,
       author = {{Moore}, Ben and {Ghigna}, Sebastiano and {Governato}, Fabio and {Lake}, George and {Quinn}, Thomas and {Stadel}, Joachim and {Tozzi}, Paolo},
        title = "{Dark Matter Substructure within Galactic Halos}",
      journal = {\apjl},
         year = 1999,
        month = oct,
       volume = {524},
       number = {1},
        pages = {L19-L22},
          doi = {10.1086/312287},
archivePrefix = {arXiv},
       eprint = {astro-ph/9907411},
 primaryClass = {astro-ph},
       adsurl = {https://ui.adsabs.harvard.edu/abs/1999ApJ...524L..19M}
}

@ARTICLE{furtak23,
       author = {{Furtak}, Lukas J. and {Zitrin}, Adi and {Weaver}, John R. and {Atek}, Hakim and {Bezanson}, Rachel and {Labb{\'e}}, Ivo and {Whitaker}, Katherine E. and {Leja}, Joel and {Price}, Sedona H. and {Brammer}, Gabriel B. and {Wang}, Bingjie and {Marchesini}, Danilo and {Pan}, Richard and {Dayal}, Pratika and {van Dokkum}, Pieter and {Feldmann}, Robert and {Fujimoto}, Seiji and {Franx}, Marijn and {Khullar}, Gourav and {Nelson}, Erica J. and {Mowla}, Lamiya A.},
        title = "{UNCOVERing the extended strong lensing structures of Abell 2744 with the deepest JWST imaging}",
      journal = {\mnras},
         year = 2023,
        month = aug,
       volume = {523},
       number = {3},
        pages = {4568-4582},
          doi = {10.1093/mnras/stad1627},
archivePrefix = {arXiv},
       eprint = {2212.04381},
 primaryClass = {astro-ph.GA},
       adsurl = {https://ui.adsabs.harvard.edu/abs/2023MNRAS.523.4568F}
}

@ARTICLE{du25,
       author = {{Du}, Xiaolong and {Gilman}, Daniel and {Treu}, Tommaso and {Benson}, Andrew and {Gannon}, Charles},
        title = "{Empirical model for the tidal evolution of dark matter substructure around strong gravitational lenses}",
      journal = {\prd},
         year = 2025,
        month = jul,
       volume = {112},
       number = {2},
          eid = {023009},
        pages = {023009},
          doi = {10.1103/6tbt-w3nv},
archivePrefix = {arXiv},
       eprint = {2503.07728},
 primaryClass = {astro-ph.CO},
       adsurl = {https://ui.adsabs.harvard.edu/abs/2025PhRvD.112b3009D}
}

@ARTICLE{gilman19,
       author = {{Gilman}, Daniel and {Birrer}, Simon and {Treu}, Tommaso and {Nierenberg}, Anna and {Benson}, Andrew},
        title = "{Probing dark matter structure down to {}10$^{7}$ solar masses: flux ratio statistics in gravitational lenses with line-of-sight haloes}",
      journal = {\mnras},
         year = 2019,
        month = aug,
       volume = {487},
       number = {4},
        pages = {5721-5738},
          doi = {10.1093/mnras/stz1593},
archivePrefix = {arXiv},
       eprint = {1901.11031},
 primaryClass = {astro-ph.CO},
       adsurl = {https://ui.adsabs.harvard.edu/abs/2019MNRAS.487.5721G}
}

@ARTICLE{hinshaw87,
       author = {{Hinshaw}, Gary and {Krauss}, Lawrence M.},
        title = "{Gravitational Lensing by Isothermal Spheres with Finite Core Radii: Galaxies and Dark Matter}",
      journal = {\apj},
         year = 1987,
        month = sep,
       volume = {320},
        pages = {468},
          doi = {10.1086/165564},
       adsurl = {https://ui.adsabs.harvard.edu/abs/1987ApJ...320..468H}
}

@ARTICLE{halkola06,
       author = {{Halkola}, A. and {Seitz}, S. and {Pannella}, M.},
        title = "{Parametric strong gravitational lensing analysis of Abell 1689}",
      journal = {\mnras},
         year = 2006,
        month = nov,
       volume = {372},
       number = {4},
        pages = {1425-1462},
          doi = {10.1111/j.1365-2966.2006.10948.x},
archivePrefix = {arXiv},
       eprint = {astro-ph/0605470},
 primaryClass = {astro-ph},
       adsurl = {https://ui.adsabs.harvard.edu/abs/2006MNRAS.372.1425H}
}

@ARTICLE{gilman20,
       author = {{Gilman}, Daniel and {Birrer}, Simon and {Nierenberg}, Anna and {Treu}, Tommaso and {Du}, Xiaolong and {Benson}, Andrew},
        title = "{Warm dark matter chills out: constraints on the halo mass function and the free-streaming length of dark matter with eight quadruple-image strong gravitational lenses}",
      journal = {\mnras},
         year = 2020,
        month = feb,
       volume = {491},
       number = {4},
        pages = {6077-6101},
          doi = {10.1093/mnras/stz3480},
archivePrefix = {arXiv},
       eprint = {1908.06983},
 primaryClass = {astro-ph.CO},
       adsurl = {https://ui.adsabs.harvard.edu/abs/2020MNRAS.491.6077G}
}

@ARTICLE{gilman21,
       author = {{Gilman}, Daniel and {Bovy}, Jo and {Treu}, Tommaso and {Nierenberg}, Anna and {Birrer}, Simon and {Benson}, Andrew and {Sameie}, Omid},
        title = "{Strong lensing signatures of self-interacting dark matter in low-mass haloes}",
      journal = {\mnras},
         year = 2021,
        month = oct,
       volume = {507},
       number = {2},
        pages = {2432-2447},
          doi = {10.1093/mnras/stab2335},
archivePrefix = {arXiv},
       eprint = {2105.05259},
 primaryClass = {astro-ph.CO},
       adsurl = {https://ui.adsabs.harvard.edu/abs/2021MNRAS.507.2432G}
}

@ARTICLE{cerny25,
       author = {{Cerny}, Catherine and {Mahler}, Guillaume and {Sharon}, Keren and {Jauzac}, Mathilde and {Khullar}, Gourav and {Beauchesne}, Benjamin and {Diego}, Jose M. and {Lagattuta}, David J. and {Limousin}, Marceau and {Patel}, Nency R. and {Richard}, Johan and {Cornil-Baiotto}, Carla and {Gladders}, Michael D. and {Werner}, Stephane and {Doppel}, Jessica E. and {Floyd}, Benjamin and {Gonzalez}, Anthony H. and {Massey}, Richard J. and {Montes}, Mireia and {Bayliss}, Matthew B. and {Bleem}, Lindsey E. and {Canning}, Rebecca E.~A. and {Edge}, Alastair C. and {McDonald}, Michael and {Natarjan}, Priyamvada and {Stark}, Anthony A. and {Gassis}, Raven},
        title = "{Strong LensIng and Cluster Evolution (SLICE) with JWST: Early Results, Lens Models, and High-Redshift Detections}",
      journal = {arXiv e-prints},
         year = 2025,
        month = mar,
          eid = {arXiv:2503.17498},
        pages = {arXiv:2503.17498},
          doi = {10.48550/arXiv.2503.17498},
archivePrefix = {arXiv},
       eprint = {2503.17498},
 primaryClass = {astro-ph.CO},
       adsurl = {https://ui.adsabs.harvard.edu/abs/2025arXiv250317498C}
}

@ARTICLE{diego23c,
       author = {{Diego}, J.~M. and {Meena}, A.~K. and {Adams}, N.~J. and {Broadhurst}, T. and {Dai}, L. and {Coe}, D. and {Frye}, B. and {Kelly}, P. and {Koekemoer}, A.~M. and {Pascale}, M. and {Willner}, S.~P. and {Zackrisson}, E. and {Zitrin}, A. and {Windhorst}, R.~A. and {Cohen}, S.~H. and {Jansen}, R.~A. and {Summers}, J. and {Tompkins}, S. and {Conselice}, C.~J. and {Driver}, S.~P. and {Yan}, H. and {Grogin}, N. and {Marshall}, M.~A. and {Pirzkal}, N. and {Robotham}, A. and {Ryan}, R.~E. and {Willmer}, C.~N.~A. and {Bradley}, L.~D. and {Caminha}, G. and {Caputi}, K. and {Carleton}, T. and {Kamieneski}, P.},
        title = "{JWST's PEARLS: A new lens model for ACT-CL J0102{\ensuremath{-}}4915, ``El Gordo,'' and the first red supergiant star at cosmological distances discovered by JWST}",
      journal = {\aap},
         year = 2023,
        month = apr,
       volume = {672},
          eid = {A3},
        pages = {A3},
          doi = {10.1051/0004-6361/202245238},
archivePrefix = {arXiv},
       eprint = {2210.06514},
 primaryClass = {astro-ph.GA},
       adsurl = {https://ui.adsabs.harvard.edu/abs/2023A&A...672A...3D}
}

@ARTICLE{hsueh20,
       author = {{Hsueh}, J. -W. and {Enzi}, W. and {Vegetti}, S. and {Auger}, M.~W. and {Fassnacht}, C.~D. and {Despali}, G. and {Koopmans}, L.~V.~E. and {McKean}, J.~P.},
        title = "{SHARP - VII. New constraints on the dark matter free-streaming properties and substructure abundance from gravitationally lensed quasars}",
      journal = {\mnras},
         year = 2020,
        month = feb,
       volume = {492},
       number = {2},
        pages = {3047-3059},
          doi = {10.1093/mnras/stz3177},
archivePrefix = {arXiv},
       eprint = {1905.04182},
 primaryClass = {astro-ph.CO},
       adsurl = {https://ui.adsabs.harvard.edu/abs/2020MNRAS.492.3047H}
}

@ARTICLE{gilman24,
       author = {{Gilman}, Daniel and {Birrer}, Simon and {Nierenberg}, Anna and {Oh}, Maverick S.~H.},
        title = "{Turbocharging constraints on dark matter substructure through a synthesis of strong lensing flux ratios and extended lensed arcs}",
      journal = {\mnras},
         year = 2024,
        month = sep,
       volume = {533},
       number = {2},
        pages = {1687-1713},
          doi = {10.1093/mnras/stae1810},
archivePrefix = {arXiv},
       eprint = {2403.03253},
 primaryClass = {astro-ph.CO},
       adsurl = {https://ui.adsabs.harvard.edu/abs/2024MNRAS.533.1687G}
}

@ARTICLE{diego16b,
       author = {{Diego}, Jose M. and {Broadhurst}, Tom and {Wong}, Jess and {Silk}, Joseph and {Lim}, Jeremy and {Zheng}, Wei and {Lam}, Daniel and {Ford}, Holland},
        title = "{A free-form mass model of the Hubble Frontier Fields cluster AS1063 (RXC J2248.7-4431) with over one hundred constraints}",
      journal = {\mnras},
         year = 2016,
        month = jul,
       volume = {459},
       number = {4},
        pages = {3447-3459},
          doi = {10.1093/mnras/stw865},
archivePrefix = {arXiv},
       eprint = {1512.07916},
 primaryClass = {astro-ph.CO},
       adsurl = {https://ui.adsabs.harvard.edu/abs/2016MNRAS.459.3447D}
}

@ARTICLE{mahler23b,
       author = {{Mahler}, Guillaume and {Jauzac}, Mathilde and {Richard}, Johan and {Beauchesne}, Benjamin and {Ebeling}, Harald and {Lagattuta}, David and {Natarajan}, Priyamvada and {Sharon}, Keren and {Atek}, Hakim and {Claeyssens}, Ad{\'e}la{\"\i}de and {Cl{\'e}ment}, Benjamin and {Eckert}, Dominique and {Edge}, Alastair and {Kneib}, Jean-Paul and {Niemiec}, Anna},
        title = "{Precision Modeling of JWST's First Cluster Lens SMACS J0723.3-7327}",
      journal = {\apj},
         year = 2023,
        month = mar,
       volume = {945},
       number = {1},
          eid = {49},
        pages = {49},
          doi = {10.3847/1538-4357/acaea9},
archivePrefix = {arXiv},
       eprint = {2207.07101},
 primaryClass = {astro-ph.GA},
       adsurl = {https://ui.adsabs.harvard.edu/abs/2023ApJ...945...49M}
}

@ARTICLE{narayan96,
       author = {{Narayan}, Ramesh and {Bartelmann}, Matthias},
        title = "{Lectures on Gravitational Lensing}",
      journal = {arXiv e-prints},
         year = 1996,
        month = jun,
          eid = {astro-ph/9606001},
        pages = {astro-ph/9606001},
          doi = {10.48550/arXiv.astro-ph/9606001},
archivePrefix = {arXiv},
       eprint = {astro-ph/9606001},
 primaryClass = {astro-ph},
       adsurl = {https://ui.adsabs.harvard.edu/abs/1996astro.ph..6001N}
}

@ARTICLE{benson12,
       author = {{Benson}, Andrew J.},
        title = "{G ALACTICUS: A semi-analytic model of galaxy formation}",
      journal = {\na},
         year = 2012,
        month = feb,
       volume = {17},
       number = {2},
        pages = {175-197},
          doi = {10.1016/j.newast.2011.07.004},
archivePrefix = {arXiv},
       eprint = {1008.1786},
 primaryClass = {astro-ph.CO},
       adsurl = {https://ui.adsabs.harvard.edu/abs/2012NewA...17..175B}
}

@ARTICLE{du24,
       author = {{Du}, Xiaolong and {Benson}, Andrew and {Zeng}, Zhichao Carton and {Treu}, Tommaso and {Peter}, Annika H.~G. and {Mace}, Charlie and {Jiang}, Fangzhou and {Yang}, Shengqi and {Gannon}, Charles and {Gilman}, Daniel and {Nierenberg}, Anna. M. and {Nadler}, Ethan O.},
        title = "{Tidal evolution of cored and cuspy dark matter halos}",
      journal = {\prd},
         year = 2024,
        month = jul,
       volume = {110},
       number = {2},
          eid = {023019},
        pages = {023019},
          doi = {10.1103/PhysRevD.110.023019},
archivePrefix = {arXiv},
       eprint = {2403.09597},
 primaryClass = {astro-ph.GA},
       adsurl = {https://ui.adsabs.harvard.edu/abs/2024PhRvD.110b3019D}
}

@ARTICLE{giocoli08,
       author = {{Giocoli}, Carlo and {Tormen}, Giuseppe and {van den Bosch}, Frank C.},
        title = "{The population of dark matter subhaloes: mass functions and average mass-loss rates}",
      journal = {\mnras},
         year = 2008,
        month = jun,
       volume = {386},
       number = {4},
        pages = {2135-2144},
          doi = {10.1111/j.1365-2966.2008.13182.x},
archivePrefix = {arXiv},
       eprint = {0712.1563},
 primaryClass = {astro-ph},
       adsurl = {https://ui.adsabs.harvard.edu/abs/2008MNRAS.386.2135G}
}

@ARTICLE{han16,
       author = {{Han}, Jiaxin and {Cole}, Shaun and {Frenk}, Carlos S. and {Jing}, Yipeng},
        title = "{A unified model for the spatial and mass distribution of subhaloes}",
      journal = {\mnras},
         year = 2016,
        month = apr,
       volume = {457},
       number = {2},
        pages = {1208-1223},
          doi = {10.1093/mnras/stv2900},
archivePrefix = {arXiv},
       eprint = {1509.02175},
 primaryClass = {astro-ph.CO},
       adsurl = {https://ui.adsabs.harvard.edu/abs/2016MNRAS.457.1208H}
}

@ARTICLE{hayashi03,
       author = {{Hayashi}, Eric and {Navarro}, Julio F. and {Taylor}, James E. and {Stadel}, Joachim and {Quinn}, Thomas},
        title = "{The Structural Evolution of Substructure}",
      journal = {\apj},
         year = 2003,
        month = feb,
       volume = {584},
       number = {2},
        pages = {541-558},
          doi = {10.1086/345788},
archivePrefix = {arXiv},
       eprint = {astro-ph/0203004},
 primaryClass = {astro-ph},
       adsurl = {https://ui.adsabs.harvard.edu/abs/2003ApJ...584..541H}
}

@ARTICLE{blandford86,
       author = {{Blandford}, Roger and {Narayan}, Ramesh},
        title = "{Fermat's Principle, Caustics, and the Classification of Gravitational Lens Images}",
      journal = {\apj},
         year = 1986,
        month = nov,
       volume = {310},
        pages = {568},
          doi = {10.1086/164709},
       adsurl = {https://ui.adsabs.harvard.edu/abs/1986ApJ...310..568B}
}

@ARTICLE{errani21,
       author = {{Errani}, Rapha{\"e}l and {Navarro}, Julio F.},
        title = "{The asymptotic tidal remnants of cold dark matter subhaloes}",
      journal = {\mnras},
         year = 2021,
        month = jul,
       volume = {505},
       number = {1},
        pages = {18-32},
          doi = {10.1093/mnras/stab1215},
archivePrefix = {arXiv},
       eprint = {2011.07077},
 primaryClass = {astro-ph.GA},
       adsurl = {https://ui.adsabs.harvard.edu/abs/2021MNRAS.505...18E}
}

@BOOK{schneider92,
       author = {{Schneider}, Peter and {Ehlers}, J{\"u}rgen and {Falco}, Emilio E.},
        title = "{Gravitational Lenses}",
         year = 1992,
          doi = {10.1007/978-3-662-03758-4},
       adsurl = {https://ui.adsabs.harvard.edu/abs/1992grle.book.....S}
}

@ARTICLE{kneib96,
       author = {{Kneib}, J. -P. and {Ellis}, R.~S. and {Smail}, I. and {Couch}, W.~J. and {Sharples}, R.~M.},
        title = "{Hubble Space Telescope Observations of the Lensing Cluster Abell 2218}",
      journal = {\apj},
         year = 1996,
        month = nov,
       volume = {471},
        pages = {643},
          doi = {10.1086/177995},
archivePrefix = {arXiv},
       eprint = {astro-ph/9511015},
 primaryClass = {astro-ph},
       adsurl = {https://ui.adsabs.harvard.edu/abs/1996ApJ...471..643K}
}

@ARTICLE{jullo07,
       author = {{Jullo}, E. and {Kneib}, J. -P. and {Limousin}, M. and {El{\'\i}asd{\'o}ttir}, {\'A}. and {Marshall}, P.~J. and {Verdugo}, T.},
        title = "{A Bayesian approach to strong lensing modelling of galaxy clusters}",
      journal = {New Journal of Physics},
         year = 2007,
        month = dec,
       volume = {9},
       number = {12},
        pages = {447},
          doi = {10.1088/1367-2630/9/12/447},
archivePrefix = {arXiv},
       eprint = {0706.0048},
 primaryClass = {astro-ph},
       adsurl = {https://ui.adsabs.harvard.edu/abs/2007NJPh....9..447J}
}

@ARTICLE{limousin25,
       author = {{Limousin}, Marceau and {Perera}, Derek and {Rihtar{\v{s}}i{\v{c}}}, Gregor and {Williams}, Liliya L.~R. and {Liesenborgs}, Jori},
        title = "{Testing light-unaffiliated mass clumps in MACS 0416 on galaxy and galaxy-cluster scales using the JWST}",
      journal = {\aap},
         year = 2025,
        month = oct,
       volume = {703},
          eid = {A10},
        pages = {A10},
          doi = {10.1051/0004-6361/202556007},
       adsurl = {https://ui.adsabs.harvard.edu/abs/2025A&A...703A..10L}
}

@ARTICLE{cha26,
       author = {{Cha}, Sangjun and {Jee}, M. James},
        title = "{MrMARTIAN: A Multiresolution Mass Reconstruction Algorithm Combining Free-form and Analytic Components}",
      journal = {\apj},
         year = 2026,
        month = jan,
       volume = {997},
       number = {1},
          eid = {18},
        pages = {18},
          doi = {10.3847/1538-4357/ae2258},
archivePrefix = {arXiv},
       eprint = {2508.13262},
 primaryClass = {astro-ph.GA},
       adsurl = {https://ui.adsabs.harvard.edu/abs/2026ApJ...997...18C}
}

@ARTICLE{perera25,
       author = {{Perera}, Derek and {Jr}, John H Miller and {Williams}, Liliya L.~R. and {Liesenborgs}, Jori and {Keen}, Allison and {Li}, Sung Kei and {Limousin}, Marceau},
        title = "{Are Models of Strong Gravitational Lensing by Clusters Converging or Diverging?}",
      journal = {The Open Journal of Astrophysics},
         year = 2025,
        month = apr,
       volume = {8},
          eid = {37},
        pages = {37},
          doi = {10.33232/001c.136341},
archivePrefix = {arXiv},
       eprint = {2411.05083},
 primaryClass = {astro-ph.GA},
       adsurl = {https://ui.adsabs.harvard.edu/abs/2025OJAp....8E..37P}
}

@ARTICLE{lovell20,
       author = {{Lovell}, Mark R.},
        title = "{Toward a General Parameterization of the Warm Dark Matter Halo Mass Function}",
      journal = {\apj},
         year = 2020,
        month = jul,
       volume = {897},
       number = {2},
          eid = {147},
        pages = {147},
          doi = {10.3847/1538-4357/ab982a},
archivePrefix = {arXiv},
       eprint = {2003.01125},
 primaryClass = {astro-ph.CO},
       adsurl = {https://ui.adsabs.harvard.edu/abs/2020ApJ...897..147L}
}

@ARTICLE{nadler23,
       author = {{Nadler}, Ethan O. and {Yang}, Daneng and {Yu}, Hai-Bo},
        title = "{A Self-interacting Dark Matter Solution to the Extreme Diversity of Low-mass Halo Properties}",
      journal = {\apjl},
         year = 2023,
        month = dec,
       volume = {958},
       number = {2},
          eid = {L39},
        pages = {L39},
          doi = {10.3847/2041-8213/ad0e09},
archivePrefix = {arXiv},
       eprint = {2306.01830},
 primaryClass = {astro-ph.GA},
       adsurl = {https://ui.adsabs.harvard.edu/abs/2023ApJ...958L..39N}
}

@ARTICLE{gilman23,
       author = {{Gilman}, Daniel and {Zhong}, Yi-Ming and {Bovy}, Jo},
        title = "{Constraining resonant dark matter self-interactions with strong gravitational lenses}",
      journal = {\prd},
         year = 2023,
        month = may,
       volume = {107},
       number = {10},
          eid = {103008},
        pages = {103008},
          doi = {10.1103/PhysRevD.107.103008},
archivePrefix = {arXiv},
       eprint = {2207.13111},
 primaryClass = {astro-ph.CO},
       adsurl = {https://ui.adsabs.harvard.edu/abs/2023PhRvD.107j3008G}
}

@ARTICLE{hou25,
       author = {{Hou}, Siyuan and {Yang}, Daneng and {Li}, Nan and {Li}, Guoliang},
        title = "{A universal analytic model for gravitational lensing by self-interacting dark matter halos}",
      journal = {\jcap},
         year = 2025,
        month = aug,
       volume = {2025},
       number = {8},
          eid = {048},
        pages = {048},
          doi = {10.1088/1475-7516/2025/08/048},
archivePrefix = {arXiv},
       eprint = {2502.14964},
 primaryClass = {astro-ph.CO},
       adsurl = {https://ui.adsabs.harvard.edu/abs/2025JCAP...08..048H}
}

@ARTICLE{tajalli25,
       author = {{Tajalli}, M. and {Vegetti}, S. and {O'Riordan}, C.~M. and {White}, S.~D.~M. and {Fassnacht}, C.~D. and {Powell}, D.~M. and {McKean}, J.~P. and {Despali}, G.},
        title = "{SHARP - IX. The dense, low-mass perturbers in B1938+666 and J0946+1006: implications for cold and self-interacting dark matter}",
      journal = {\mnras},
         year = 2025,
        month = aug,
          doi = {10.1093/mnras/staf1357},
archivePrefix = {arXiv},
       eprint = {2505.07944},
 primaryClass = {astro-ph.CO},
       adsurl = {https://ui.adsabs.harvard.edu/abs/2025MNRAS.tmp.1312T}
}

@ARTICLE{despali17,
       author = {{Despali}, Giulia and {Vegetti}, Simona},
        title = "{The impact of baryonic physics on the subhalo mass function and implications for gravitational lensing}",
      journal = {\mnras},
         year = 2017,
        month = aug,
       volume = {469},
       number = {2},
        pages = {1997-2010},
          doi = {10.1093/mnras/stx966},
archivePrefix = {arXiv},
       eprint = {1608.06938},
 primaryClass = {astro-ph.GA},
       adsurl = {https://ui.adsabs.harvard.edu/abs/2017MNRAS.469.1997D}
}

@ARTICLE{vegetti14,
       author = {{Vegetti}, S. and {Koopmans}, L.~V.~E. and {Auger}, M.~W. and {Treu}, T. and {Bolton}, A.~S.},
        title = "{Inference of the cold dark matter substructure mass function at z = 0.2 using strong gravitational lenses}",
      journal = {\mnras},
         year = 2014,
        month = aug,
       volume = {442},
       number = {3},
        pages = {2017-2035},
          doi = {10.1093/mnras/stu943},
archivePrefix = {arXiv},
       eprint = {1405.3666},
 primaryClass = {astro-ph.GA},
       adsurl = {https://ui.adsabs.harvard.edu/abs/2014MNRAS.442.2017V}
}

@ARTICLE{planck20,
       author = {{Planck Collaboration} and {Aghanim}, N. and {Akrami}, Y. and {Ashdown}, M. and {Aumont}, J. and {Baccigalupi}, C. and {Ballardini}, M. and {Banday}, A.~J. and {Barreiro}, R.~B. and {Bartolo}, N. and {Basak}, S. and {Battye}, R. and {Benabed}, K. and {Bernard}, J. -P. and {Bersanelli}, M. and {Bielewicz}, P. and {Bock}, J.~J. and {Bond}, J.~R. and {Borrill}, J. and {Bouchet}, F.~R. and {Boulanger}, F. and {Bucher}, M. and {Burigana}, C. and {Butler}, R.~C. and {Calabrese}, E. and {Cardoso}, J. -F. and {Carron}, J. and {Challinor}, A. and {Chiang}, H.~C. and {Chluba}, J. and {Colombo}, L.~P.~L. and {Combet}, C. and {Contreras}, D. and {Crill}, B.~P. and {Cuttaia}, F. and {de Bernardis}, P. and {de Zotti}, G. and {Delabrouille}, J. and {Delouis}, J. -M. and {Di Valentino}, E. and {Diego}, J.~M. and {Dor{\'e}}, O. and {Douspis}, M. and {Ducout}, A. and {Dupac}, X. and {Dusini}, S. and {Efstathiou}, G. and {Elsner}, F. and {En{\ss}lin}, T.~A. and {Eriksen}, H.~K. and {Fantaye}, Y. and {Farhang}, M. and {Fergusson}, J. and {Fernandez-Cobos}, R. and {Finelli}, F. and {Forastieri}, F. and {Frailis}, M. and {Fraisse}, A.~A. and {Franceschi}, E. and {Frolov}, A. and {Galeotta}, S. and {Galli}, S. and {Ganga}, K. and {G{\'e}nova-Santos}, R.~T. and {Gerbino}, M. and {Ghosh}, T. and {Gonz{\'a}lez-Nuevo}, J. and {G{\'o}rski}, K.~M. and {Gratton}, S. and {Gruppuso}, A. and {Gudmundsson}, J.~E. and {Hamann}, J. and {Handley}, W. and {Hansen}, F.~K. and {Herranz}, D. and {Hildebrandt}, S.~R. and {Hivon}, E. and {Huang}, Z. and {Jaffe}, A.~H. and {Jones}, W.~C. and {Karakci}, A. and {Keih{\"a}nen}, E. and {Keskitalo}, R. and {Kiiveri}, K. and {Kim}, J. and {Kisner}, T.~S. and {Knox}, L. and {Krachmalnicoff}, N. and {Kunz}, M. and {Kurki-Suonio}, H. and {Lagache}, G. and {Lamarre}, J. -M. and {Lasenby}, A. and {Lattanzi}, M. and {Lawrence}, C.~R. and {Le Jeune}, M. and {Lemos}, P. and {Lesgourgues}, J. and {Levrier}, F. and {Lewis}, A. and {Liguori}, M. and {Lilje}, P.~B. and {Lilley}, M. and {Lindholm}, V. and {L{\'o}pez-Caniego}, M. and {Lubin}, P.~M. and {Ma}, Y. -Z. and {Mac{\'\i}as-P{\'e}rez}, J.~F. and {Maggio}, G. and {Maino}, D. and {Mandolesi}, N. and {Mangilli}, A. and {Marcos-Caballero}, A. and {Maris}, M. and {Martin}, P.~G. and {Martinelli}, M. and {Mart{\'\i}nez-Gonz{\'a}lez}, E. and {Matarrese}, S. and {Mauri}, N. and {McEwen}, J.~D. and {Meinhold}, P.~R. and {Melchiorri}, A. and {Mennella}, A. and {Migliaccio}, M. and {Millea}, M. and {Mitra}, S. and {Miville-Desch{\^e}nes}, M. -A. and {Molinari}, D. and {Montier}, L. and {Morgante}, G. and {Moss}, A. and {Natoli}, P. and {N{\o}rgaard-Nielsen}, H.~U. and {Pagano}, L. and {Paoletti}, D. and {Partridge}, B. and {Patanchon}, G. and {Peiris}, H.~V. and {Perrotta}, F. and {Pettorino}, V. and {Piacentini}, F. and {Polastri}, L. and {Polenta}, G. and {Puget}, J. -L. and {Rachen}, J.~P. and {Reinecke}, M. and {Remazeilles}, M. and {Renzi}, A. and {Rocha}, G. and {Rosset}, C. and {Roudier}, G. and {Rubi{\~n}o-Mart{\'\i}n}, J.~A. and {Ruiz-Granados}, B. and {Salvati}, L. and {Sandri}, M. and {Savelainen}, M. and {Scott}, D. and {Shellard}, E.~P.~S. and {Sirignano}, C. and {Sirri}, G. and {Spencer}, L.~D. and {Sunyaev}, R. and {Suur-Uski}, A. -S. and {Tauber}, J.~A. and {Tavagnacco}, D. and {Tenti}, M. and {Toffolatti}, L. and {Tomasi}, M. and {Trombetti}, T. and {Valenziano}, L. and {Valiviita}, J. and {Van Tent}, B. and {Vibert}, L. and {Vielva}, P. and {Villa}, F. and {Vittorio}, N. and {Wandelt}, B.~D. and {Wehus}, I.~K. and {White}, M. and {White}, S.~D.~M. and {Zacchei}, A. and {Zonca}, A.},
        title = "{Planck 2018 results. VI. Cosmological parameters}",
      journal = {\aap},
         year = 2020,
        month = sep,
       volume = {641},
          eid = {A6},
        pages = {A6},
          doi = {10.1051/0004-6361/201833910},
archivePrefix = {arXiv},
       eprint = {1807.06209},
 primaryClass = {astro-ph.CO},
       adsurl = {https://ui.adsabs.harvard.edu/abs/2020A&A...641A...6P}
}

@ARTICLE{delpopolo17,
       author = {{Del Popolo}, Antonino and {Le Delliou}, Morgan},
        title = "{Small Scale Problems of the {\ensuremath{\Lambda}}CDM Model: A Short Review}",
      journal = {Galaxies},
         year = 2017,
        month = feb,
       volume = {5},
       number = {1},
          eid = {17},
        pages = {17},
          doi = {10.3390/galaxies5010017},
archivePrefix = {arXiv},
       eprint = {1606.07790},
 primaryClass = {astro-ph.CO},
       adsurl = {https://ui.adsabs.harvard.edu/abs/2017Galax...5...17D}
}

@ARTICLE{birrer17,
       author = {{Birrer}, Simon and {Amara}, Adam and {Refregier}, Alexandre},
        title = "{Lensing substructure quantification in RXJ1131-1231: a 2 keV lower bound on dark matter thermal relic mass}",
      journal = {\jcap},
         year = 2017,
        month = may,
       volume = {2017},
       number = {5},
          eid = {037},
        pages = {037},
          doi = {10.1088/1475-7516/2017/05/037},
archivePrefix = {arXiv},
       eprint = {1702.00009},
 primaryClass = {astro-ph.CO},
       adsurl = {https://ui.adsabs.harvard.edu/abs/2017JCAP...05..037B}
}

@ARTICLE{dalal02,
       author = {{Dalal}, N. and {Kochanek}, C.~S.},
        title = "{Direct Detection of Cold Dark Matter Substructure}",
      journal = {\apj},
         year = 2002,
        month = jun,
       volume = {572},
       number = {1},
        pages = {25-33},
          doi = {10.1086/340303},
archivePrefix = {arXiv},
       eprint = {astro-ph/0111456},
 primaryClass = {astro-ph},
       adsurl = {https://ui.adsabs.harvard.edu/abs/2002ApJ...572...25D}
}

@ARTICLE{he22,
       author = {{He}, Qiuhan and {Robertson}, Andrew and {Nightingale}, James and {Cole}, Shaun and {Frenk}, Carlos S. and {Massey}, Richard and {Amvrosiadis}, Aristeidis and {Li}, Ran and {Cao}, Xiaoyue and {Etherington}, Amy},
        title = "{A forward-modelling method to infer the dark matter particle mass from strong gravitational lenses}",
      journal = {\mnras},
         year = 2022,
        month = apr,
       volume = {511},
       number = {2},
        pages = {3046-3062},
          doi = {10.1093/mnras/stac191},
archivePrefix = {arXiv},
       eprint = {2010.13221},
 primaryClass = {astro-ph.CO},
       adsurl = {https://ui.adsabs.harvard.edu/abs/2022MNRAS.511.3046H}
}

@ARTICLE{natarajan04,
       author = {{Natarajan}, Priyamvada and {Springel}, Volker},
        title = "{Abundance of Substructure in Clusters of Galaxies}",
      journal = {\apjl},
         year = 2004,
        month = dec,
       volume = {617},
       number = {1},
        pages = {L13-L16},
          doi = {10.1086/427079},
archivePrefix = {arXiv},
       eprint = {astro-ph/0411515},
 primaryClass = {astro-ph},
       adsurl = {https://ui.adsabs.harvard.edu/abs/2004ApJ...617L..13N}
}

@ARTICLE{umetsu16,
       author = {{Umetsu}, Keiichi and {Zitrin}, Adi and {Gruen}, Daniel and {Merten}, Julian and {Donahue}, Megan and {Postman}, Marc},
        title = "{CLASH: Joint Analysis of Strong-lensing, Weak-lensing Shear, and Magnification Data for 20 Galaxy Clusters}",
      journal = {\apj},
         year = 2016,
        month = apr,
       volume = {821},
       number = {2},
          eid = {116},
        pages = {116},
          doi = {10.3847/0004-637X/821/2/116},
archivePrefix = {arXiv},
       eprint = {1507.04385},
 primaryClass = {astro-ph.CO},
       adsurl = {https://ui.adsabs.harvard.edu/abs/2016ApJ...821..116U}
}

@ARTICLE{kravtsov04,
       author = {{Kravtsov}, Andrey V. and {Berlind}, Andreas A. and {Wechsler}, Risa H. and {Klypin}, Anatoly A. and {Gottl{\"o}ber}, Stefan and {Allgood}, Brandon and {Primack}, Joel R.},
        title = "{The Dark Side of the Halo Occupation Distribution}",
      journal = {\apj},
         year = 2004,
        month = jul,
       volume = {609},
       number = {1},
        pages = {35-49},
          doi = {10.1086/420959},
archivePrefix = {arXiv},
       eprint = {astro-ph/0308519},
 primaryClass = {astro-ph},
       adsurl = {https://ui.adsabs.harvard.edu/abs/2004ApJ...609...35K}
}

@ARTICLE{taylor01,
       author = {{Taylor}, James E. and {Babul}, Arif},
        title = "{The Dynamics of Sinking Satellites around Disk Galaxies: A Poor Man's Alternative to High-Resolution Numerical Simulations}",
      journal = {\apj},
         year = 2001,
        month = oct,
       volume = {559},
       number = {2},
        pages = {716-735},
          doi = {10.1086/322276},
archivePrefix = {arXiv},
       eprint = {astro-ph/0012305},
 primaryClass = {astro-ph},
       adsurl = {https://ui.adsabs.harvard.edu/abs/2001ApJ...559..716T}
}

@ARTICLE{koopmans05,
       author = {{Koopmans}, L.~V.~E.},
        title = "{Gravitational imaging of cold dark matter substructures}",
      journal = {\mnras},
         year = 2005,
        month = nov,
       volume = {363},
       number = {4},
        pages = {1136-1144},
          doi = {10.1111/j.1365-2966.2005.09523.x},
archivePrefix = {arXiv},
       eprint = {astro-ph/0501324},
 primaryClass = {astro-ph},
       adsurl = {https://ui.adsabs.harvard.edu/abs/2005MNRAS.363.1136K}
}

@ARTICLE{vegetti09,
       author = {{Vegetti}, S. and {Koopmans}, L.~V.~E.},
        title = "{Bayesian strong gravitational-lens modelling on adaptive grids: objective detection of mass substructure in Galaxies}",
      journal = {\mnras},
         year = 2009,
        month = jan,
       volume = {392},
       number = {3},
        pages = {945-963},
          doi = {10.1111/j.1365-2966.2008.14005.x},
archivePrefix = {arXiv},
       eprint = {0805.0201},
 primaryClass = {astro-ph},
       adsurl = {https://ui.adsabs.harvard.edu/abs/2009MNRAS.392..945V}
}

@ARTICLE{ritondale19,
       author = {{Ritondale}, E. and {Vegetti}, S. and {Despali}, G. and {Auger}, M.~W. and {Koopmans}, L.~V.~E. and {McKean}, J.~P.},
        title = "{Low-mass halo perturbations in strong gravitational lenses at redshift z {\ensuremath{\sim}} 0.5 are consistent with CDM}",
      journal = {\mnras},
         year = 2019,
        month = may,
       volume = {485},
       number = {2},
        pages = {2179-2193},
          doi = {10.1093/mnras/stz464},
archivePrefix = {arXiv},
       eprint = {1811.03627},
 primaryClass = {astro-ph.CO},
       adsurl = {https://ui.adsabs.harvard.edu/abs/2019MNRAS.485.2179R}
}

@ARTICLE{vegetti18,
       author = {{Vegetti}, S. and {Despali}, G. and {Lovell}, M.~R. and {Enzi}, W.},
        title = "{Constraining sterile neutrino cosmologies with strong gravitational lensing observations at redshift z {\ensuremath{\sim}} 0.2}",
      journal = {\mnras},
         year = 2018,
        month = dec,
       volume = {481},
       number = {3},
        pages = {3661-3669},
          doi = {10.1093/mnras/sty2393},
archivePrefix = {arXiv},
       eprint = {1801.01505},
 primaryClass = {astro-ph.CO},
       adsurl = {https://ui.adsabs.harvard.edu/abs/2018MNRAS.481.3661V}
}

@ARTICLE{keeley24,
       author = {{Keeley}, Ryan E. and {Nierenberg}, A.~M. and {Gilman}, D. and {Gannon}, C. and {Birrer}, S. and {Treu}, T. and {Benson}, A.~J. and {Du}, X. and {Abazajian}, K.~N. and {Anguita}, T. and {Bennert}, V.~N. and {Djorgovski}, S.~G. and {Gupta}, K.~K. and {Hoenig}, S.~F. and {Kusenko}, A. and {Lemon}, C. and {Malkan}, M. and {Motta}, V. and {Moustakas}, L.~A. and {Oh}, Maverick S.~H. and {Sluse}, D. and {Stern}, D. and {Wechsler}, R.~H.},
        title = "{JWST lensed quasar dark matter survey - II. Strongest gravitational lensing limit on the dark matter free streaming length to date}",
      journal = {\mnras},
         year = 2024,
        month = dec,
       volume = {535},
       number = {2},
        pages = {1652-1671},
          doi = {10.1093/mnras/stae2458},
archivePrefix = {arXiv},
       eprint = {2405.01620},
 primaryClass = {astro-ph.CO},
       adsurl = {https://ui.adsabs.harvard.edu/abs/2024MNRAS.535.1652K}
}

@ARTICLE{palencia25,
       author = {{Palencia}, J.~M. and {Diego}, J.~M. and {Dai}, L. and {Pascale}, M. and {Windhorst}, R. and {Koekemoer}, A.~M. and {Li}, Sung Kei and {Kavanagh}, B.~J. and {Sun}, Fengwu and {Alfred}, Amruth and {Meena}, Ashish K. and {Broadhurst}, Thomas J. and {Kelly}, Patrick L. and {Perera}, Derek and {Williams}, Hayley and {Zitrin}, Adi},
        title = "{Microlensing at cosmological distances: Event rate predictions in the Warhol arc of MACS 0416}",
      journal = {\aap},
         year = 2025,
        month = jul,
       volume = {699},
          eid = {A295},
        pages = {A295},
          doi = {10.1051/0004-6361/202555447},
archivePrefix = {arXiv},
       eprint = {2504.07039},
 primaryClass = {astro-ph.CO},
       adsurl = {https://ui.adsabs.harvard.edu/abs/2025A&A...699A.295P}
}

@ARTICLE{bullock17,
       author = {{Bullock}, James S. and {Boylan-Kolchin}, Michael},
        title = "{Small-Scale Challenges to the {\ensuremath{\Lambda}}CDM Paradigm}",
      journal = {\araa},
         year = 2017,
        month = aug,
       volume = {55},
       number = {1},
        pages = {343-387},
          doi = {10.1146/annurev-astro-091916-055313},
archivePrefix = {arXiv},
       eprint = {1707.04256},
 primaryClass = {astro-ph.CO},
       adsurl = {https://ui.adsabs.harvard.edu/abs/2017ARA&A..55..343B}
}

@ARTICLE{sales22,
       author = {{Sales}, Laura V. and {Wetzel}, Andrew and {Fattahi}, Azadeh},
        title = "{Baryonic solutions and challenges for cosmological models of dwarf galaxies}",
      journal = {Nature Astronomy},
         year = 2022,
        month = jun,
       volume = {6},
        pages = {897-910},
          doi = {10.1038/s41550-022-01689-w},
archivePrefix = {arXiv},
       eprint = {2206.05295},
 primaryClass = {astro-ph.GA},
       adsurl = {https://ui.adsabs.harvard.edu/abs/2022NatAs...6..897S}
}

@ARTICLE{penarrubia08,
       author = {{Pe{\~n}arrubia}, Jorge and {Navarro}, Julio F. and {McConnachie}, Alan W.},
        title = "{The Tidal Evolution of Local Group Dwarf Spheroidals}",
      journal = {\apj},
         year = 2008,
        month = jan,
       volume = {673},
       number = {1},
        pages = {226-240},
          doi = {10.1086/523686},
archivePrefix = {arXiv},
       eprint = {0708.3087},
 primaryClass = {astro-ph},
       adsurl = {https://ui.adsabs.harvard.edu/abs/2008ApJ...673..226P}
}

@ARTICLE{fudamoto2025,
       author = {{Fudamoto}, Yoshinobu and {Sun}, Fengwu and {Diego}, Jose M. and {Dai}, Liang and {Oguri}, Masamune and {Zitrin}, Adi and {Zackrisson}, Erik and {Jauzac}, Mathilde and {Lagattuta}, David J. and {Egami}, Eiichi and {Iani}, Edoardo and {Windhorst}, Rogier A. and {Abe}, Katsuya T. and {Bauer}, Franz Erik and {Bian}, Fuyan and {Bhatawdekar}, Rachana and {Broadhurst}, Thomas J. and {Cai}, Zheng and {Chen}, Chian-Chou and {Chen}, Wenlei and {Cohen}, Seth H. and {Conselice}, Christopher J. and {Espada}, Daniel and {Foo}, Nicholas and {Frye}, Brenda L. and {Fujimoto}, Seiji and {Furtak}, Lukas J. and {Golubchik}, Miriam and {Hsiao}, Tiger Yu-Yang and {Jolly}, Jean-Baptiste and {Kawai}, Hiroki and {Kelly}, Patrick L. and {Koekemoer}, Anton M. and {Kohno}, Kotaro and {Kokorev}, Vasily and {Li}, Mingyu and {Li}, Zihao and {Lin}, Xiaojing and {Magdis}, Georgios E. and {Meena}, Ashish K. and {Niemiec}, Anna and {Nabizadeh}, Armin and {Richard}, Johan and {Steinhardt}, Charles L. and {Wu}, Yunjing and {Zhu}, Yongda and {Zou}, Siwei},
        title = "{Identification of more than 40 gravitationally magnified stars in a galaxy at redshift 0.725}",
      journal = {Nature Astronomy},
         year = 2025,
        month = mar,
       volume = {9},
        pages = {428-437},
          doi = {10.1038/s41550-024-02432-3},
archivePrefix = {arXiv},
       eprint = {2404.08045},
 primaryClass = {astro-ph.GA},
       adsurl = {https://ui.adsabs.harvard.edu/abs/2025NatAs...9..428F}
}

@ARTICLE{pascale25,
       author = {{Pascale}, Massimo and {Dai}, Liang and {Frye}, Brenda L. and {Beverage}, Aliza G.},
        title = "{Is Earendel a Star Cluster?: Metal-poor Globular Cluster Progenitors at z {\ensuremath{\sim}} 6}",
      journal = {\apjl},
         year = 2025,
        month = aug,
       volume = {988},
       number = {2},
          eid = {L76},
        pages = {L76},
          doi = {10.3847/2041-8213/aded93},
archivePrefix = {arXiv},
       eprint = {2507.05483},
 primaryClass = {astro-ph.GA},
       adsurl = {https://ui.adsabs.harvard.edu/abs/2025ApJ...988L..76P}
}

@ARTICLE{bradley25,
       author = {{Bradley}, Larry D. and {Adamo}, Angela and {Vanzella}, Eros and {Sharon}, Keren and {Brammer}, Gabriel and {Coe}, Dan and {Diego}, Jose M. and {Kokorev}, Vasily and {Mahler}, Guillaume and {Oguri}, Masamune and {Abdurro'uf} and {Bhatawdekar}, Rachana and {Christensen}, Lise and {Fujimoto}, Seiji and {Hashimoto}, Takuya and {Hsiao}, Tiger Y. -Y. and {Inoue}, Akio K. and {Jim{\'e}nez-Teja}, Yolanda and {Messa}, Matteo and {Norman}, Colin and {Ricotti}, Massimo and {Tamura}, Yoichi and {Windhorst}, Rogier A. and {Xu}, Xinfeng and {Zitrin}, Adi},
        title = "{Unveiling the Cosmic Gems Arc at z {\ensuremath{\sim}} 10 with JWST NIRCam}",
      journal = {\apj},
         year = 2025,
        month = sep,
       volume = {991},
       number = {1},
          eid = {32},
        pages = {32},
          doi = {10.3847/1538-4357/adf638},
archivePrefix = {arXiv},
       eprint = {2404.10770},
 primaryClass = {astro-ph.GA},
       adsurl = {https://ui.adsabs.harvard.edu/abs/2025ApJ...991...32B}
}

@ARTICLE{umetsu14,
       author = {{Umetsu}, Keiichi and {Medezinski}, Elinor and {Nonino}, Mario and {Merten}, Julian and {Postman}, Marc and {Meneghetti}, Massimo and {Donahue}, Megan and {Czakon}, Nicole and {Molino}, Alberto and {Seitz}, Stella and {Gruen}, Daniel and {Lemze}, Doron and {Balestra}, Italo and {Ben{\'\i}tez}, Narciso and {Biviano}, Andrea and {Broadhurst}, Tom and {Ford}, Holland and {Grillo}, Claudio and {Koekemoer}, Anton and {Melchior}, Peter and {Mercurio}, Amata and {Moustakas}, John and {Rosati}, Piero and {Zitrin}, Adi},
        title = "{CLASH: Weak-lensing Shear-and-magnification Analysis of 20 Galaxy Clusters}",
      journal = {\apj},
         year = 2014,
        month = nov,
       volume = {795},
       number = {2},
          eid = {163},
        pages = {163},
          doi = {10.1088/0004-637X/795/2/163},
archivePrefix = {arXiv},
       eprint = {1404.1375},
 primaryClass = {astro-ph.CO},
       adsurl = {https://ui.adsabs.harvard.edu/abs/2014ApJ...795..163U}
}

@ARTICLE{doppel25,
       author = {{Doppel}, Jessica E. and {Jauzac}, Mathilde and {Lagattuta}, David J. and {Fattahi}, Azadeh and {Mahler}, Guillaume},
        title = "{Tiny galaxies and dark substructures: exploring the ``dark'' subhaloes in TNG50}",
      journal = {arXiv e-prints},
         year = 2025,
        month = jun,
          eid = {arXiv:2506.09122},
        pages = {arXiv:2506.09122},
          doi = {10.48550/arXiv.2506.09122},
archivePrefix = {arXiv},
       eprint = {2506.09122},
 primaryClass = {astro-ph.GA},
       adsurl = {https://ui.adsabs.harvard.edu/abs/2025arXiv250609122D}
}




\appendix

\section{Testing the Assumption of a Linear Critical Curve}\label{txt:lineartest}

\begin{figure}
    \centering
    \includegraphics[trim={5.1cm 0.35cm 5.1cm 0.35cm},clip,width=0.49\textwidth]{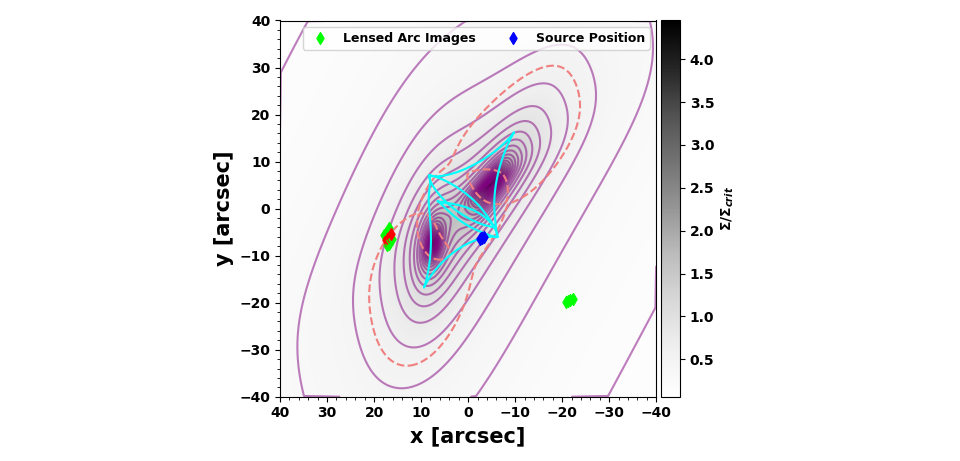}
\caption{ Same as Figure \ref{fig:mainlens}, but for a source that forms nearer to the macrolens caustic cusp. The local curvature of the critical curve for this arc is greater than the fiducial model since we are far from the fold. The test arc in this case contains four lensed knots that span 1.75" along the critical curve, representing a more extreme case.}  
\label{fig:mainlens_cuspsource}
\end{figure}

\begin{figure}
    \centering
    \includegraphics[trim={5.1cm 0.35cm 5.1cm 0.35cm},clip,width=0.49\textwidth]{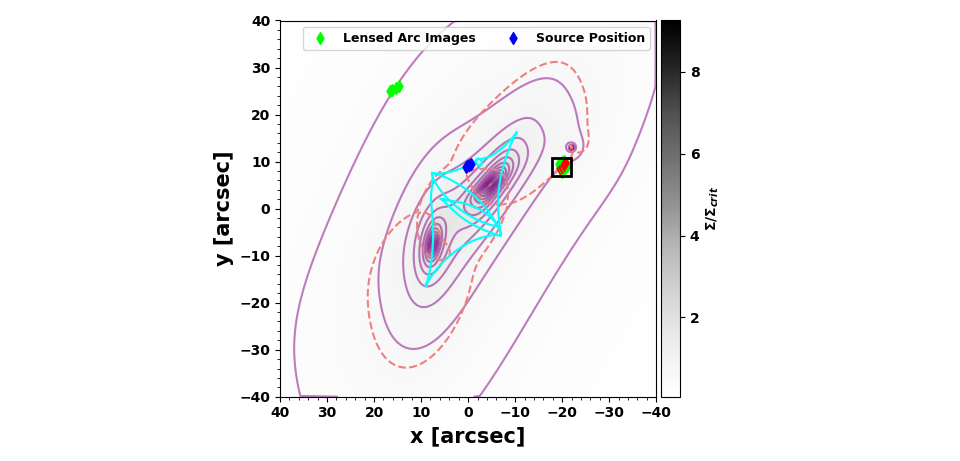}
    \includegraphics[trim={5.1cm 0.35cm 5.1cm 0.35cm},clip,width=0.49\textwidth]{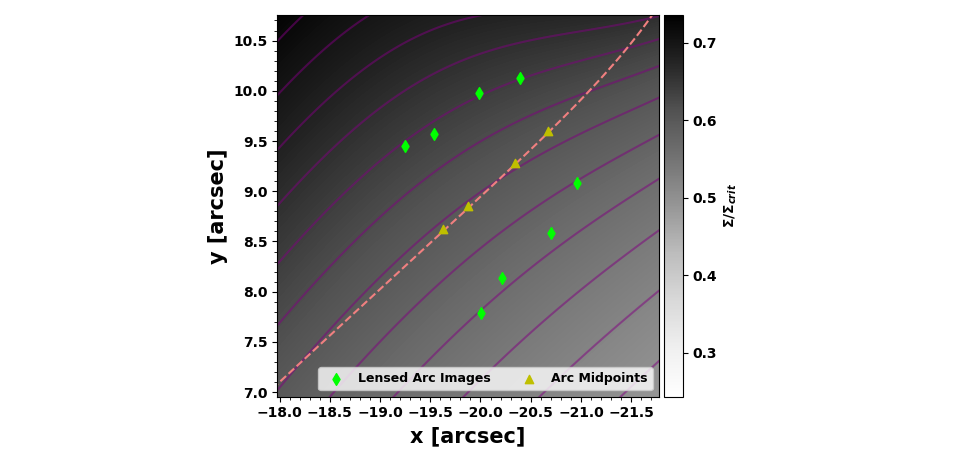}
\caption{ {\it Top:} Same as Figure \ref{fig:mainlens}, with a cluster member galaxy located $\sim3$" from the subhalo window. This represents a small macrolens perturbation to the window where we simulate subhalo populations. {\it Bottom:} Same as bottom panel of Figure \ref{fig:perpendiculararc}, but with the macrolens including a cluster member galaxy. The cluster member is placed just outside this window (in the top right), and produces a noticeable change in the local macrolens density profile.}  
\label{fig:arcwgal}
\end{figure}

Throughout this paper, we assume that on small scales for a smooth cluster-scale lens model, the critical curve is approximately linear. This assumption is motivated by previous analyses \citep{venumadhav17,dai18} and is crucial to our analysis since the image midpoints will form along a straight line. It is important to determine the limits of this assumption, as this will determine whether it is appropriate to apply this assumption to observed arcs. This Appendix conducts two main tests: (1) A test of the limits of $\xi$ for deviations from a straight critical curve, and (2) A test of how much a deviation from linearity causes a bias in the inference of $f_{\rm sub}$. In the second test, we intend to show that any potential bias in the inference as a result of a nonlinear critical curve occur for arcs far more nonlinear than any considered in this work.

We begin with Test (1). The two main contributors against the linear approximation are intrinsic curvature of the cluster-scale critical curve and local perturbations from cluster member galaxies. Their effects will differ for perpendicular and parallel arcs, due to the linearity of the midpoints depending strongly on how far along the critical curve an arc spans. We test these two contributors in this Appendix based on how well $\xi$ (equation \ref{eq:asymmetry}) as a metric captures the linearity. We note that for the tests presented in this section, all measurements are exact in order to rigorously test the limit of linearity.

We first start with our fiducial model as described in Section \ref{txt:lensedarc}. The perpendicular arc spans 0.23" along the critical curve, and the unperturbed images have an asymmetry metric $\xi = -7.42$. Likewise, the parallel arc spans 1.02" along the critical curve with $\xi = -4.64$. As expected, the unperturbed asymmetry is larger for the parallel arc. 

An easy way to determine if an arc is impacted by intrinsic curvature is by fitting the midpoints to both a linear and quadratic curve. The quadratic fit will estimate if the midpoints are subject to curvature. From these fits, we can conduct a $\chi^2$ test. If the $p$-value for the linear fit falls below 0.05, then we can conclude that a linear assumption is invalid. Otherwise, we are able to conclude that a linear assumption is sufficient. It is important to note that for larger arcs a quadratic fit will almost always be better than a linear one. However, the test we describe aims to determine if a linear fit is sufficient, and not necessarily better. In this way, we are determining if $\xi$ as a summary statistic remains sufficiently informative for a given arc. Later, in Test (2), we will examine the potential biases from nonlinear critical curves

Doing this for the fiducial model, we find the linear $p$-values to be 0.99 and 0.97 for the perpendicular and parallel arcs, respectively\footnote{We note that the number of lensed knots for this test needs to be increased to at least four in order to have nonzero degrees of freedom. In practice, this has virtually no effect on this test since there are no perturbations from subhalos.}. For the both arcs, a linear fit is better than a quadratic by a significant margin. Therefore, we can safely conclude that for the fiducial model, a linear assumption is valid. 

The results for the fiducial model also confirm our underlying restriction that we can only apply our results to arcs near the caustic fold. However, in practice, the true macrolens model is unknown, so there is some uncertainty as to where the caustic fold lies. In principle, one can roughly determine whether an arc lies on the caustic fold based on the configuration of the arc and the location of its third counterimage. Ignoring this for now, let's now assume an arc closer to the caustic cusp (see Figure \ref{fig:mainlens_cuspsource}). The arc in this case will now form in a more intrinsically curved portion of the macrolens. Since we have already established that smaller arcs along the critical curve satisfy the linear assumption, we now test a more extreme arc that spans 1.75" along the critical curve. The measured asymmetry is $\xi = -3.31$. The linear and quadratic $p$-values are 0.16 and 0.77, respectively. Therefore, we conclude that while a quadratic offers a much better fit to this arc, given the intrinsic curvature of the critical curve, a linear fit is still sufficient to approximate the arc. This test represents the most extreme case considered in this paper and depicts the rough limit of the linear assumption to be $\xi \sim-3$.

Lastly, we test the effect of cluster member galaxies on the linear assumption. We note that cluster members that form very near to the arc will obviously distort the critical curve and induce asymmetries \citep[e.g.][]{diego23c}. Instead of these obvious cases, we instead want to test the subtle presence of cluster members and how much they contribute to deviations from linearity. To test this, we add a small cluster member galaxy $\sim3$" from the fiducial arc, as shown in Figure \ref{fig:arcwgal}. The galaxy is located just outside of the subhalo window, such that it its presence is still noticeable in the background density profile, but not dominating the local mass distribution. As can be seen, the galaxy produces its own microcaustic near to the source position. The cluster member is modeled with a Singular Isothermal Sphere (SIS):
\begin{equation}
    \Sigma\left(\boldsymbol{\theta}\right) = \frac{\sigma_v^2}{2GD_d\sqrt{\theta_1^2 + \theta_2^2}}\label{eq:SIS}
\end{equation}
where $\sigma_v = 200$ km s$^{-1}$, giving a total mass of $\sim5 \times 10^{12} M_{\odot}$.

The test arc shown in Figure \ref{fig:arcwgal} spans 1.44" along the critical curve, with asymmetry of $\xi = -4.02$. The $p$-value for a linear fit is 0.75, once again indicating that a linear fit is sufficient.

We emphasize that the presence of nearby cluster member galaxies and their effect on the linear assumption should be accounted for on a case-by-case basis. If a lensed arc of interest is nearby to any galaxies, we suggest repeating tests such as this one to ensure that the effect of the galaxy is minimal on the critical curve curvature.

To summarize the Appendix thus far, we have shown that $\xi$ as a summary statistic is sufficient to capture the linearity of lensed arcs including in cases where the arc forms away from a fold or near a cluster member galaxy. Now, we shift our focus to determining how a systematic bias in our inference can manifest from nonlinear critical curves (Test (2)).

\begin{figure}
    \centering
    \includegraphics[trim={5.1cm 0.35cm 5.1cm 0.35cm},clip,width=0.49\textwidth]{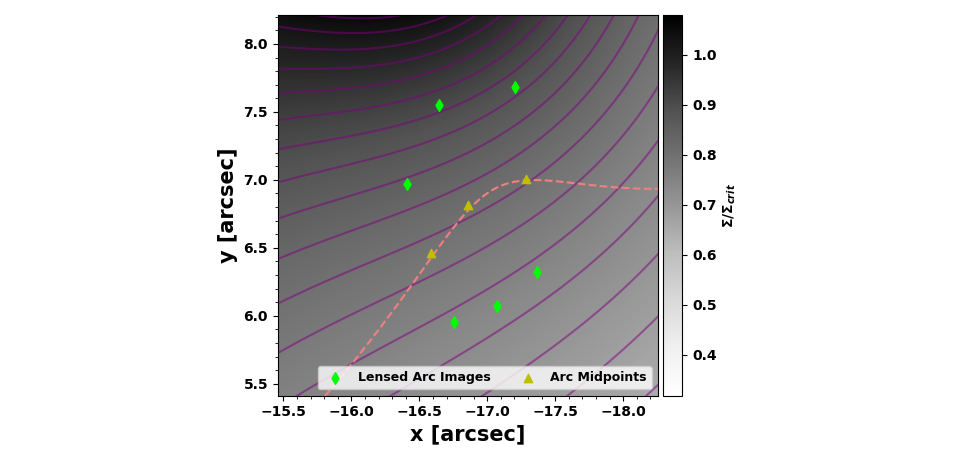}
    \includegraphics[trim={0cm 0.35cm 0cm 0.35cm},clip,width=0.49\textwidth]{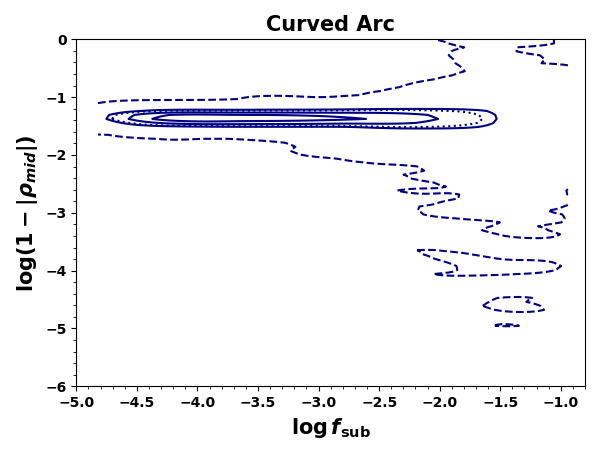}
\caption{{\it Top:} Same as the bottom panel of Figure \ref{fig:arcwgal}, with a cluster member galaxy producing a more pronounced curvature in the critical curve. This represents a significant macrolens perturbation to the window where we simulate subhalo populations. {\it Bottom:} Same as Figure \ref{fig:parameterspace}, but for the arc shown in the top panel. The parameter space probed by a significantly curved critical curve is noticeably different. The dashed contours indicate the 99.7\% confidence interval, shown to highlight the behavior of the parameter space for samples with high $f_{\rm sub}$.}  
\label{fig:curvedarc}
\end{figure}

Focusing on Test (2) now, we determine the level of bias that is caused by an intrinsically curved critical curve. To evaluate this, we compute the parameter space that is probed by our method for an obviously nonlinear critical curve. If the parameter space is different from Figure \ref{fig:parameterspace}, we can conclude that bias is introduced when the critical curve is nonlinear. Figure \ref{fig:curvedarc} shows the arc that we consider. In this case, the critical curve is significantly perturbed by a large galaxy outside of the simulation window, producing a pronounced kink in the local critical curve where the arc forms. The galaxy is more diffuse and thus behaves in the simulation window as a larger scale perturbation than the smaller galaxy considered in Figure \ref{fig:arcwgal}. The measured $\xi$ is -1.38.

The bottom panel of Figure \ref{fig:curvedarc} shows the parameter space that is probed when applying our method to this arc. In comparison with the linear case shown in Figure \ref{fig:parameterspace}, there is a significant change in the parameter space for a curved arc. There are two primary conclusions that we draw from this. The first is that since the macrolens critical curve contains high curvature prior to the addition of subhalos, populations of subhalos with low $f_{\rm sub}$ barely affect the asymmetry. Thus, the spread in $\xi$ at low $f_{\rm sub}$ is dominated by astrometric noise. The second conclusion we draw is that the constraining power of the method is effectively destroyed. This is because at high $f_{\rm sub}$, there is a roughly equal chance that subhalos will produce a more symmetric arc rather than an asymmetric arc. Therefore, any measurement of $\xi$ that is away from the initial asymmetry of the arc will constrain the same value of $f_{\rm sub}$. This demonstrates that the linear assumption is required for our method to yield meaningful results and that significant bias is introduced if the macrolens critical curve is nonlinear. 

We emphasize that this example is for an outlier case and that all the arcs we consider in this paper are much closer to linear than this. To quantify this, we introduce a metric based on the change in the direction of the tangential stretch eigenvector. We define $\left|\frac{d\phi}{ds}\right|$ to be the absolute value of the derivative of the angle $\phi$ of the tangential stretch eigenvector with respect to the critical curve position $s$. Evaluating this in the region spanned by a lensed arc allows us to take the mean $\left|\overline{\frac{d\phi}{ds}}\right|$ as a quantifiable metric of how curved an arc is. If a critical curve is linear, $\left|\overline{\frac{d\phi}{ds}}\right| = 0$, while deviations from linearity result in increasingly large values. For the highly curved arc in Figure \ref{fig:curvedarc}, $\left|\overline{\frac{d\phi}{ds}}\right| = 0.17$. For AS1063 System 1 and Warhol, $\left|\overline{\frac{d\phi}{ds}}\right|$ is 0.02 and 0.01, respectively, which is closer to our fiducial model $\left|\overline{\frac{d\phi}{ds}}\right|$ of 0.03. Since the two real arcs we consider in this paper are straighter than the highly curved arc by an order of magnitude, and very close to that of our fiducial model, we conclude that the level of bias in our inference is minimal.

From the two tests conducted in this Appendix, we conclude the following:
\begin{itemize}
    \item $\xi$ as a metric for the asymmetry of lensed arcs is moderately resistant to small deviations from the traditional symmetry setup. We show in Test (1) that in cases where the arc forms away from the caustic fold or is perturbed by a cluster member, $\xi$ is still able to sufficiently capture linearity. This shows that small perturbations to linearity in the critical curve do not destroy the use of $\xi$ as a summary statistic. This is an important result, as it broadens the usability of $\xi$ to a range of lensed arcs in the future that may not be as ideal as our fiducial model.
    \item Bias in our inference will occur for large deviations away from a linear critical curve. In cases with a highly nonlinear macrolens critical curve, the constraining power of the method will be destroyed. This level of bias, however, will only occur for arcs that exhibit significantly more curvature than any of the arcs considered in our paper. Since AS1063 System 1 and Warhol both have similar levels of curvature to our fiducial model, we do not expect bias in our inference of $f_{\rm sub}$.
    
\end{itemize}

\section{Higher Order Lensing Effects Near the Critical Curve}\label{txt:higherorder}

In Section \ref{txt:cclenstheory} we demonstrated that sources near the caustic fold will produce images that are symmetric across the critical curve (equation \ref{eq:midpointposition}), and thus contain image midpoints that trace the critical curve. This result is valid in the limit that images are very close to the critical curve. It is therefore necessary to examine at what point this approximation breaks down, as larger arcs that contain knots further away from the critical curve may no longer contain midpoints along the critical curve. We examine this limit analytically, beginning first by expanding equation \ref{eq:order2expansion} to the third order:
\begin{multline}\label{eq:order3expansion}
\boldsymbol{\beta_i}\left(\boldsymbol{\theta_c} +\delta\boldsymbol{\theta}\right) = \boldsymbol{\beta_i}(\boldsymbol{\theta_c}) + \sum_j \frac{\partial\boldsymbol{\beta_i}}{\partial\boldsymbol{\theta_j}}\bigg\rvert_{\boldsymbol{\theta_c}}\delta\boldsymbol{\theta_j} + \frac{1}{2}\sum_{j,k}\frac{\partial^2\boldsymbol{\beta_i}}{\partial\boldsymbol{\theta}_j\partial\boldsymbol{\theta}_k}\bigg\rvert_{\boldsymbol{\theta_c}}\delta\boldsymbol{\theta_j}\delta\boldsymbol{\theta_k} \\
+ \frac{1}{6}\sum_{j,k,l}\frac{\partial^3\boldsymbol{\beta_i}}{\partial\boldsymbol{\theta}_j\partial\boldsymbol{\theta}_k\partial\boldsymbol{\theta}_l}\bigg\rvert_{\boldsymbol{\theta_c}}\delta\boldsymbol{\theta_j}\delta\boldsymbol{\theta_k}\delta\boldsymbol{\theta_l}
\end{multline}
To solve, we proceed the same as we do in the Section \ref{txt:cclenstheory}, writing to leading order:
\begin{equation}
    \beta_2 = \frac{1}{2}\frac{\partial^2\beta_2}{\partial\theta_2^2}\bigg\rvert_{0}\delta\theta_2^2 + \frac{1}{6}\frac{\partial^3\beta_2}{\partial\theta_2^3}\bigg\rvert_{0}\delta\theta_2^3\label{eq:beta2higherorder}
\end{equation}
where this equation is the higher order equivalent of equation \ref{eq:beta2}. 

The goal is to solve equation \ref{eq:beta2higherorder} to determine the new form of equation \ref{eq:midpointposition}, which we call $\delta\theta_2^{\rm tot}$. Since we are interested in how the symmetry is broken, we define: $\delta\theta_2^{\rm tot} = \delta\theta^{\rm init}_2 + \delta\theta^{'}_2$ where $\delta\theta^{\rm init}_2$ is the initial result from equation \ref{eq:midpointposition}. This reduces the problem to solving for $\delta\theta^{'}_2$, which is the correction term from including the third order expansion. To simplify notation, let $b = \frac{\partial^2\beta_2}{\partial\theta_2^2}\bigg\rvert_{0}$ and $c = \frac{\partial^3\beta_2}{\partial\theta_2^3}\bigg\rvert_{0}$, where it is implicitly assumed that $b >> c$. This is likely a good assumption for smooth potential variations, like those considered here, but would break down for small compact variations, like those from subhalos.

Plugging $\delta\theta_2^{\rm tot}$ into the third order expansion gives:
\begin{equation}
    \beta_2 = \frac{1}{6}c\left(\delta\theta^{\rm init}_2 + \delta\theta^{'}_2\right)^3 + \frac{1}{2}b\left(\delta\theta^{\rm init}_2 + \delta\theta^{'}_2\right)^2
\end{equation}
Since $\delta\theta^{'}_2$ is very small, we continue by only keeping leading order terms of $\delta\theta^{\rm init}_2$ and $\delta\theta^{'}_2$. We are assuming $\delta\theta^{\rm init}_2 >>\delta\theta^{'}_2$, such that $\beta_2$ remains approximately equal to the second order result (equation \ref{eq:beta2} in Section \ref{txt:cclenstheory}), allowing $\beta_2$ to cancel with the $\frac{1}{2}b\left(\delta\theta^{\rm init}_2\right)^2$ term:
\begin{equation}
    \frac{1}{6}c\left(\delta\theta^{\rm init}_2\right)^3 + \frac{1}{2}c\left(\delta\theta^{\rm init}_2\right)^2\delta\theta^{'}_2 +b\delta\theta^{\rm init}_2\delta\theta^{'}_2 \approx 0
\end{equation}
Solving this with our approximations gives (with the denominator dominated by the term including $b$):
\begin{equation}
    \delta\theta^{'}_2 = -\frac{\frac{1}{6}c\left(\delta\theta^{\rm init}_2\right)^2}{\frac{1}{2}c\delta\theta^{\rm init}_2 + b} \approx -\frac{c}{6b}\left(\delta\theta^{\rm init}_2\right)^2
\end{equation}
This result for $\delta\theta^{'}_2$ is interpreted as the midpoint shift off the critical curve position. Importantly, the shift scales quadratically with the image distance from the critical curve $\delta\theta^{\rm init}_2$, meaning that this midpoint shift becomes important for increasingly larger distances from the critical curve. 

In order to determine at what distance this midpoint shift becomes important, it is necessary to calculate the magnitude of $\delta\theta^{'}_2$ for a cluster scale lens. This is analytically difficult as the derivative terms $b$ and $c$ cannot be solved generally. To simplify this problem analytically such that we can get a rough scale of the distances at which the midpoint shift becomes important, we consider the following setup.

We assume that the cluster lens can be approximated as an SIS with a power-law lens potential: $\psi(r) \propto r^{\alpha}$. This allows the derivative terms to be solved analytically because the critical curve is a circle located at the Einstein radius $\theta_E$. This is a subtly important simplification because it allows the critical direction to be radial. If circular symmetry is not present, a change of variable is required, complicating the problem. We note that the usage of an SIS is an imperfect approximation because an SIS will not form an image pair that straddles the critical curve. However, since the SIS is only used in this calculation to calculate the derivative terms analytically (i.e. the SIS is not used to calculate the image positions $\delta\theta^{\rm init}_2$) and scale $\delta\theta^{'}_2$ with some property of the lens (in this case $\theta_E$), we justify this choice as sufficient.

With our simplifications and justifications stated, an SIS power-law lens potential gives:
\begin{equation}
    |\delta\theta^{'}_2| \approx \frac{\alpha - 3}{6\theta_E}\left(\delta\theta^{\rm init}_2\right)^2 \propto \frac{\left(\delta\theta^{\rm init}_2\right)^2}{\theta_E}
\end{equation}
For an exact SIS, $\alpha=1$, so $ |\delta\theta^{'}_2| \approx (\theta_E)^{-1}{\left(\delta\theta^{\rm init}_2\right)^2}/3$. To apply this as an approximation to the observed arcs all we need to do is calculate $\theta_E$. This result is intuitive, as the mass of the lens increases the shift in the midpoint position off the critical curve is reduced. For the fiducial perpendicular arc we examine in Section \ref{txt:lensedarc}, this calculation predicts $\delta\theta^{'}_2 = 0.02$". We measure the exact shift to be 0.04", which is reasonably close in spite of the simplifications. The disagreement is due to combination of simplifying assumptions, source displacement from the exact fold position, and complexity in the lens potential. Nonetheless, this gives a sense of the scale of the midpoint shift, where it is on the same order as the astrometric uncertainty. Importantly, relative to $\delta\theta^{\rm init}_2$ for this arc, which is observed to be 1.33'', $\delta\theta^{'}_2/\delta\theta^{\rm init}_2 = 0.03$, making up a very small shift relative to the size of the arc. 

For the real arcs we study in Section \ref{txt:real}, we can repeat the calculation with greater interest since these arcs are much larger than the fiducial ones. For Warhol and AS1063 Sys 1, we calculate $\delta\theta^{'}_2$ to be 0.03" and 0.1", respectively. Likewise, the relative shift to the arc size $\delta\theta^{'}_2/\delta\theta^{\rm init}_2$ is 0.02 and 0.03 for Warhol and AS1063 Sys 1, respectively, once again indicating a small shift relative to the arc size. AS1063 Sys 1 is of greater interest since the shift is expected to be larger than the astrometric uncertainty, corroborating the complications previously discussed in Section \ref{txt:obsresults}.

We conclude that the midpoint shift off the critical curve is primarily due to a third order effect. The magnitude of the shift scales with the image distance from the critical curve and inversely scales with the mass of the lens. This effect can be captured in the astrometric uncertainty for our fiducial arcs, allowing us to conclude that it is a negligible effect to the efficacy of our method. In practice, however, greater care must be taken to acknowledge this midpoint shift, as evidenced by AS1063 Sys 1. We also emphasize that the shape of the lens potential can contribute to a highly nontrivial expected shift. Likewise, a constant offset could be a result of the source lying off the exact caustic fold position. While these will bias the estimation of the exact critical curve position (something not directly observed), it will not bias $\xi$, as this is sensitive only to uncorrelated shifts in the midpoint positions.

\section{The Effect of Individual Subhalos on Asymmetry}\label{txt:individualsubhalos}

As can be seen in Figure \ref{fig:asymrealizations}, some realizations of asymmetry appear to be dominated by individual massive subhalos. This motivates exploring how the summary statistic $\xi$ is affected by the presence of individual massive subhalos. It can be shown that there is indeed a weak positive correlation between the most massive subhalo of a given realization and $\xi$. This correlation is intuitively expected, however, it remains subdominant due to considerable scatter in the relation. Furthermore, our statistical framework already accounts for the relative likelihood that an asymmetry comes from a single massive subhalo by averaging over many realizations. Therefore, this point is rendered a conceptual point rather than a true statistical effect in our method.

The more important effect with regards to our statistical method is the population level maximum subhalo mass. Our initial choice to simulate subhalos between $10^6 - 10^{10} M_{\odot}$ is consistent with previous works. However, recent work has explored the validity of this assumption. For example, \cite{doppel25} finds with TNG50 that truly observationally dark subhalos are most likely to have mass $\lesssim10^{8.32} M_{\odot}$. This perhaps motivates reducing the upper bound of the mass range used for our simulation. Here, we vary the upper bounds of the simulation mass range for subhalos to study its effect on $\xi$.

\begin{figure}
    \centering
    \includegraphics[trim={0cm 0.35cm 0cm 0.35cm},clip,width=0.49\textwidth]{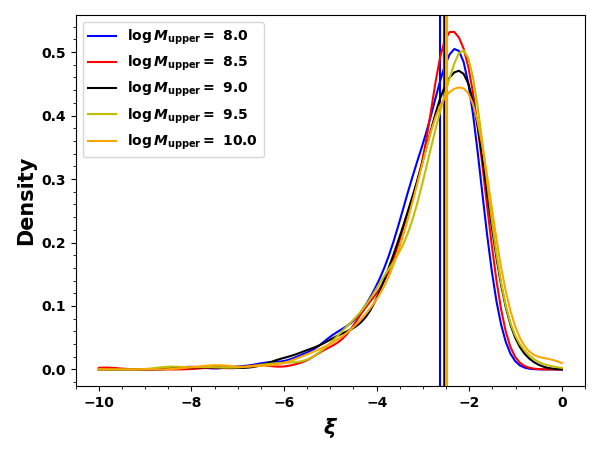}
\caption{ Density distribution of the measured $\xi$ with $\delta_{xy}=0.01$" for various upper mass range bounds. The lower mass range bound is $10^6 M_{\odot}$ in each case. Vertical lines indicate the median $\xi$ for each distribution.}  
\label{fig:massrange}
\end{figure}

We vary the upper mass range bound $M_{\rm upper}$, while keeping the lower mass range bound fixed. We sample 1000 realizations per $M_{\rm upper}$ bin, and only reduce $M_{\rm upper}$ in intervals of 0.5 (in log space). We do not consider increasing $M_{\rm upper}$ since subhalos with masses $>10^{10} M_{\odot}$ are unlikely and are more likely to have an electromagnetic counterpart. Since we are interested in how this would change our results (which are derived from $\xi$), we include $\delta_{xy}=0.01$" for this test. Figure \ref{fig:massrange} shows these results. As can be seen, the distribution of $\xi$ is relatively constant across all $M_{\rm upper}$. There is a very small trend that can be observed where the median $\xi$ for each distribution slightly increases with $M_{\rm upper}$, however, it is statistically insignificant in this case since the fractional change in the median $\xi$ between each $M_{\rm upper}$ bin, relative to the $95\%$ confidence interval, is $\sim2\%$. For these reasons, we conclude that changing $M_{\rm upper}$ would have a negligible effect on our results. Likewise, varying the lower mass range bound, while keeping the upper bound constant, also does not effect our results.

We conclude that the most massive subhalos in the simulation are a subdominant effect on $\xi$ that are already accounted for in our statistical model. In addition, properly accounting for the relative likelihood change from individual massive subhalos by varying the upper mass range bounds that we sample subhalos between has a very negligible effect on the results. We mention that in the future, greater care should be taken when determining the upper and lower mass range bounds in order to ensure that physically realistic subhalo masses are being used. This will be especially important once $\delta_{xy}$ becomes minimized for most systems.

\section{Testing the Convergence of the Posterior Distributions}\label{txt:convergencetest}

We conduct a convergence test to determine if our results are dependent on the value of $N$, the number of realizations. To do so, we randomly discard one-third of the samples and recalculate the posterior with the remaining samples with the same acceptance criterion of the best 100. If the resampled posterior from a smaller $N$ remains mostly unchanged from before, then we can conclude that the posterior is converged. Because we are randomly discarding samples, there will be an associated scatter in the resampled posterior shape. To account for this, we resample 50 times and calculate the posterior medians. If the range in resampled posterior medians is low relative to the 68\% confidence interval of the original posterior sample (from $N = 20000$), then we conclude that the posterior is sufficiently converged relative to the uncertainty. When we divide these two quantities, we will refer to this as the relative median scatter, where a relative median scatter of 1.0 indicates that the resampled posterior median range is equal to the original 68\% confidence interval. We note that this definition serves as an upper limit, as it can be susceptible to outliers.

\begin{figure}
    \centering
    \includegraphics[trim={0cm 0cm 0cm 0cm},clip,width=0.49\textwidth]{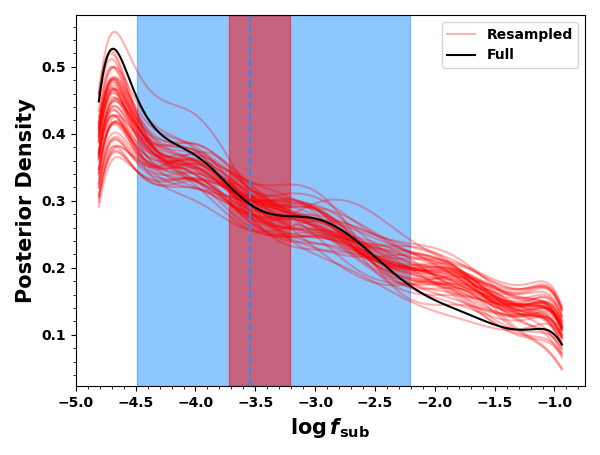}
    \includegraphics[trim={0cm 0cm 0cm 0cm},clip,width=0.49\textwidth]{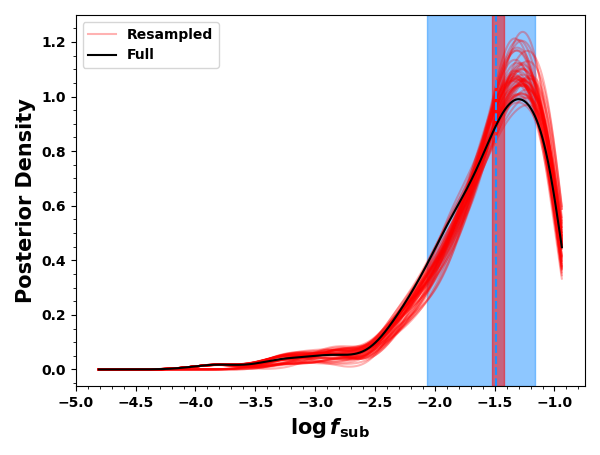}
\caption{Convergence test posterior results for the two perpendicular mock arcs shown in Figure \ref{fig:mockindividual} of the main text. The top and bottom panels correspond to the same arcs as those in the left and right columns of Figure \ref{fig:mockindividual}, respectively. The black lines indicate the original posterior, the same as the dashed black line in Figure \ref{fig:mockindividual}. The red lines show resampled posteriors after discarding one-third of the original posterior samples. The blue dashed line and shaded region indicate the original posterior medians and 68\% confidence interval. The red shaded region indicates the full posterior median range from the resampled posteriors.}
\label{fig:convergence}
\end{figure}

We perform this test for the same mock arcs that we analyzed in Figure \ref{fig:mockindividual} of the main text. These results are shown in Figure \ref{fig:convergence}, where the top and bottom panels correspond to the left and right columns of Figure \ref{fig:mockindividual}. For clarity, we refer to these as the left and right mock arcs for this discussion. As can be seen, the resampled posteriors are relatively unchanged. We quantify this by dividing the resampled posterior median range (length of the red shaded region of Figure \ref{fig:convergence}) with the original 68\% confidence interval (length of the blue shaded region). We find that the left and right mock arcs exhibit a relative median scatter of 0.2 and 0.1, respectively. Because this scatter is much lower than the original posterior 68\% confidence interval, we conclude that the original posteriors are stable and converged relative to the uncertainty. Repeating our statistical method with larger values of $N$ will only improve the precision of the results, and we encourage doing so in future works, provided the appropriate computational resources.

\bsp	
\label{lastpage}
\end{document}